\documentclass[twocolumn]{aastex61}
\pdfoutput=1 
\usepackage{amsmath,amstext}
\usepackage[T1]{fontenc}
\usepackage{apjfonts} 
\usepackage[figure,figure*]{hypcap}
\usepackage{color, soul}
\usepackage[normalem]{ulem}
\usepackage{xspace}
\usepackage{enumerate}
\usepackage{hyperref}
\usepackage{chngcntr}
\usepackage[normalem]{ulem}

\definecolor{royalpurple}{cmyk}{0.50, 0.60, 0.0, 0.3}

\newcommand*{\rom}[1]{\expandafter\@slowromancap\romannumeral #1@}
\newcommand{\rnum}[1]{\uppercase\expandafter{\romannumeral #1\relax}}
\newcommand{\chandra}{\textit{Chandra}\xspace}
\newcommand{\suzaku}{\textit{Suzaku}\xspace}
\newcommand{\nustar}{\textit{NuSTAR}\xspace}
\newcommand{\asca}{\textit{ASCA}\xspace}

\newcommand{\kev}{\mathrm{keV}\xspace}
\newcommand{\kms}{\mathrm{km\,s^{-1}}\xspace}

\newcommand{\pc}{\mathrm{pc}\xspace}

\newcommand{\degree}{$^{\circ}$\xspace}

\newcommand{\nee}{$n_\mathit{e}$}
\newcommand{\nh}{$N_\mathit{H}$\xspace}
\newcommand{\xspec}{\texttt{Xspec}\xspace}
\newcommand{\kt}{$kT_\mathit{e}$\xspace}

\received{August 21, 2019}
\revised{March 22, 2020}
\accepted{April 15, 2020}

\shorttitle{X-Ray Spectroscopy of N132D}
\shortauthors{Sharda, Gaetz, Kashyap and Plucinsky}

\begin{document}

\title{Spatially Resolved \textit{Chandra} Spectroscopy of the Large Magellanic Cloud Supernova Remnant N132D}

\author[0000-0003-3347-7094]{Piyush Sharda}\thanks{piyush.sharda@anu.edu.au}
\affiliation{Harvard-Smithsonian Center for Astrophysics, 60 Garden St., Cambridge, MA 02140, USA}
\affiliation{Research School of Astronomy and Astrophysics, Australian National University, Canberra, ACT 2611, Australia}
\affiliation{Department of Physics, Birla Institute of Technology and Science, Pilani, Rajasthan 333031, India}
\affiliation{Australian Research Council Centre of Excellence for All Sky Astrophysics in 3 Dimensions (ASTRO 3D), Australia}
\author[0000-0002-5115-1533]{Terrance J. Gaetz}\thanks{tgaetz@cfa.harvard.edu}
\affiliation{Harvard-Smithsonian Center for Astrophysics, 60 Garden St., Cambridge, MA 02140, USA}
\author[0000-0002-3869-7996]{Vinay L. Kashyap}
\affiliation{Harvard-Smithsonian Center for Astrophysics, 60 Garden St., Cambridge, MA 02140, USA}
\author[0000-0003-1415-5823]{Paul P. Plucinsky}
\affiliation{Harvard-Smithsonian Center for Astrophysics, 60 Garden St., Cambridge, MA 02140, USA}

\begin{abstract}
We perform detailed spectroscopy of the X-ray brightest supernova remnant (SNR) in the Large Magellanic Cloud (LMC), N132D, using \textit{Chandra} archival observations. By analyzing the spectra of the entire well-defined rim, we determine the mean abundances for O, Ne, Mg, Si, S and Fe for the local LMC environment. We find evidence of enhanced O on the north-western and S on the north-eastern blast wave. By analyzing spectra interior to the remnant, we confirm the presence of a Si-rich relatively hot plasma ($\ga 1.5\,\kev$) that is also responsible for the Fe\,K emission. \textit{Chandra} images show that the Fe\,K emission is distributed throughout the interior of the southern half of the remnant but does not extend out to the blast wave. We estimate the progenitor mass to be $15\pm5\,M_{\odot}$ using abundance ratios in different regions that collectively cover a large fraction of the remnant, as well as from the radius of the forward shock compared with models of an explosion in a cavity created by stellar winds. We fit ionizing and recombining plasma models to the Fe\,K emission and find that the current data cannot distinguish between the two, hence the origin of the high-temperature plasma remains uncertain. Our analysis is consistent with N132D being the result of a core-collapse supernova in a cavity created by its intermediate mass progenitor.
\end{abstract}

\keywords{supernova:individual (N132D), Xrays:individual (N132D), ISM:supernova remnants, plasmas, shock waves, ISM:abundances.}

\section{Introduction}
\label{s:intro}
Magellanic cloud supernova remnant (MCSNR) J0525-6938 (commonly referred to as N132D, following the catalog by \citealt{1956ApJS....2..315H}) is the X-ray-brightest SNR in the Large Magellanic Cloud (LMC; \citealt{1979ApJ...234L..77L}) with an X-ray luminosity of $L_\mathrm{X} \sim 3\times10^{37}\,\mathrm{erg\,s^{-1}}$ \citep{2016A&A...585A.162M}\footnote{The quoted X-ray luminosity is uncorrected for LMC absorption.}. It was first classified as a core-collapse supernova (CCSN) by \cite{1966MNRAS.131..371W}, and has been subsequently studied in great detail over the last few decades \citep{1997A&A...324L..45F,2008AdSpR..41..416X,2018ApJ...854...71B,2020arXiv200400016L}. Based on optical observations, it has been classified as an Oxygen-rich remnant \citep{1976ApJ...207..394D,1978ApJ...223..109L,1980ApJ...237..765L}, thought to have exploded inside a low density cavity in the interstellar medium (ISM, \citealt{1987ApJ...314..103H}). \cite{1995ApJ...439..365S} discuss the origin of this cavity, which might have formed due to a wind bubble mechanism common to Wolf-Rayet stars \citep{2007ApJ...667..226D}. It has been proposed by \cite{2000ApJ...537..667B} that this remnant might be the outcome of a Type Ib supernova (core collapse) and is believed to be roughly $2500\,\mathrm{yr}$ old \citep{1995AJ....109.2104M,1998ApJ...505..732H,2003ApJ...595..227C,2011Ap&SS.331..521V,2020arXiv200400016L}.

There are several characteristics of this remnant that make it a useful laboratory to study SNRs interacting with molecular clouds. Analysis of \nustar (Nuclear Spectroscopic Telescope Array) and \suzaku observations of N132D by \cite{2018ApJ...854...71B} reinforces the claim by \cite{1995AJ....109..200D} that this remnant is in the transition stage from a young to a middle-aged remnant. The integrated radio luminosity of N132D at $1\,\mathrm{Ghz}$ is 50 per cent of Cas A, an SNR which is $\sim5.5\times$ smaller in diameter than N132D \citep{1995AJ....109..200D}. High Energy Spectroscopic System (H.E.S.S.) observations of N132D classify this radio loud SNR as one of the strongest emitters of $\gamma$ rays in the LMC \citep{2015Sci...347..406H,2016A&A...586A..71A}. It has been estimated N132D has converted up to 17\% of its explosion energy into accelerating cosmic rays \citep{2015Sci...347..406H}. N132D is also the brightest SNR amongst all the known SNRs in the $1-100\,\mathrm{GeV}$ band \citep{2016ApJS..224....8A}. There is evidence for active star formation in the vicinity of N132D, as observed in the H$\alpha$ images from Magellanic Cloud Emission-Line Survey (MCELS, \citealt{1999IAUS..190...28S,2004AAS...20510108S}), however, no young stellar objects (YSOs) have been detected in the molecular cloud interacting with the SNR (\citealt{2010AJ....140..584D}; see also, \citealt{1976ApJ...207..394D}).

\textit{Chandra X-ray Observatory} (\chandra) observations \citep{2007ApJ...671L..45B} reveal a well-structured rim running along the southern part of the remnant (see Figure \ref{fig:allusefulregions}). This well-defined rim is associated with dense molecular clouds in this direction \citep{1997ApJ...480..607B,2015ASPC..499..257S} and is also present in the infrared (IR) observations of dust continuum emission in N132D taken by \textit{Spitzer} \citep{2006ApJ...652L..33W}. Using IR data from \textit{Spitzer} and \textit{Herschel Space Observatory} \citep{2015ApJ...799...50L}, it has been proposed that the X-ray emitting hot plasma has destroyed almost half of the dust grains in the remnant \citep{2006ApJ...653..267T,2012ApJ...754..132T,2013ApJ...779..134S,2018ApJS..237...10D,2019ApJ...882..135Z}. The X-ray emission also shows a bright arc-shaped structure close to the outermost shell in the south and south-east that may be attributed to the reverse shock encountering the ejecta or face-on filaments produced by the forward shock interacting with density enhancements in the surrounding medium. Towards the north, there are filament-like structures protruding outwards that are relatively faint in X-rays as compared to the rest of the remnant. Given that these structures are at the edge of the \ion{H}{1} cloud \citep{2003ApJS..148..473K} that encompasses the remnant \citep[see their Figure 12]{2016A&A...585A.162M}, they may have resulted as a consequence of strong shocks breaking out of the cavity into the ambient ISM. 

Although N132D is the brightest SNR in the LMC in X-ray, a full spectral analysis of the archival \chandra data \citep{2007ApJ...671L..45B} have not yet been performed. In this work, we carry out a spatially-resolved analysis of the well-defined rim as well as several interesting regions in the interior of the remnant that collectively cover around one-third of the remnant in projection. We assume the distance to N132D to be $50\,\mathrm{kpc}$ in all calculations hereafter \citep{2003AJ....125.1309C,2013Natur.495...76P,2019Natur.567..200P}. At this distance, $1\arcsec = 0.24\,\mathrm{pc}$. We describe the data reduction and processing in Section \ref{s:data_xray}, and source and background models used for spectral analysis of all the regions in Section \ref{s:analysis}. Section \ref{s:results} gives the resulting fits. We discuss the results in Section \ref{s:discuss}, and summarize our analysis in Section \ref{s:summary}.

\section{X-Ray Data and Reduction}
\label{s:data_xray}
We use X-ray observations of SNR N132D obtained with the S3 chip in \textit{Chandra's} Advanced Charged Couple Device (CCD) Imaging Spectrometer (ACIS-S) detector array \citep{1998SPIE.3444..210B}. N132D was observed for $89\,\mathrm{ks}$ by \chandra \citep{2007ApJ...671L..45B} in three ACIS-S observations in the Very Faint mode (see Table \ref{tab:n132dlog}). These X-ray observations showed the $\mathrm{pc}$-scale substructure in the previously-known roughly elliptical shape ($\sim\,14.8\,\times\,10.9~\pc$) in exquisite detail. We find no flaring in the data after examining the light curves of the observations. However, the X-ray data suffer from pileup in certain regions \citep{1999A&AS..135..371B,2001ApJ...562..575D}. We show a map of the pileup in the remnant in Appendix \ref{s:app_nbandandpileup}. For certain bright areas in the regions in the interior where pileup is greater than 10 per cent, we exclude them from the fit. We utilize the X-ray analysis package \chandra Interactive Analysis of Observations (CIAO version 4.9, \citealt{2006SPIE.6270E..1VF}) and \chandra Calibration Database (CALDB, version 4.7.3, \citealt{2007ChNew..14...33G}). We use \xspec version 12.9.1k \citep{1996ASPC..101...17A} to perform X-ray spectroscopy in various regions in the remnant. The line emission data is taken from AtomDB version 3.0.7 \citep{2013AIPC.1545..252F} whereas the non-equilibrium ionization (NEI) models come from NEI version 3.0.4. We use the cosmic abundance set by \cite{2000ApJ...542..914W} as the baseline abundance level for all our analysis. 

\begin{deluxetable}{l c c c c r}
\tabletypesize{\scriptsize}
\tablenum{1}
\label{tab:n132dlog}
\tablecolumns{6}
\tablecaption{\chandra ACIS-S observation log of SNR N132D.}
\tablehead{\colhead{ObsID} & \colhead{Observation Date} & \colhead{Exposure ($\mathrm{ks}$)} & \colhead{RA} & \colhead{Dec} & \colhead{Roll}}
\startdata
05532&Jan 09, 2006&44.59 &81.2595\degree &-69.6437\degree&330.2\degree\\
07259&Jan 10, 2006&24.85&81.2595\degree &-69.6437\degree &330.2\degree\\
07266&Jan 15, 2006&19.90&81.2595\degree &-69.6437\degree &330.2\degree\\
\hline
\enddata
\end{deluxetable}

\section{Spectral Analysis}
\label{s:analysis}
\begin{figure}
\centering
\includegraphics[width=1.0\linewidth]{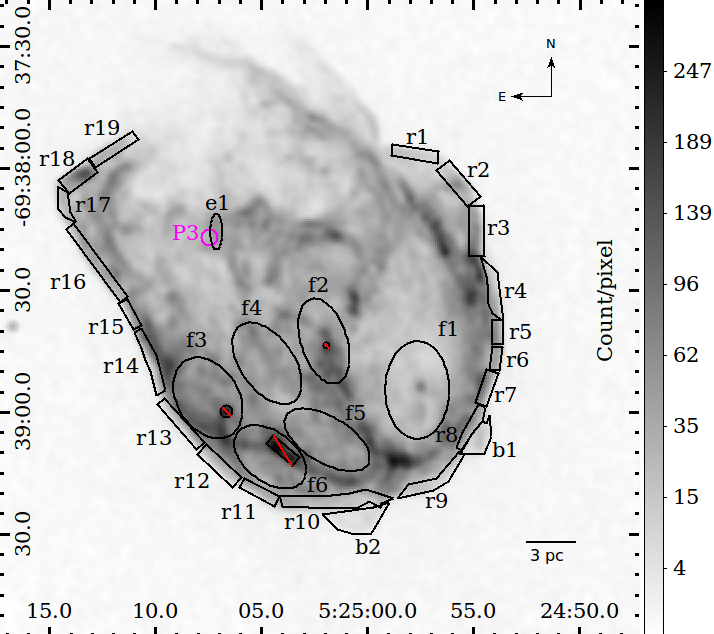}
\caption{\chandra ACIS-S image of counts per pixel in N132D in the $0.35-7.0\,\mathrm{keV}$ band with $x$ and $y$ axes showing the Right Ascension (RA) and Declination (Dec), respectively. One pixel is 0.5\arcsec, where 1\arcsec = $0.24\,\mathrm{pc}$. The grey scale has been inverted so that darker shade corresponds to higher counts. All regions studied in this work are indicated by black polygons with labels. $\mathrm{r1-r19}$ are the rim regions and $\mathrm{b1-b2}$ are the two likely protrusions beyond the blast wave. In the interior, we study regions $\mathrm{e1}$ (which has enriched abundances of O, Ne, Mg, Si and Fe) and $\mathrm{f1-f6}$ (where weak Fe K emission is detected). Black areas with a red strikethrough line within regions $\mathrm{f2,\,f3}$ and $\mathrm{f6}$ are the bright patches affected by pileup which are excluded from the spectroscopic analysis. $\mathrm{P3}$ is an O-rich region found in optical studies of N132D by \cite{1996AJ....112.2350M,1995AJ....109.2104M,2000ApJ...537..667B}.}
\label{fig:allusefulregions}
\end{figure}

We first analyze the well-defined rim of SNR N132D to get a picture of emission from the forward shock. We number the rim regions $\mathrm{r1-r19}$ in the clockwise direction, as we show in Figure \ref{fig:allusefulregions}. We also identify and analyze two ``blobs'' (labeled $\mathrm{b1}$ and $\mathrm{b2}$) which are likely protruding ahead of the forward shock. We then search the entire remnant for regions that show possibly enhanced abundances of one or more elements, through visual inspections of narrow band images centered on line features of O, Ne, Mg, Si, S and Fe (see Appendix \ref{s:app_nbandandpileup}), as well as hardness ratio images in the soft ($0.3-0.9\,\kev$), medium ($0.9-2\,\kev$), and hard ($2-7\,\kev$) bands. We select interior regions $\mathrm{e1,\,f1,\,f2,\,f3,\,f4,\,f5}$ and $\mathrm{f6}$ for further study (see Sections \ref{s:ejectaregion} and \ref{s:fekregion} for additional details). We note that regions $\mathrm{f2,\,f3}$ and $\mathrm{f6}$ contain bright areas in projection that are significantly affected by pileup, as shown in Figure \ref{fig:pileup}. We exclude such areas when performing X-ray spectroscopy on these regions. Table \ref{tab:regions} lists the classification of each region together with the location of its center and area; following subsections describe the background and source models we use to fit the background and source spectra, respectively.

\begin{deluxetable}{l c c c c r}
\tabletypesize{\scriptsize}
\tablenum{2}
\label{tab:regions}
\tablecolumns{5}
\tablecaption{Classification of all regions (shown in Figure \ref{fig:allusefulregions}) we study in this work. The RA and Dec coordinates mark the centers of each region. All the rim regions have at least 3500 X-ray photon counts within $0.3-7.0\,\mathrm{keV}$.}
\tablehead{\colhead{Region} & \colhead{Location} & \colhead{RA} & \colhead{Dec} & \colhead{Area ($\mathrm{pc}^2$)}}
\startdata
$\mathrm{r1}$&Rim&5:24:57.753&-69:37:56.60&1.93\\
$\mathrm{r2}$&Rim&5:24:55.700&-69:38:03.86&2.84\\
$\mathrm{r3}$&Rim&5:24:54.851&-69:38:15.55&2.51\\
$\mathrm{r4}$&Rim&5:24:54.186&-69:38:30.42&2.99\\
$\mathrm{r5}$&Rim&5:24:53.824&-69:38:40.57&0.96\\
$\mathrm{r6}$&Rim&5:24:53.929&-69:38:46.91&0.82\\
$\mathrm{r7}$&Rim&5:24:54.364&-69:38:54.15&1.49\\
$\mathrm{r8}$&Rim&5:24:55.145&-69:39:04.67&1.81\\
$\mathrm{r9}$&Rim&5:24:56.799&-69:39:16.15&6.37\\
$\mathrm{r10}$&Rim&5:25:01.193&-69:39:21.65&6.29\\
$\mathrm{r11}$&Rim&5:25:05.082&-69:39:19.87&1.48\\
$\mathrm{r12}$&Rim&5:25:06.968&-69:39:13.36&2.22\\
$\mathrm{r13}$&Rim&5:25:08.805&-69:39:03.01&2.10\\
$\mathrm{r14}$&Rim&5:25:10.241&-69:38:47.50&2.73\\
$\mathrm{r15}$&Rim&5:25:11.173&-69:38:36.08&1.01\\
$\mathrm{r16}$&Rim&5:25:12.741&-69:38:23.41&2.75\\
$\mathrm{r17}$&Rim&5:25:14.211&-69:38:09.52&1.29\\
$\mathrm{r18}$&Rim&5:25:13.631&-69:38:02.10&2.17\\
$\mathrm{r19}$&Rim&5:25:11.931&-69:37:55.47&1.91\\
$\mathrm{b1}$&Blob&5:24:54.730&-69:39:05.84&4.05\\
$\mathrm{b2}$&Blob&5:25:00.428&-69:39:26.84&6.36\\
$\mathrm{e1}$&Interior&5:25:07.105&-69:38:15.80&1.21\\
$\mathrm{f1}$&Interior&5:24:58.031&-69:38:42.24&17.51\\
$\mathrm{f2}$&Interior&5:25:03.585&-69:38:44.65&11.20\\
$\mathrm{f3}$&Interior&5:25:07.744&-69:38:56.73&14.49\\
$\mathrm{f4}$&Interior&5:25:04.716&-69:38:48.14&13.74\\
$\mathrm{f5}$&Interior&5:25:01.888&-69:39:06.96&12.23\\
$\mathrm{f6}$&Interior&5:25:04.575&-69:39:11.08&10.16\\
\hline
\enddata
\end{deluxetable}

\subsection{Background Model}
The background region we select is a 1.62\arcmin $\,$square located at RA = 05:24:36.963, Dec = -69:37:05.68 at a distance of 2.74\arcmin $\,$from the remnant. We do not subtract the background spectrum from each source spectrum, rather, we model it separately because of the low number of counts at energies $> 2.5\,\kev$. With low counts, the subtraction of Poisson distributions results in a distribution which is non-Poissonian and far from Gaussian; in addition, the number of counts after subtraction can be negative (see, for example, \citealt{2001ApJ...548..224V,2017MNRAS.472..308G}). 

We differentiate the background model into sky (imaged through the X-ray optics) and detector (not imaged through the optics) components. For the detector background model, we analyze the so-called ``stowed'' background data in the Very Faint mode to construct a spectral model for the S3 CCD, similar to the approach used for the ACIS-I CCDs by \cite{2014A&A...566A..25B}. We download the background data set acis7D2005-09-01bgstow\_ctiN0002.fits from the CALDB. We then run \texttt{acis\_process\_events} to populate the \texttt{TDETX} and \texttt{TDETY} columns. After copying over the status column from acis7D2005-09-01bgstow\_ctiN0002.fits to the processed file (since \texttt{acis\_process\_events} zeroed the status column), we apply the CIAO tool \texttt{reproject\_events}, using ObsID 05332 as the match file to project the background events to the sky. We extract the detector background from the same region as used for the sky background (see below), and generate a weighted \texttt{RMF} using a \texttt{WMAP} in \texttt{TDET} coordinates. The detector background model consists of a broken powerlaw (\texttt{bkn2pow}) to represent most of the spectrum from $0.3-11.0\,\kev$, with a broad Gaussian to account for the high energy ACIS-S3 background continuum. We include Gaussian lines for the instrumental fluorescence lines (Al\,K$\alpha$, Si\,K$\alpha$, Au\,M complex, Ni\,K$\alpha$, and Au\,L$\beta$). We initially adopt the line energies from \cite{1967RvMP...39...78B}, subsequently thawing the line energy for Si\,K$\alpha$, the Au\,M complex, Ni\,K$\alpha$ and the Au\,L$\beta$ complex. We also thaw the line width for Au\,M and Au\,L$\beta$ complexes. Once a good fit is found, we freeze all of the parameters, and thaw a multiplicative \texttt{const} parameter (initially frozen at 1.0) which provides an overall normalization scaling.

The sky model consists of an absorption (\texttt{tbabs}) plus two thermal plasma \texttt{apec} ($\sim 0.2\,\kev$ and $\sim 0.8\,\kev$) components and a powerlaw. The $0.19\,\kev$ \texttt{apec} model primarily represents emission from the local hot bubble (LHB) and the $0.77\,\kev$ model represents other Galactic and LMC emission along the line of sight and the Galactic Halo \citep{1998ApJ...493..715S,2008A&A...478..615S}. There may be emission from the LHB that contributes to the emission we model with the $0.77\,\kev$ plasma model and there may be emission from the Halo that contributes to the emission we model with the $0.19\,\kev$ plasma model \citep{1997ApJ...485..125S,2001ApJ...554..684K,2002ApJ...576..188M,2019arXiv191012754K}. This is not an issue for us as we require an empirical model for the background. We use the \texttt{powerlaw} component with a fixed slope of 1.46 \citep{,1997MNRAS.285..449C,2004xmmg.rept.....S,2010ApJS..188...46K} to model the cosmic X-ray background from unresolved point sources including active galactic nuclei (AGNs). We fit the sky model using the absorbed thermal models and powerlaw, together with the detector background model (described above). In the fitting, we allow the detector background \texttt{const} parameter to vary, but otherwise, we fix the shape of the detector background; the fit is performed over $0.30\,\kev$ to $11.0\,\kev$. Once a good fit is obtained, we freeze the sky model parameters, and allow a multiplicative \texttt{const} factor (initially frozen at 1.0) to vary.

\begin{figure}
\centering
\includegraphics[width=0.60\linewidth, angle=270]{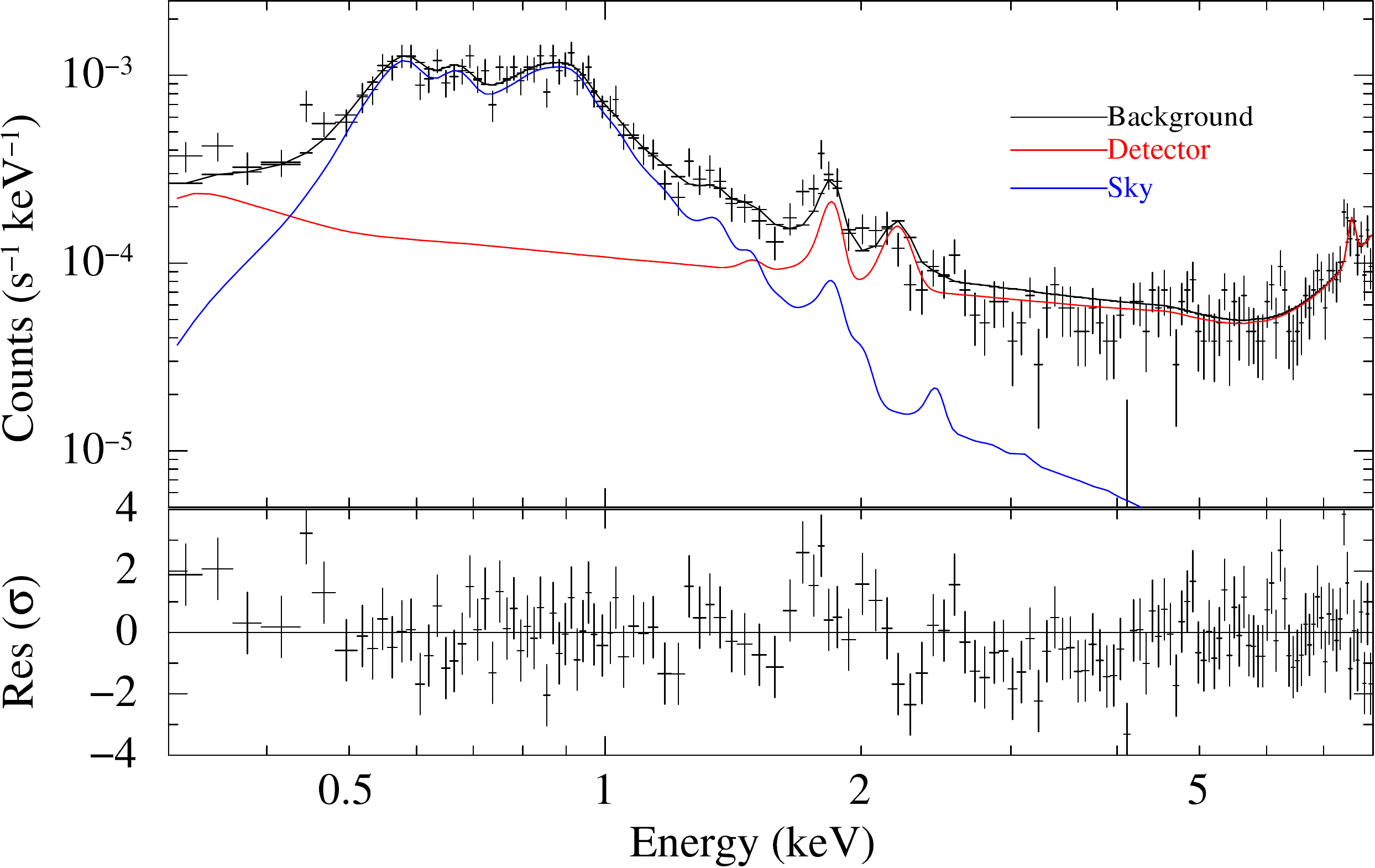}
\caption{Background spectrum and model between $0.3-8.0\,\kev$. Background model (black) is a mixture of detector (red) and sky (blue) components. The detector component consists of a broken power law (\texttt{bkn2pow}) and multiple Gaussians whereas the sky component consists of two thermal (\texttt{apec}) models and a \texttt{powerlaw} model. The thermal models dominate from $0.5-1.0\,\kev$ and the detector background dominates above $2.0\,\kev$. The background becomes significant above $5.5\,\kev$.}
\label{fig:background}
\end{figure}

When fitting a source spectrum, we freeze the parameters that affect the detector and sky background model shapes while allowing the overall normalizations to vary through multiplicative constants for the detector and sky backgrounds (see, for example, \citealt{2016A&A...585A.162M,2017MNRAS.472..308G}). Table \ref{tab:fitbkgrnd} presents the background model and Figure \ref{fig:background} shows the background fit. As we show later in Sections \ref{s:results_rim} and \ref{s:results_e1}, the background is significantly lower than the source spectra in the interior as well as on the rim, respectively, for most of the band-pass except at the highest energies (E > $5.5\,\kev$).

\begin{deluxetable}{l l c l}
\tabletypesize{\scriptsize}
\tablenum{3}
\label{tab:fitbkgrnd}
\tablecolumns{4}
\tablecaption{Best-fit parameters of the background model, consisting of the detector and the sky components. Errors represent the 68 per cent confidence intervals which correspond to $1\sigma$ in the Gaussian case. Parameters without errors were frozen in the fit.}
\tablehead{\colhead{Component} & \colhead{Parameter} & \colhead{Units} & \colhead{Value}}
\startdata
Detector&\texttt{bkn2pow}   PhoIndex1 & ...    & $1.60^{+1.17}_{-0.58}$ \\
&\texttt{bkn2pow}   BreakE1   & $\kev$ & $0.50^{+0.03}_{-0.03}$ \\
&\texttt{bkn2pow}   PhoIndex2 & ...    & $0.46^{+0.03}_{-0.03}$ \\
&\texttt{bkn2pow}   BreakE2   & $\kev$ & $4.58^{+0.70}_{-0.49}$ \\
&\texttt{bkn2pow}   PhoIndex3 & ...    & $1.51^{+0.70}_{-0.54}$ \\
&\texttt{bkn2pow}   Norm      & $\mathrm{photons\,keV^{-1}\,cm^{-2}\,s^{-1}}$ & $1.433^{+0.001}_{-0.001}\times 10^{-3}$ \\
&\texttt{gaussian}1 LineE     & $\kev$ & $11.314^{+0.226}_{-0.188}$ \\
&\texttt{gaussian}1 Sigma     & $\kev$ & $1.954^{+0.145}_{-1.23}$ \\
&\texttt{gaussian}1 Norm      & $\mathrm{photons\,cm^{-2}\,s^{-1}}$ & $6.893^{+0.008}_{-0.007}\times 10^{-2}$ \\
&\texttt{gaussian}2 LineE     & $\kev$ & $1.487$ \\
&\texttt{gaussian}2 Sigma     & $\kev$ & $0.0$   \\
&\texttt{gaussian}2 Norm      & $\mathrm{photons\,cm^{-2}\,s^{-1}}$ & $3.868^{+3.680}_{-3.476}\times 10^{-5}$ \\
&\texttt{gaussian}3 LineE     & $\kev$ & $1.860^{+0.007}_{-0.012}$ \\
&\texttt{gaussian}3 Sigma     & $\kev$ & $0.0$ \\
&\texttt{gaussian}3 Norm      & $\mathrm{photons\,cm^{-2}\,s^{-1}}$ & $4.650^{+0.573}_{0.055}\times 10^{-4}$ \\
&\texttt{gaussian}4 LineE     & $\kev$ & $2.212^{+0.012}_{-0.123}$ \\
&\texttt{gaussian}4 Sigma     & $\kev$ & $0.060^{+0.028}_{-0.019}$ \\
&\texttt{gaussian}4 Norm      & $\mathrm{photons\,cm^{-2}\,s^{-1}}$ & $5.070^{+0.808}_{-0.698}\times 10^{-4}$ \\
&\texttt{gaussian}5 LineE     & $\kev$ & $7.555^{+0.019}_{-0.008}$ \\
&\texttt{gaussian}5 Sigma     & $\kev$ & $0.0$ \\
&\texttt{gaussian}5 Norm      & $\mathrm{photons\,cm^{-2}\,s^{-1}}$ & $3.648^{+0.580}_{-0.555}\times 10^{-4}$ \\
&\texttt{gaussian}6 LineE     & $\kev$ & $9.853^{+0.030}_{-0.017}$ \\
&\texttt{gaussian}6 Sigma     & $\kev$ & $0.050^{+0.030}_{-0.050}$ \\
&\texttt{gaussian}6 Norm      & $\mathrm{photons\,cm^{-2}\,s^{-1}}$ & $8.780^{+1.649}_{-1.552}\times 10^{-4}$ \\
\hline
Sky     &\texttt{TBabs}     \nh       & $\mathrm{cm^{-2}}$ & $0.186^{+0.132}_{-0.052}\times 10^{21}$ \\
&\texttt{apec}1     \kt       & $\kev$             & $0.175^{+0.009}_{-0.023}$ \\
&\texttt{apec}1     Norm      & $\mathrm{cm^{-5}}$ & $1.578^{+3.001}_{-0.427}\times 10^{-4}$ \\
&\texttt{apec}2     \kt       & $\kev$             & $0.768^{+0.040}_{-0.051}$ \\
&\texttt{apec}2     Norm      & $\mathrm{cm^{-5}}$ & $3.097^{+1.002}_{-0.431}\times 10{-5}$ \\
&\texttt{powerlaw}  PhoIndex  & ...                & $1.46$\\
&\texttt{powerlaw}  Norm      & $\mathrm{photons\,keV^{-1}\,cm^{-2}\,s^{-1}}$ & $3.404^{+1.049}_{-1.346}\times 10^{-6}$ \\
\enddata
\end{deluxetable}

\subsection{Source Models}
For all of the (source) regions we analyze in this work, we take a two component absorption model to account for Galactic (\texttt{tbabs}) and LMC (\texttt{tbvarabs}) absorption by gas, molecules and grains along the line of sight. Following \cite{1990ARA&A..28..215D}, we fix the Galactic Hydrogen column density $N_\mathit{H,\mathrm{Gal}}$ to $5.5\times10^{20}\,\mathrm{cm^{-2}}$ with solar abundances \citep{2000ApJ...542..914W} whereas we allow the LMC hydrogen column density to vary. For all spectral fits, we set the initial guess for LMC elemental abundances to be $0.4\,\times$ solar on the \citet{2000ApJ...542..914W} scale, in line with the estimated metallicity of the LMC \citep{1982ApJ...252..461D,1992ApJ...384..508R,1997macl.book.....W}. Due to the limited number of counts in the energy range $0.3 - 0.5\,\kev$ in our spectra, where emission from C and N is prominent, we tie the C and N abundances to O in the source models. Similarly, due to poor constraints on the abundances of S, Ar, and Ca and the possibility of the L-shell emission of these elements affecting fits at lower energies, we tie them together. We cannot constrain the abundance of Ni with the current data due to low number statistics, and tie it to that of Fe. For all the fits, we utilize the energy range between $0.3-7.0\,\kev$, except for the regions where we study Fe K emission and extend the fit to $7.5\,\kev$ (see Section \ref{s:fekregion} for details).

With the angular resolution of \chandra, we can separate the forward shock from the rest of the remnant along the rim. We fit the rim regions with a plane-parallel shock model (\texttt{vpshock}; see \citealt{2001ApJ...548..820B}), because we expect to find a shock running into relatively cold and mostly neutral material. This model loses its accuracy when the conditions in the emitting region depart significantly from its assumptions, for example, the temperature and/or density vary across the region, or the material is already heated by previous shocks or thermal conduction \citep{1983ApJS...51..115H,1991SSRv...58..259J}. Consequently, we add a non-equilibrium ionization (NEI) component (\texttt{vnei}) to the model to explain emission from plasma heated to some temperature and evolved for a particular time ($\tau$), while not including emission from earlier times (see, for example, \citealt{1994ApJ...437..770M,2001ApJ...548..820B,2007ApJ...661..879E}). It also allows for the possibility of the detection of ejecta fragments if we allow the abundances of the \texttt{vnei} component to vary. In cases where the source model consists of more than one component, we start the fit by fixing the abundances of one or more \texttt{vnei} components to be the same as that of \texttt{vpshock} component. 

Apart from the \texttt{vpshock} and \texttt{vnei} models, we also investigate the case of a recombining plasma which may be responsible for emission in the Fe K complex. In the case of a recombining plasma, the ionization temperature of ions exceeds the electron temperature \citep{1974ApJ...188..335M,1977PASJ...29..813I}. We use the non-equilibrium plasma model \texttt{vrnei} which is a modified version of \texttt{vnei} in which the initial temperature (\texttt{kT\_init}) can be specified; the model starts in collisional ionization equilibrium at \texttt{kT\_init}, the temperature is changed to \texttt{kT}, the ionization state evolves at constant \texttt{kT} and density. A \texttt{vrnei} with \texttt{kT\_init} set to $0.0808\,\kev$ is equivalent to \texttt{vnei}. If \texttt{kT\_init} exceeds \texttt{kT}, the model evolves by recombining. As with \texttt{vnei}, the emission is calculated at a specific value of $\tau$. We describe this further in Section \ref{s:fekregion}. In the following subsections, we lay out the fitting algorithms for the different regions we analyze.

\subsubsection{Rim Regions}
\label{s:rimspectral}
We define rectangular regions on the rim wherever possible; some regions are distorted in shape to account for the locally non-uniform curvature of the remnant. All the regions around the rim ($\mathrm{r1-r19,\,b1-b2}$) have nearly the same width ($0.6-0.7\,\pc$) and have at least 3500 counts in the $0.3-7.0\,\mathrm{keV}$ bandpass. We follow the following procedure to fit the rim regions and the blobs:

\begin{enumerate}[1]
\item{Fit the spectrum of a region on the rim with a source model (\texttt{tbabs} $\times$ \texttt{tbvarabs} $\times$ \texttt{vpshock}), with abundances fixed at $0.4\times$ solar \citep{2007ApJ...671L..45B}.}
\item{If the fit is acceptable in step 1 (following the criteria we describe in Section \ref{s:fiteval}), note the abundances.}
\item{If the fit is not acceptable, allow the abundances of O, Ne, Mg, Si, S, and Fe to vary one by one. If it is acceptable after the abundances have been allowed to vary, note the best-fit abundances and error bars.}
\item{Fit all the regions on the rim in the same manner. After this step, all regions would have been fitted once with \texttt{vpshock}.}
\item{Find an average abundance for each element from regions where the fit in steps 1 or 3 was acceptable.}
\item{Refit the regions where the fit was not acceptable in steps 1 or 3 with the mean abundances calculated in step 5.}
\item{If the fit is still not acceptable in step $6$, add an NEI component (\texttt{vnei}) to the source model and refit.}
\end{enumerate}

As we show in Section \ref{s:results_rim}, for the two regions on the rim where a single \texttt{vpshock} did not generate an acceptable fit, the two-component model satisfactorily fits the spectra. Thus, we do not go beyond step 7 to fit any region on the rim. Finally, to calculate the mean local LMC abundances for all elements, we add an additional step in the algorithm in which we fit all of the rim spectra with the abundances of O, Ne, Mg, Si, S, and Fe free. This is necessary to get meaningful uncertainties on the average abundances which would otherwise be underestimated if some elemental abundances were held fixed in some regions \citep{2019A&A...631A.127M}.
 
\subsubsection{Region $\mathrm{e1}$}
\label{s:ejectaregion}
We examine the O-rich ring seen in the optical in N132D \citep{1995AJ....109.2104M,2000ApJ...537..667B}, also called the \textit{Lasker's Bowl}, as an interior region that might exhibit enriched abundances in the X-ray spectral data. The presence of ejecta-rich knots in X-rays in this ring was previously reported by \citet[see their Figure 2]{2007ApJ...671L..45B}. We select region $\mathrm{e1}$ on this ring which overlaps with both the ejecta-rich knots marked in the X-ray data and the O-rich ejecta seen in the optical. The spatial coincidence of optical O-rich ejecta and X-ray enhancements in O, Ne, Mg, Si, and Fe emission point to $\mathrm{e1}$ being a complex region in which multiple components with different plasma conditions are contributing. Moreover, its location also overlaps with a region which shows O and Ne rich ejecta in the $14-36\,\mu\mathrm{m}$ infrared map of N132D \citep[see their region I in Figure 1]{2012ApJ...754..132T}. To fit this region, we use an NEI component (\texttt{vnei}), and add a \texttt{vpshock} component to account for the shell emission. 

\subsubsection{Regions with Fe K Emission}
\label{s:fekregion}
\begin{figure*}
\includegraphics[width=1.0\linewidth]{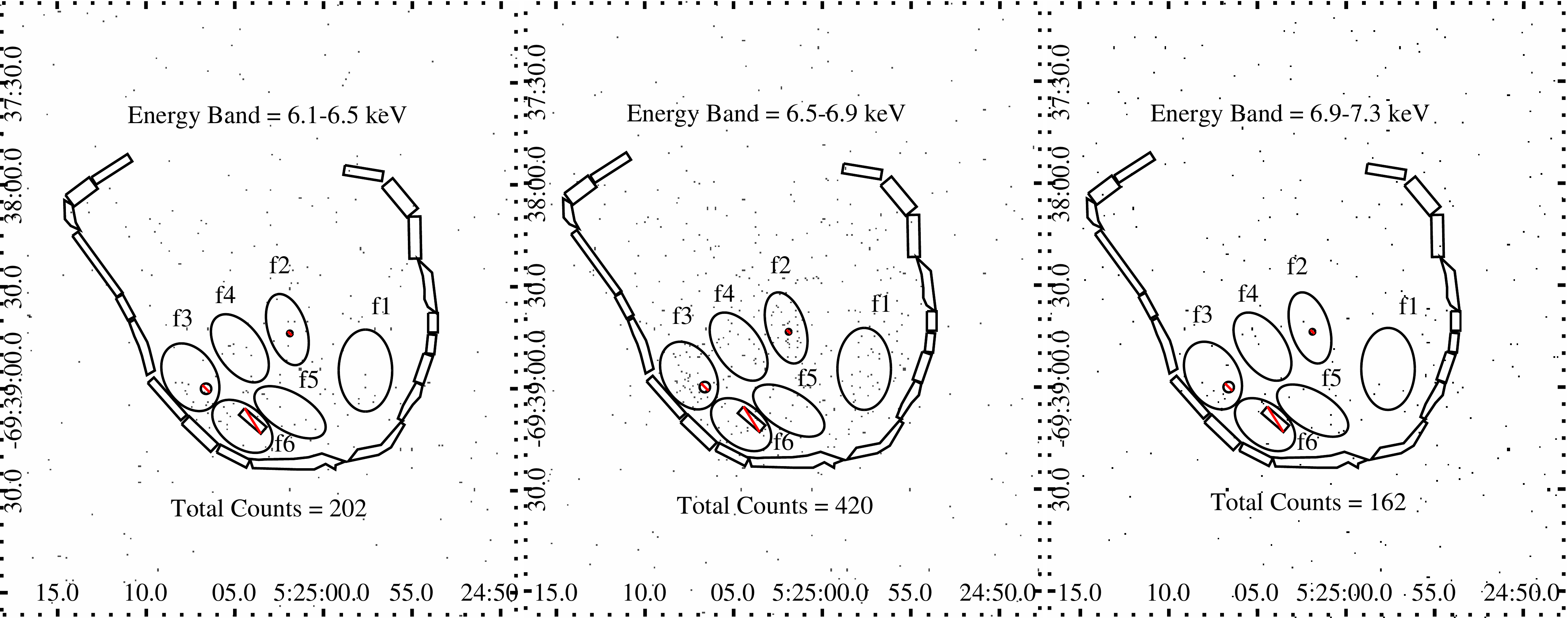}
\caption{Spatial distribution of counts in the energy range $6.1-7.3\,\kev$ in N132D. From left to right, images are shown for energy ranges $6.1-6.5, 6.5-6.9$ and $6.9-7.3\,\kev$, respectively. Rim regions mark the extent of the remnant. The middle panel shows emission from He-like Fe K$\alpha$ ($\sim6.7\,\kev$), while the adjacent lower and higher energy bands indicate the continuum levels. The middle energy band shows a significant excess compared to the continuum.}
\label{fig:fekcounts_spatialdistr}
\end{figure*}

N132D is one of the few extragalactic SNRs for which direct measurements of the spatial distribution of Fe-rich ejecta can be made. The Fe K$\alpha$ complex ranges from $6.4\,\kev$ for neutral Fe to $7.0\,\kev$ for \ion{Fe}{26}. The spectrum of the entire remnant indicates a peak in emission at $\sim 6.7\,\kev$ (presumably \ion{Fe}{25} emission). A center-filled excess of Fe K emission was detected in the observations of N132D taken by \textit{XMM-Newton} \citep{2001A&A...365L.242B}, however, \chandra data reveals that the extent of this emission is spread largely across the southern part of the remnant. As we show in Figure \ref{fig:fekcounts_spatialdistr}, we create three $0.4\,\kev$ wide passbands to sample this Fe K emission and the surrounding continuum: $6.1-6.5, 6.5-6.9$ and $6.9-7.3\,\kev$ (see also, Figure \ref{fig:narrowbands}). We then select six large regions ($\mathrm{f1 - f6}$) to study the Fe K emission feature in this remnant. We select enough regions such that they collectively sample the majority of the Fe K counts observed in the spectrum, and exclude areas where the pileup fraction is high, as shown in Figure \ref{fig:pileup}. The analysis of a single spectrum from the entire southern half of the remnant combines data from regions with different plasma conditions such that a complex, multi-component model is necessary to represent the data. Thus, it is more meaningful to analyze the spectra on smaller spatial scales in which the inherent variations in the plasma conditions are smaller.

For regions where we study emission from Fe K lines, we fit the spectra in the energy range $0.3 - 7.5\,\kev$ to sufficiently sample the continuum on either side of the feature at $6.7\,\kev$. We present analyses based on both ionizing and recombining plasma models for regions $\mathrm{f1 - f6}$. SNRs interacting with molecular clouds are frequently associated with recombining plasma, although the mechanism which produces the recombining plasma is not clear. One possibility is thermal conduction between the remnant shell and the cloud as suggested by \cite{1998ApJ...503L.167R}. A commonly quoted evidence for this scenario is the anti-correlation between electron temperature and recombining timescale (e.g., \citealt{2018PASJ...70..110K,2018PASJ...70...35O,2020ApJ...890...62O}). On the other hand, an overionization of the plasma is possible if the shock evaporates the cloud \citep{1991ApJ...373..543W}. \cite{1989MNRAS.236..885I} and \cite{2009ApJ...705L...6Y} suggest that an overionized plasma may be produced by rapid adiabatic expansion if the shock propagates from a region of high density to a region of low density \citep{2012PASJ...64...24S}. In this scenario, a positive correlation is observed between the electron temperature and recombining timescale (e.g., \citealt{2018ApJ...868L..35Y}). Detailed simulations of the X-ray emission from an SNR shock interacting with a distribution of clouds in the ISM conducted by \cite{2019ApJ...875...81Z} show that both thermal conduction and adiabatic expansion are likely to produce radiative recombination emission at different locations in the remnant (see also, \citealt{2011MNRAS.415..244Z}).

Non-equilibrium ionization in SNRs typically manifests itself as 1. an ionizing plasma, or 2. a recombining plasma. In the former case, the plasma is underionized; the ionization stages and line ratios reflect ionization temperature $kT_\mathrm{z} < kT_\mathrm{e}$. The plasma evolves via ionization, and the radiative recombination continuum (RRC) features are weak. In the latter case, the plasma is overionized, the ionization stages and line ratios reflect $kT_\mathrm{z} > kT_\mathrm{e}$. The plasma evolves via recombination, and has strong RRC features, with the continuum featuring sawtooth-like excesses extending upward in energy from the ionization potential, and line ratios exceeding the expectations for collisional ionization equilibrium due to radiative cascade populating higher levels.

The basic ionizing plasma \textit{vs.} recombining plasma features for SNR spectra have been long understood \citep{1977PASJ...29..813I,1981A&AS...45...11M, 1990ApJ...353..245G,1994ApJ...437..770M}. The shocking of low temperature material results in ionization to more excited states. Eventually adiabatic cooling dominates as the remnant expands. Ultimately the plasma becomes overionized, with a recombining plasma. The surprising aspect of recent discoveries of recombining plasmas was that the plasmas are strongly recombining, with $kT_\mathrm{z}$ greatly exceeding $kT_\mathrm{e}$ with significant radiative RRCs, and significant line ratio and ionization state anomalies. \citet{2002ApJ...572..897K} proposed an overionized plasma based on anomalous line ratios in Advanced Satellite for Cosmology and Astrophysics (\asca) observations of SNR IC\,443. The existence of strongly recombining plasmas in SNRs was established by \citet{2009ApJ...705L...6Y} with the discovery of radiative recombination continua (RRCs) of H-like Si and S in \suzaku observations of SNR IC\,443, and that of H-like Fe in \suzaku observations of SNR N49B by \cite{2009ApJ...706L..71O}.

The strength of the RRC emission depends on the electron temperature, the ion temperature and the ionization timescale (see \citet{2009ApJ...705L...6Y} for a discussion). The presence of hot, He-like Fe plasmas in N132D is suggestive of recombining plasma. The Fe K RRC feature in the \chandra spectra is difficult to disentangle from systematic instrument characteristics like decreasing effective area and increasing detector background at $\gtrsim\,7\,\kev$, and the ability to detect excess K$\beta$ over K$\alpha$ is also limited by the CCD spectral resolution and the low sensitivity achieved in the $\sim 90\,\mathrm{ks}$ of available data. In principle, \cite{2018PASJ...70...16H} observations of N132D with low background could potentially be used to constrain the H-like K$\alpha$ to He-like K$\alpha$ ratio that can provide evidence for a recombining plasma \citep{2002ApJ...572..897K,2010SSRv..157..103P,2013ApJ...777..145L}, however, the low number of counts in the \textit{Hitomi} spectrum makes such an analysis challenging, and is beyond the scope of this paper. As we show later in Section \ref{s:results_fek}, it is thus not possible to provide a definitive case for the existence of a hot Fe K emitting recombining plasma in the remnant. Nevertheless, we examine this case as a possible alternative to the ionizing case.

For the case of an ionizing plasma, we introduce a two component \texttt{vnei}, where the cooler component explains the soft X-ray spectrum and Fe L emission, and the hotter component explains the hard X-ray spectrum and Fe K emission. For the case of a recombining plasma, we use the recombining collisional plasma model \texttt{vrnei}, together with a \texttt{vnei} which can account for the low temperature plasma. Both models also contain a \texttt{vpshock} component to represent the shell emission along the line of sight. As we show in Section \ref{s:results_fek}, such three-component models (\texttt{vnei/vrnei + vnei + vpshock}) are necessary to account for the Fe K emission in these regions.

\subsection{Fit Evaluation}
\label{s:fiteval}
We use the C-statistic (which approximates the Poisson log-likelihood) to evaluate the spectral fits since it does not introduce a bias in the case of a low (or null) number of counts per spectral bin \citep{1979ApJ...228..939C,1989ApJ...342.1207N,2007A&A...472...21L}. We further use the goodness-of-fit criterion developed for the C-statistic by \cite{2017A&A...605A..51K}, by comparing the observed value of the C-statistic (\texttt{cstat (O)}) with the expected value (\texttt{cstat(E)}) and expected variance determined from the predicted model counts in each bin, using the numerical estimates derived by \citet{2017A&A...605A..51K}. We show both the expected value and the width of the distributions that would result when the fit is good. We adopt the following criterion to determine if a fit is acceptable if $\mathtt{cstat(E)}-2.6\,\sigma_\mathrm{E} < \mathtt{cstat (O)} < \mathtt{cstat(E)} + 2.6\,\sigma_\mathrm{E}$, where $\sigma_\mathrm{E}^2$ is the expected variance of \texttt{cstat(E)} and we choose the bounds such that the probability that \texttt{cstat(O)} falls outside the range is 1 per cent. 

\section{Results}
\label{s:results}
\subsection{Rim Regions and Blobs}
\label{s:results_rim}
Using the fit evaluation criteria we outline in Section \ref{s:fiteval}, we find that spectral analysis of 17 out of the 19 rim regions produce an acceptable fit with the single component \texttt{vpshock} model, which we summarize in Table \ref{tab:rimfit1}. For the two regions where the single model fit fails, we re-do the fitting while adding a \texttt{vnei} component. We present the results for the two-component model in Table \ref{tab:rimfit2}. 

Table \ref{tab:scatter} shows the mean abundance values ($\mu$, with $1\sigma_{\mu}$ errors) we calculate for the rim and the scatter in each parameter. We emphasize that the mean values we calculate are from fitting the abundances of O, Ne, Mg, Si, S, and Fe in all the rim regions; this prevents bias in the estimate of the mean due to some fits having some parameters frozen. We follow the method of Multiple Imputations \citep{2011ApJ...731..126L} to find the error on the mean ($\sigma_\mu$) and the scatter, which takes into account the statistical as well as systematic uncertainties (in other words, within and between variance). We present the details of this method in Appendix \ref{s:append_errors}. If the scatter in an elemental abundance along the rim is < 1 (implying that there is more systematic than statistical error), we consider its variation to be insignificant. If the best-fit abundance and associated $1\sigma$ error in any region on the rim is more than $\mu + \sigma_{\mu}$ or less than $\mu - \sigma_{\mu}$, we classify it as being enhanced or reduced, respectively.

\begin{figure}
\includegraphics[width=0.65\linewidth, angle=270]{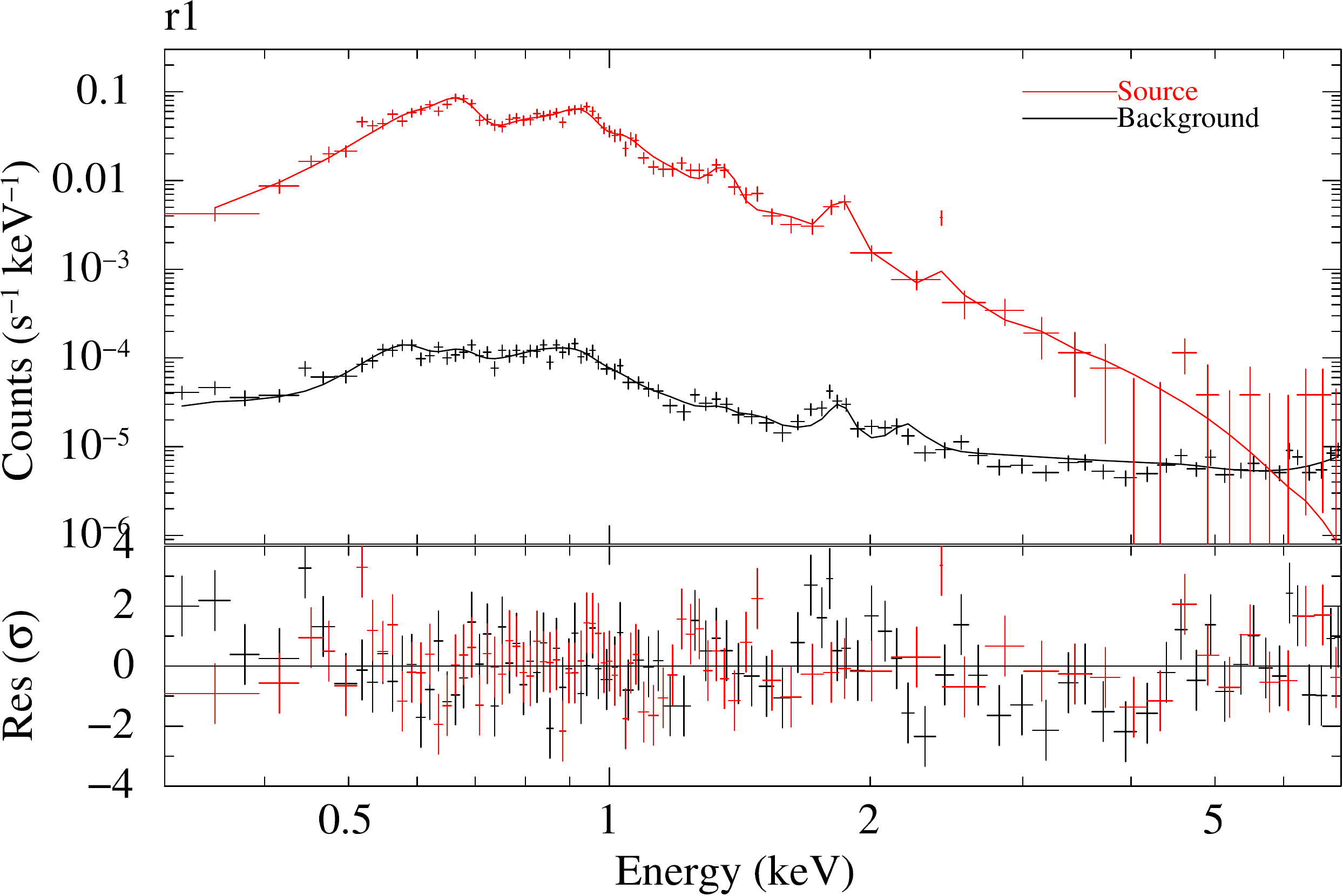}
\caption{Spectral fit for region $\mathrm{r1}$ on the rim with the single \texttt{vpshock} model. Also plotted is the background model fit that shows that the background is significantly less than the source spectrum below $5.5\,\kev$. Note that the data have been rebinned for plotting purposes. All other spectral fits for the rim regions and the blobs are present in Appendix \ref{s:app_rimfits}.}
\label{fig:rim1}
\end{figure}

A single component \texttt{vpshock} model provides an adequate fit (evaluated using the criteria described in Section \ref{s:fiteval}) for regions r1, r2, r4$-$r9, r11$-$r19, b1, and b2. Figure \ref{fig:rim1} shows the spectral fit with this model for region $\mathrm{r1}$, along with the background to emphasize that the background counts are significantly less than the source counts (see Appendix \ref{s:app_rimfits} for all other spectral fits of the rim regions and the blobs). However, some peculiarities are noticeable in the fits: regions $\mathrm{r1}$ and $\mathrm{r2}$ show systematic residual deviations around $1.5-2\,\kev$ and $0.5-0.6\,\kev$ respectively; $\mathrm{r6}$ and $\mathrm{r7}$ show excess Fe; $\mathrm{r11}$ underpredicts the flux near $1.2\,\kev$; $\mathrm{r13-r15}$ require a lower abundance of Mg; $\mathrm{r14}$ is also underabundant in O, Ne, and Si; $\mathrm{r16}$ shows higher than mean levels of S, and $\mathrm{r17}$ shows enhanced S; $\mathrm{b1}$ is consistent with excess Fe; and $\mathrm{b2}$ appears to contain ambient ISM material. Regions $\mathrm{r3}$ and $\mathrm{r10}$ are poorly fit with this model, and we re-fit them with the more complex \texttt{vpshock+vnei} model. We find that these regions show an additional plasma component with a higher temperature than the shell emission, which has been recently excited given their low ionization timescales. Unlike the single-component fits, the difference between \texttt{cstat (O)} and \texttt{cstat(E)} is well within the $2.6\,\sigma_{\mathrm{E}}$ limit. 

\begin{figure*}
\centering
\includegraphics[width=1.0\linewidth]{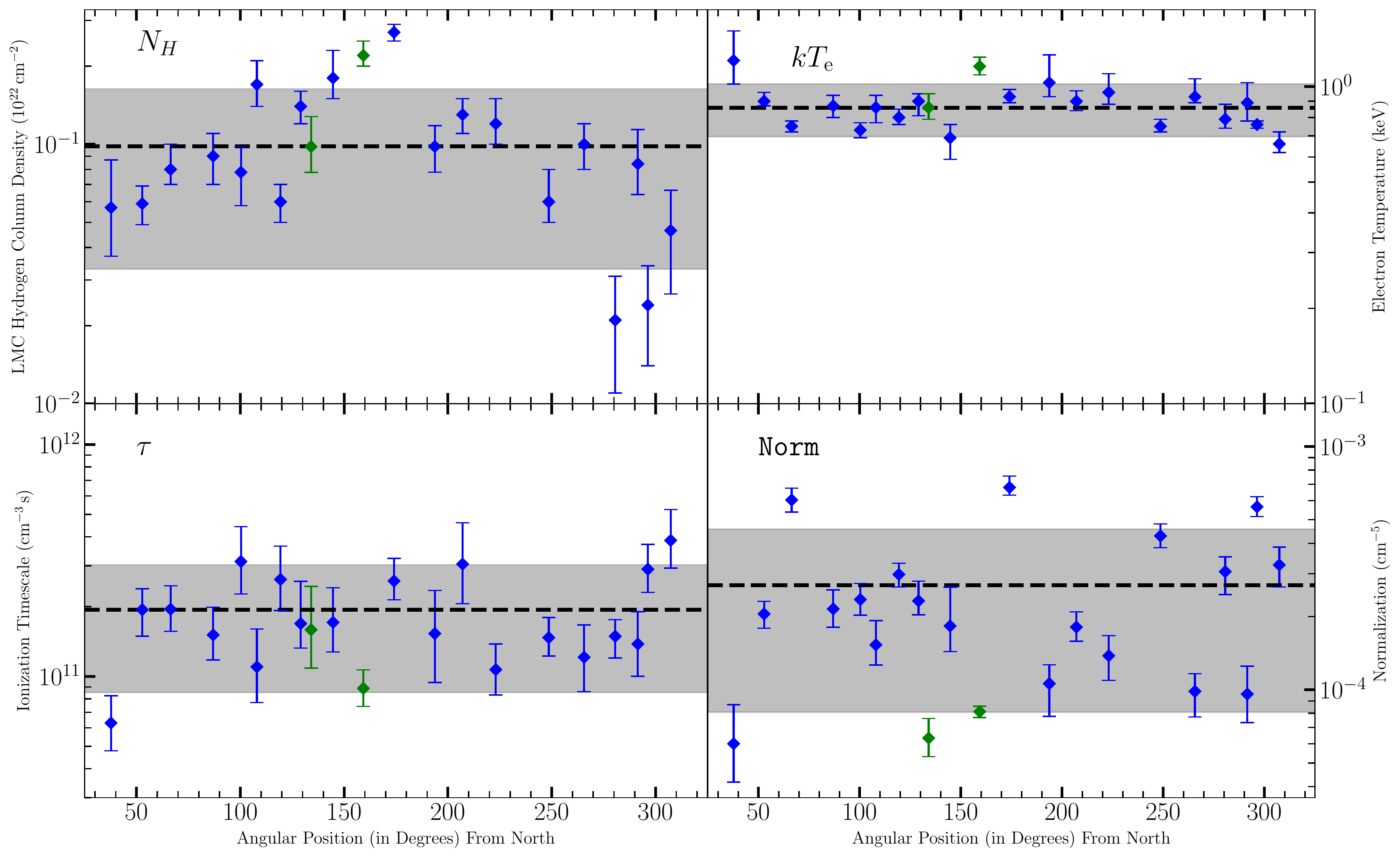}
\caption{Trends along the rim for a plane-parallel shock model (\texttt{vpshock}). Horizontal axis depicts the angular position of each region (in degrees) clockwise from the North. Blue markers denote values for the rim regions, and green markers denote values for the blobs (see Figure \ref{fig:allusefulregions} for information on the location of each region). Dashed line depicts the mean local LMC value for each parameter (see Table \ref{tab:scatter}) and the shaded area corresponds to the total (statistical+systematic) uncertainty on the mean.}
\label{fig:rim_trends1}
\end{figure*}

Figure \ref{fig:rim_trends1} depicts the trends seen along the rim in the parameters of interest, for the single \texttt{vpshock} model. We see the column density along the line of sight (\nh) to be higher in the southern part of the remnant than in the western and eastern parts which corresponds to presumably denser material (molecular clouds) present in that direction, as has been observed in the NANTEN CO survey \citep{2008ApJS..178...56F}, the Magellanic Mopra Assessment (MAGMA) survey of CO in the LMC with the Mopra telescope \citep{2011ApJS..197...16W}, and high resolution ALMA observations of N132D \citep{2019asrc.confE.123S}. In fact, many southern rim regions spatially coincide with the locations of shocked ISM clouds found by \citet[see their Figure 2]{2018ApJS..237...10D} in the optical. The ionization timescale ($\tau$) is roughly uniform over the shell and its values are indicative of a non-equilibrium plasma.

\begin{figure*}
\centering
\includegraphics[width=1.0\linewidth]{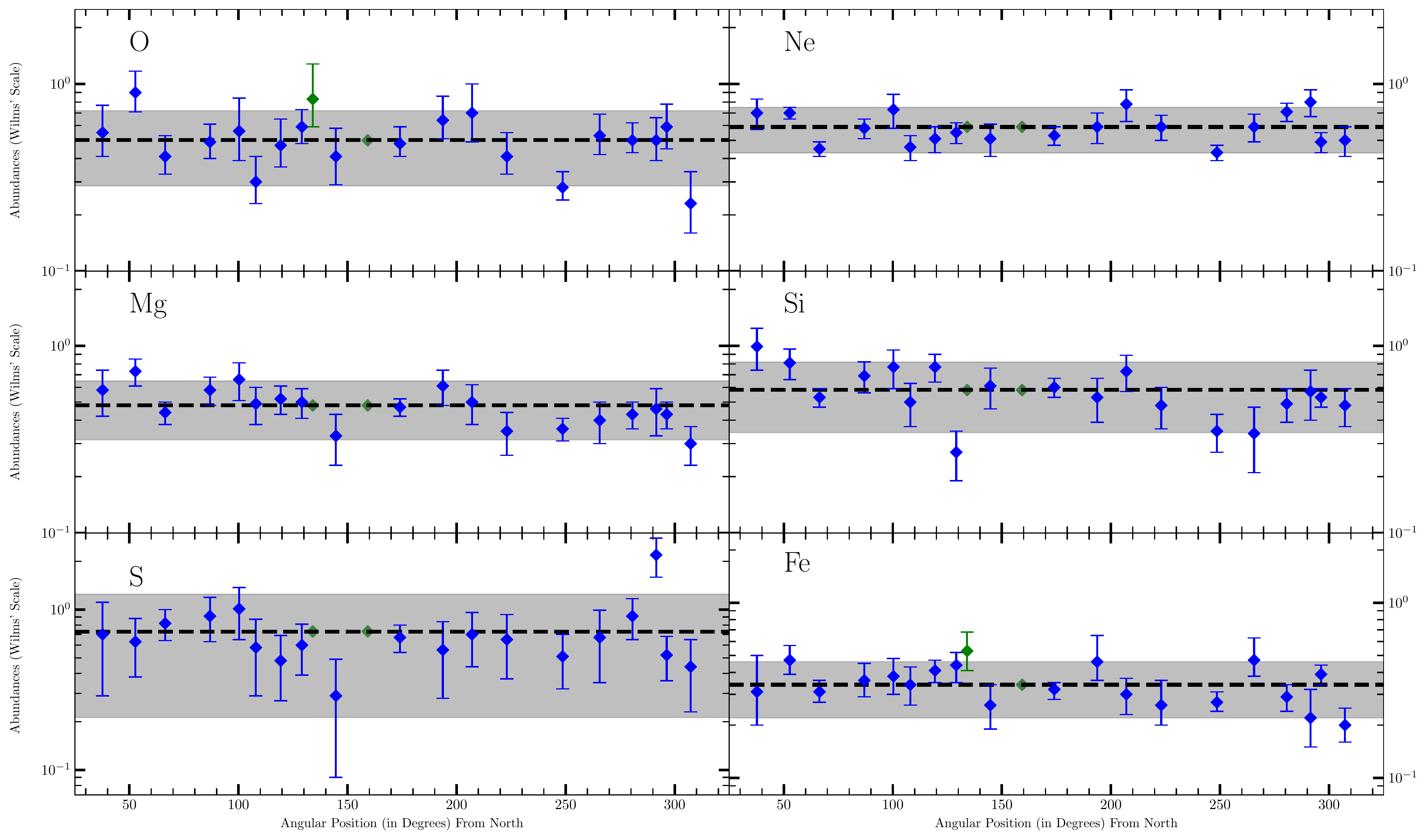}
\caption{Same as Figure \ref{fig:rim_trends1}, showing abundance pattern across the rim. Vertical axis is the abundance value relative to cosmic, on \cite{2000ApJ...542..914W} scale. Blue markers denote values for the rim regions, and green markers denote values for the blobs. The shaded area depicts the $\mu\pm\sigma_{\mu}$ domain for each abundance, where $\mu$ is the average value local to N132D (denoted by dashed lines) and $\sigma_{\mu}$ is the total (statistical+systematic) uncertainty on the mean that we use as a criteria to classify an abundance as enhanced/reduced (see Appendix \ref{s:append_errors}). The mean local LMC value ($\mu$) for each parameter is available in Table \ref{tab:scatter}, along with a comparison with other works in Table \ref{tab:lmcavg_compare}.}
\label{fig:rim_trends2}
\end{figure*}

Figure \ref{fig:rim_trends2} shows the abundance pattern across the rim for O, Ne, Mg, Si, S and Fe. Note that the fit results plotted in Figure \ref{fig:rim_trends2} are not the same as in Table \ref{tab:rimfit1} as we explain above. The shaded areas correspond to $1\sigma_{\mu}$ deviations from the mean abundance value, where $\sigma_{\mu}$ accounts for the statistical as well as systematic uncertainty around the mean. The thick dashed lines mark the mean value. In summary:

\begin{enumerate}
\item{\textrm{O}: The abundance of O is uniform along the rim, except in region $\mathrm{r2}$, where it is enhanced.}
\item{\textrm{Ne}: The abundance of Ne is within $\sigma_{\mu}$ of the mean throughout the rim.}
\item{\textrm{Mg}: The scatter in the abundance of Mg is < 1, implying that the variance between the different measurements is less than that within the measurements.}
\item{\textrm{Si}: The abundance of Si is also uniform across the rim, however, the scatter is more than one, implying the presence of localized variations. Further, region $\mathrm{r8}$ is marginally consistent with the average.} 
\item{\textrm{S}: We are cautious while thawing the abundance of S in the fits, because of the caveats listed in Section \ref{s:analysis}. Although it is poorly constrained on the rim regions due to low counts, region $\mathrm{r17}$ shows a significant enhancement. The scatter in S > 1, again implying the presence of localized variations.}
\item{\textrm{Fe}: The abundance of Fe is uniform over the rim. Like Mg, the scatter in the abundance of Fe < 1.}
\end{enumerate}

Based on our spectral analysis and the evaluation criteria for enhanced/reduced abundance measurements, we find that the abundances are largely uniform around the rim. The two exceptions to this are the enhanced O on the north-western rim (region $\mathrm{r2}$) and S on the north-eastern rim (region $\mathrm{r17}$).

\begin{deluxetable*}{l c c c c c c c c c c c r}
\tabletypesize{\scriptsize}
\tablenum{4}
\tablecolumns{13}
\tablecaption{Fit results of plane-parallel shock model (\texttt{vpshock}) on rim regions and blobs (see Figure \ref{fig:allusefulregions} for the location of each region). The abundances are with respect to the \texttt{Wilms} scale \citep{2000ApJ...542..914W}. Errors are the 68 per cent confidence intervals, equivalent to Gaussian $1\sigma$ around the quoted abundance.}
\tablehead{\colhead{Region ID} & \colhead{\nh} & \colhead{\kt} & \colhead{$\tau$ } & \colhead{\texttt{norm}} & \colhead{O} & \colhead{Ne}& \colhead{Mg}& \colhead{Si}& \colhead{S}& \colhead{Fe} & \colhead{\texttt{cstat(O)} / \texttt{dof}} & \colhead{\texttt{cstat(E)}$\pm\,\sigma_{\mathrm{E}}$}}
\colnumbers
\startdata
...&$10^{22}\,\mathrm{cm}^{-2}$&$\kev$&$10^{11}\,\mathrm{cm}^{-3}\,\mathrm{s}$&$10^{-4}\,\mathrm{cm}^{-5}$&...&...&...&...&...&...&...&...\\
\hline
$\mathrm{r1}$&0.05$^{+0.02}_{-0.02}$&1.22$^{+0.11}_{-0.08}$&0.60$^{+0.10}_{-0.09}$&0.66$^{+0.03}_{-0.04}$&0.46&0.59&0.44&0.84$^{+0.17}_{-0.15}$&0.40&0.29&816 / 911&$778\pm41$\\
$\mathrm{r2}$&0.06$^{+0.01}_{-0.01}$&0.92$^{+0.07}_{-0.04}$&1.78$^{+0.45}_{-0.38}$&2.17$^{+0.19}_{-0.29}$&0.78$^{+0.18}_{-0.13}$&0.59&0.64$^{+0.11}_{-0.09}$&0.70$^{+0.10}_{-0.11}$&0.40&0.45$^{+0.10}_{-0.08}$&869 / 908&$802\pm42$\\
$\mathrm{r3^*}$&0.08$^{+0.01}_{-0.01}$&0.75$^{+0.03}_{-0.03}$&2.12$^{+0.25}_{-0.23}$&6.03$^{+0.39}_{-0.38}$&0.46&0.46$^{+0.03}_{-0.04}$&0.44&0.52&0.81$^{+0.20}_{-0.09}$&0.32$^{+0.03}_{-0.02}$&939 / 909&$803\pm42$\\
$\mathrm{r4}$&0.09$^{+0.02}_{-0.02}$&0.82$^{+0.03}_{-0.03}$&1.79$^{+0.21}_{-0.20}$&2.57$^{+0.13}_{-0.11}$&0.46&0.54$^{+0.05}_{-0.05}$&0.49$^{+0.06}_{-0.06}$&0.52&0.82$^{+0.28}_{-0.24}$&0.29&826 / 909&$782\pm42$\\
$\mathrm{r5}$&0.09$^{+0.02}_{-0.02}$&0.71$^{+0.02}_{-0.02}$&3.29$^{+0.38}_{-0.33}$&2.35$^{+0.08}_{-0.10}$&0.46&0.59&0.44&0.52&0.75$^{+0.30}_{-0.25}$&0.29&823 / 911&$750\pm41$\\
$\mathrm{r6}$&0.15$^{+0.03}_{-0.02}$&0.82$^{+0.11}_{-0.05}$&1.82$^{+0.48}_{-0.64}$&1.45$^{+0.19}_{-0.29}$&0.46&0.59&0.59$^{+0.12}_{-0.08}$&0.52&0.40&0.37$^{+0.10}_{-0.05}$&744 / 910&$747\pm40$\\
$\mathrm{r7}$&0.06$^{+0.01}_{-0.01}$&0.80$^{+0.04}_{-0.04}$&2.70$^{+0.42}_{-0.40}$&3.16$^{+0.22}_{-0.26}$&0.46&0.48$^{+0.04}_{-0.07}$&0.44&0.70$^{+0.10}_{-0.05}$&0.40&0.39$^{+0.05}_{-0.04}$&792 / 909&$783\pm41$\\
$\mathrm{r8}$&0.15$^{+0.02}_{-0.02}$&0.82$^{+0.09}_{-0.05}$&1.60$^{+0.30}_{-0.37}$&2.71$^{+0.30}_{-0.39}$&0.46&0.59&0.44&0.52&0.66$^{+0.24}_{-0.20}$&0.34$^{+0.06}_{-0.04}$&883 / 910&$778\pm42$\\
$\mathrm{r9}$&0.18$^{+0.03}_{-0.02}$&0.70$^{+0.03}_{-0.04}$&1.67$^{+0.33}_{-0.26}$&1.63$^{+0.12}_{-0.85}$&0.46&0.59&0.44&0.71$^{+0.15}_{-0.14}$&0.40&0.29&760 / 911&$741\pm41$\\
$\mathrm{r10^*}$&0.27$^{+0.02}_{-0.03}$&0.93$^{+0.02}_{-0.01}$&2.61$^{+0.27}_{-0.40}$&7.20$^{+0.22}_{-0.25}$&0.46&0.51$^{+0.03}_{-0.03}$&0.44&0.52&0.40&0.29&967 / 911&$855\pm42$\\
$\mathrm{r11}$&0.10$^{+0.02}_{-0.02}$&1.04$^{+0.12}_{-0.07}$&1.61$^{+0.46}_{-0.27}$&1.61$^{+0.22}_{-0.13}$&0.56$^{+0.09}_{-0.08}$&0.59&0.51$^{+0.08}_{-0.07}$&0.52&0.40&0.40$^{+0.04}_{-0.06}$&872 / 909&$779\pm41$\\
$\mathrm{r12}$&0.12$^{+0.02}_{-0.02}$&0.97$^{+0.04}_{-0.03}$&1.88$^{+0.26}_{-0.23}$&1.82$^{+0.07}_{-0.08}$&0.46&0.59&0.44&0.63$^{+0.10}_{-0.10}$&0.40&0.29&822 / 911&$797\pm42$\\
$\mathrm{r13}$&0.11$^{+0.02}_{-0.02}$&1.02$^{+0.07}_{-0.07}$&0.99$^{+0.36}_{-0.20}$&1.25$^{+0.09}_{-0.08}$&0.46&0.59&0.38$^{+0.06}_{-0.06}$&0.52&0.40&0.29&849 / 911&$784\pm40$\\
$\mathrm{r14}$&0.05$^{+0.01}_{-0.01}$&0.77$^{+0.04}_{-0.03}$&1.42$^{+0.31}_{-0.25}$&4.05$^{+0.15}_{-0.14}$&0.29$^{+0.04}_{-0.03}$&0.43$^{+0.03}_{-0.03}$&0.37$^{+0.04}_{-0.04}$&0.36$^{+0.07}_{-0.07}$&0.40&0.29&827 / 908&$789\pm41$\\
$\mathrm{r15}$&0.10$^{+0.02}_{-0.02}$&0.95$^{+0.10}_{-0.08}$&1.03$^{+0.31}_{-0.23}$&1.00$^{+0.15}_{-0.14}$&0.46&0.59&0.38$^{+0.08}_{-0.08}$&0.33$^{+0.12}_{-0.11}$&0.40&0.44$^{+0.08}_{-0.07}$&789 / 909&$758\pm40$\\
$\mathrm{r16}$&0.02$^{+0.02}_{-0.01}$&0.79$^{+0.04}_{-0.03}$&1.47$^{+0.16}_{-0.23}$&3.11$^{+0.32}_{-0.27}$&0.46&0.69$^{+0.04}_{-0.03}$&0.44&0.52&0.86$^{+0.26}_{-0.23}$&0.29&827 / 910&$786\pm41$\\
$\mathrm{r17}$&0.07$^{+0.02}_{-0.02}$&1.04$^{+0.08}_{-0.07}$&0.96$^{+0.20}_{-0.18}$&0.81$^{+0.07}_{-0.06}$&0.46&0.73$^{+0.07}_{-0.06}$&0.44&0.52&1.84$^{+0.56}_{-0.47}$&0.29&767 / 910&$766\pm41$\\
$\mathrm{r18}$&0.02$^{+0.01}_{-0.01}$&0.76$^{+0.03}_{-0.02}$&2.26$^{+0.24}_{-0.22}$&5.84$^{+0.34}_{-0.35}$&0.46&0.45$^{+0.03}_{-0.03}$&0.44&0.52&0.40&0.36$^{+0.03}_{-0.02}$&883 / 910&$804\pm42$\\
$\mathrm{r19}$&0.02$^{+0.01}_{-0.01}$&0.71$^{+0.02}_{-0.01}$&5.48$^{+0.64}_{-0.58}$&2.46$^{+0.07}_{-0.08}$&0.46&0.69$^{+0.06}_{-0.06}$&0.44&0.52&0.40&0.29&800 / 911&$758\pm41$\\
\hline
$\mathrm{b1}$&0.10$^{+0.03}_{-0.02}$&0.86$^{+0.09}_{-0.08}$&1.59$^{+0.86}_{-0.50}$&0.63$^{+0.13}_{-0.10}$&0.83$^{+0.45}_{-0.24}$&0.59&0.48&0.58&0.73&0.53$^{+0.15}_{-0.12}$&770 / 910&$730\pm39$\\
$\mathrm{b2}$&0.22$^{+0.03}_{-0.03}$&1.12$^{+0.08}_{-0.10}$&0.95$^{+0.30}_{-0.15}$&0.88$^{+0.07}_{-0.05}$&0.46&0.59&0.44&0.52&0.40&0.29&848 / 912&$797\pm40$\\
\hline
\enddata
\let\amp=&
\catcode`\&=12
\tablecomments{Columns $2-5$ describe the LMC hydrogen column density, electron temperature, ionization timescale and normalization parameter, respectively. Columns $6-11$ describe the best-fit abundances of each element. Column 12 lists \texttt{cstat(O)} which is the observed C-statistic we obtain from \xspec; \texttt{cstat(E)} listed in column 13 is the expected C-statistic from the \cite{2017A&A...605A..51K} formulation. The fits are considered unsuccessful if $|\texttt{cstat(O)-cstat(E)}| > 2.6\sigma_{\mathrm{E}}$ where $\sigma^{2}_{\mathrm{E}}$ is the expected variance of \texttt{cstat(E)}. Regions with asterisk ($\mathrm{r3}$ and $\mathrm{r10}$) are those which could not be fit with a single \texttt{vpshock} model; they were re-fit with a more complex model as described in Table \ref{tab:rimfit2}. In region $\mathrm{r11}$, the best-fit shows an enhanced abundance of N (which is otherwise tied to O): N = $0.80\pm0.11$. The mean values derived from fitting all the abundances in all the regions are present in Table \ref{tab:scatter}. Spectral fits for all regions are present in Appendix \ref{s:app_rimfits}.}
\label{tab:rimfit1}
\end{deluxetable*}

\subsection{Interior Regions}
\label{s:results_ejecta}
In this section, we describe results from the spectral analysis of the interior regions $\mathrm{e1}$ and $\mathrm{f1}-\mathrm{f6}$ that were selected from narrow-band and hardness ratio images as having enhanced abundances and signatures of Fe K emission, respectively. 

\subsubsection{Region with Enriched Abundances}
\label{s:results_e1}
\begin{figure}
\includegraphics[width=0.60\columnwidth, angle=270]{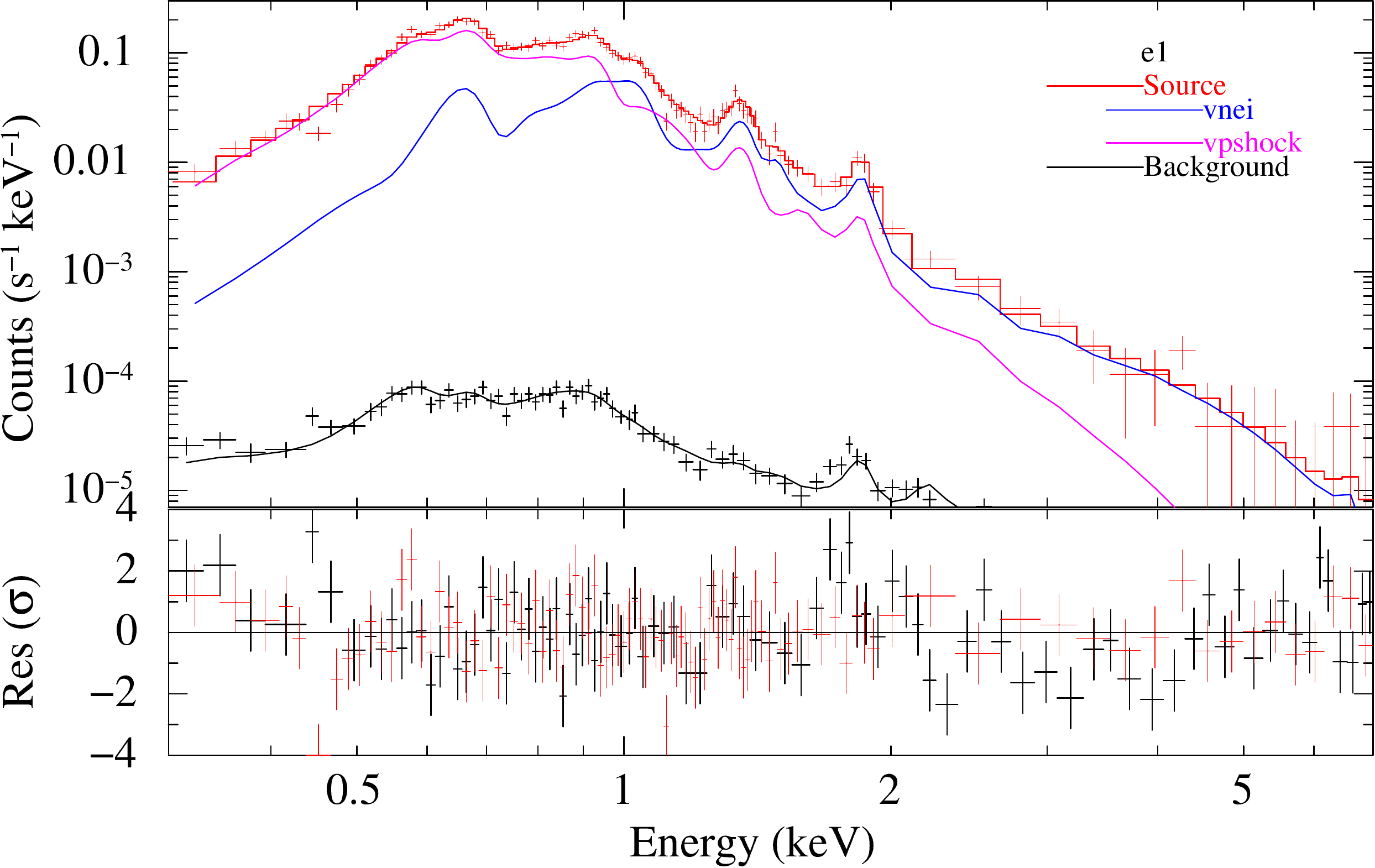}
\caption{Fit and residuals for region $\mathrm{e1}$ with the (\texttt{vpshock} + \texttt{vnei}) model. The best-fit parameters are listed in Table \ref{tab:rimfit2}. This region shows enhanced abundances ($\ga 2.5 \times$ mean) of all elements (except S) and contains one or more X-ray bright ejecta fragments. It partially overlaps with the ejecta fragment seen in the optical \protect\citep{1996AJ....112.2350M,1995AJ....109.2104M,2000ApJ...537..667B} as shown in Figure \ref{fig:allusefulregions}. Also plotted is the background that is significantly lesser than the source at energies < $5.5\,\kev$. Note that the spectral counts have been rebinned for display purposes.}
\label{fig:ejectafit}
\end{figure}

Table \ref{tab:rimfit2} shows the fit results for region $\mathrm{e1}$ and Figure \ref{fig:ejectafit} shows the source spectra with the best-fit model. The results reveal enriched abundances ($\ga\,2.5\,\times$ mean) of all elements (except S) in this region, consistent with the excess flux at different line energies we observe in the narrow band images. Adding a single NEI component to the model fits the observed spectrum well with an electron temperature of $\sim2.0\,\kev$. The higher temperature of the \texttt{vnei} as compared to the shell emission from the rim implies the presence of one or multiple shock heated ejecta clumps in this region. The shorter ionization timescale indicates that the ejecta-rich clump(s) present in this region have been recently heated by the shock. The best-fit abundances have large uncertainties due to low number of counts. Nevertheless, they are significantly higher than the LMC abundances. We use the best-fitting parameters from the fit for this region to deduce the mass of the progenitor in Section \ref{s:ejecta}. The coeval presence of optical and X-ray emitting ejecta in a region has also been observed in SNR G292.0+1.8 and Cas A, where the optical emission is proposed to come from dense ejecta-rich knots and the X-ray emission from a lower density plasma \citep{2005ApJ...635..365G,2014ApJ...789..138P}.

\begin{deluxetable*}{c c c c c | c c c c c c c c c | c c}
\rotate
\tabletypesize{\scriptsize}
\tablenum{5}
\tablecolumns{16}
\tablecaption{As in Table \ref{tab:rimfit1}, for fit results of \texttt{vpshock+vnei} model on rim regions (where the single \texttt{vpshock} did not yield an acceptable fit) and region $\mathrm{e1}$. The spectral fits for the rim regions are present in Appendix \ref{s:app_rimfits} and for region $\mathrm{e1}$ in Figure \ref{fig:ejectafit}.}
\tablehead{\colhead{Region ID} & \colhead{\nh} & \colhead{\kt} & \colhead{$\tau$} & \colhead{\texttt{norm}} \vline & \colhead{\kt} & \colhead{$\tau$} & \colhead{\texttt{norm}} & \colhead{O} & \colhead{Ne}& \colhead{Mg}& \colhead{Si}& \colhead{S}& \colhead{Fe} \vline &\colhead{\texttt{cstat (O)} / \texttt{dof}} & \colhead{\texttt{cstat(E)}$\pm\,\sigma_{\mathrm{E}}$}}
\colnumbers
\startdata
...&$10^{22}\,\mathrm{cm}^{-2}$&$\kev$&$10^{11}\,\mathrm{cm}^{-3}\,\mathrm{s}$&$10^{-4}\,\mathrm{cm}^{-5}$&$\kev$&$10^{11}\,\mathrm{cm}^{-3}\,\mathrm{s}$&$10^{-4}\,\mathrm{cm}^{-5}$&...&...&...&...&...&...&...&...\\
\hline
$\mathrm{r3}$&0.10$^{+0.02}_{-0.02}$&0.64$^{+0.02}_{-0.03}$&6.14$^{+2.39}_{-1.48}$&6.41$^{+0.52}_{-0.74}$&1.64$^{+0.59}_{-0.47}$&0.05$^{+0.03}_{-0.02}$&0.56$^{+0.32}_{-0.19}$&0.46&0.59&0.48$^{+0.06}_{-0.06}$&0.57$^{+0.13}_{-0.12}$&0.40&0.29&888 / 907&$829\pm42$\\
$\mathrm{r10}$&0.24$^{+0.03}_{-0.03}$&0.76$^{+0.08}_{-0.06}$&3.21$^{+0.85}_{-0.61}$&5.78$^{+1.36}_{-0.94}$&1.54$^{+0.67}_{-0.33}$&0.69$^{+0.30}_{-0.21}$&1.55$^{+0.90}_{-0.85}$&0.46&0.59&0.44&0.52&0.40&0.29&943 / 909&$878\pm43$\\
$\mathrm{e1}$&0.09$^{+0.02}_{-0.01}$&0.61$^{+0.35}_{-0.09}$&0.70$^{+0.20}_{-0.20}$&1.47$^{+0.32}_{-0.19}$&2.02$^{+1.29}_{-0.68}$&0.42$^{+0.38}_{-0.08}$&0.25$^{+0.07}_{-0.16}$&1.63$^{+0.64}_{-0.29}$&1.59$^{+1.00}_{-0.42}$&2.50$^{+1.27}_{-0.35}$&1.70$^{+0.86}_{-0.21}$&0.40&0.96$^{+0.47}_{-0.37}$&858 / 904&$803\pm41$\\
\hline
\enddata
\tablecomments{Parameters in columns $3-5$ consist of the \texttt{vpshock} component, those in columns $6-14$ belong to the \texttt{vpshock} component. Elemental abundances of \texttt{vpshock} are frozen to local LMC averages.}
\label{tab:rimfit2}
\end{deluxetable*}

\begin{table}
\centering
\tablenum{6}
\caption{Mean ($\mu$) local LMC parameters, obtained from the fits to the rim regions. Error ($\sigma_{\mu}$) denotes the statistical as well as systematic uncertainty on the mean. Abundances are quoted with respect to the \texttt{Wilms} scale. Uniformity of a parameter over the shell is judged by its scatter value. Both $\sigma_{\mu}$ and scatter are calculated from the method of Multiple Imputations (see Appendix \ref{s:append_errors}).}
\begin{tabular}{|lccr|}
\hline
Parameter&Units&$\mu\,\pm\,\sigma_{\mu}$&Scatter\\
\hline
\hline
\nh&$10^{22}\,\mathrm{cm}^{-2}$&$0.10\pm0.07$&2.74\\
\kt&$\kev$&$0.86\pm0.16$&1.37\\
$\tau$&$10^{11}\,\mathrm{cm}^{-3}\,\mathrm{s}$&$1.94\pm1.09$&1.26\\
O&...&$0.50\pm0.22$&1.00\\
Ne&...&$0.59\pm0.16$&1.01\\
Mg&...&$0.48\pm0.17$&0.93\\
Si&...&$0.58\pm0.24$&1.14\\
S&...&$0.73\pm0.52$&1.18\\
Fe&...&$0.34\pm0.12$&0.96\\
\texttt{norm}&$10^{-4}\,\mathrm{cm}^{-5}$&$2.69\pm1.88$&4.10\\
\hline
\end{tabular}\\
\label{tab:scatter}
\end{table}

\subsubsection{Regions with Fe K Emission}
\label{s:results_fek}
We use a \texttt{vnei} + \texttt{vnei} + \texttt{vpshock} model (referred to as the \textit{ionizing} model) to fit the spectra in regions $\mathrm{f1-f6}$. We also use a \textit{recombining} plasma model (\texttt{vrnei} + \texttt{vnei} + \texttt{vpshock}) as an alternate explanation to look for possible signatures of a recombining plasma in N132D. These models contain three components: one to account for shell emission along the line of sight (\texttt{vpshock}), second to account for the cooler soft X-ray emission and some of the high energy continuum (\texttt{vnei}2 and \texttt{vnei} in ionizing and recombining models, respectively), and third to account for the hotter Fe K emission and the remainder of the hard X-ray spectra (\texttt{vnei}1 and \texttt{vrnei} in ionizing and recombining models, respectively). We find that a three component fit is essential because no combination of a two component model of an NEI plasma is able to model the Fe K feature while simultaneously explaining the Fe L shell emission around $1\,\kev$. Evidence for the need of a third, hotter component is established when we artificially increase the abundance of Fe in the NEI component of a two-component (\texttt{vnei/vrnei} + \texttt{vpshock}) model to reproduce the observed flux in the Fe K feature. This experiment of increasing the Fe abundance in order to get enough flux in the Fe K$\alpha$ line overproduces the Fe L emission at $\sim1\,\kev$. Thus, we establish that at least two separate plasma conditions are needed to explain the Fe L ($\sim 1\,\kev$) and Fe K ($\sim 6.7\,\kev$) emissions, which has also been noted before for this remnant \citep{2016A&A...585A.162M,2018ApJ...854...71B}. 

\begin{figure}
{\includegraphics[angle=270, width=0.95\linewidth]{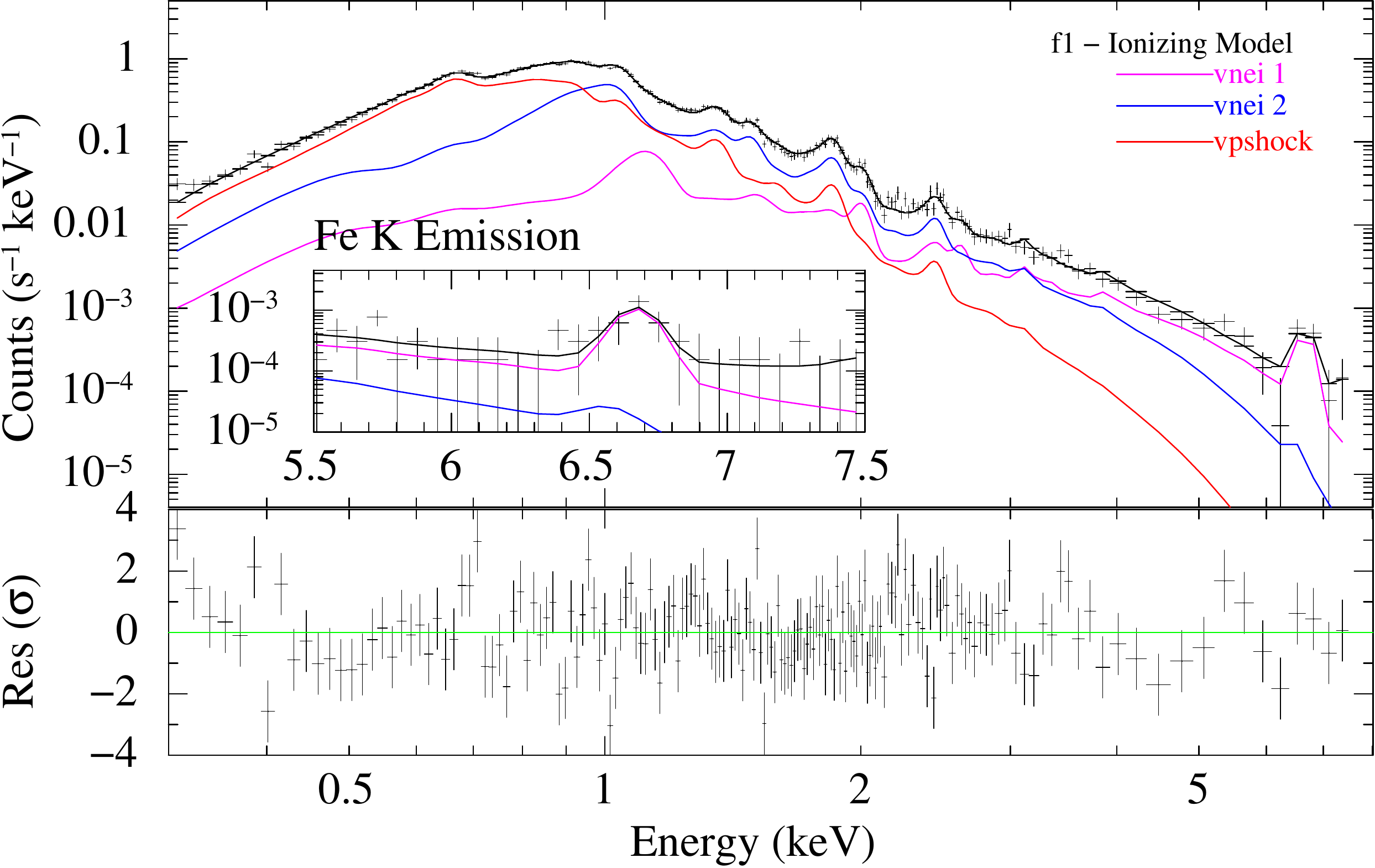}}
{\includegraphics[angle=270, width=0.95\linewidth]{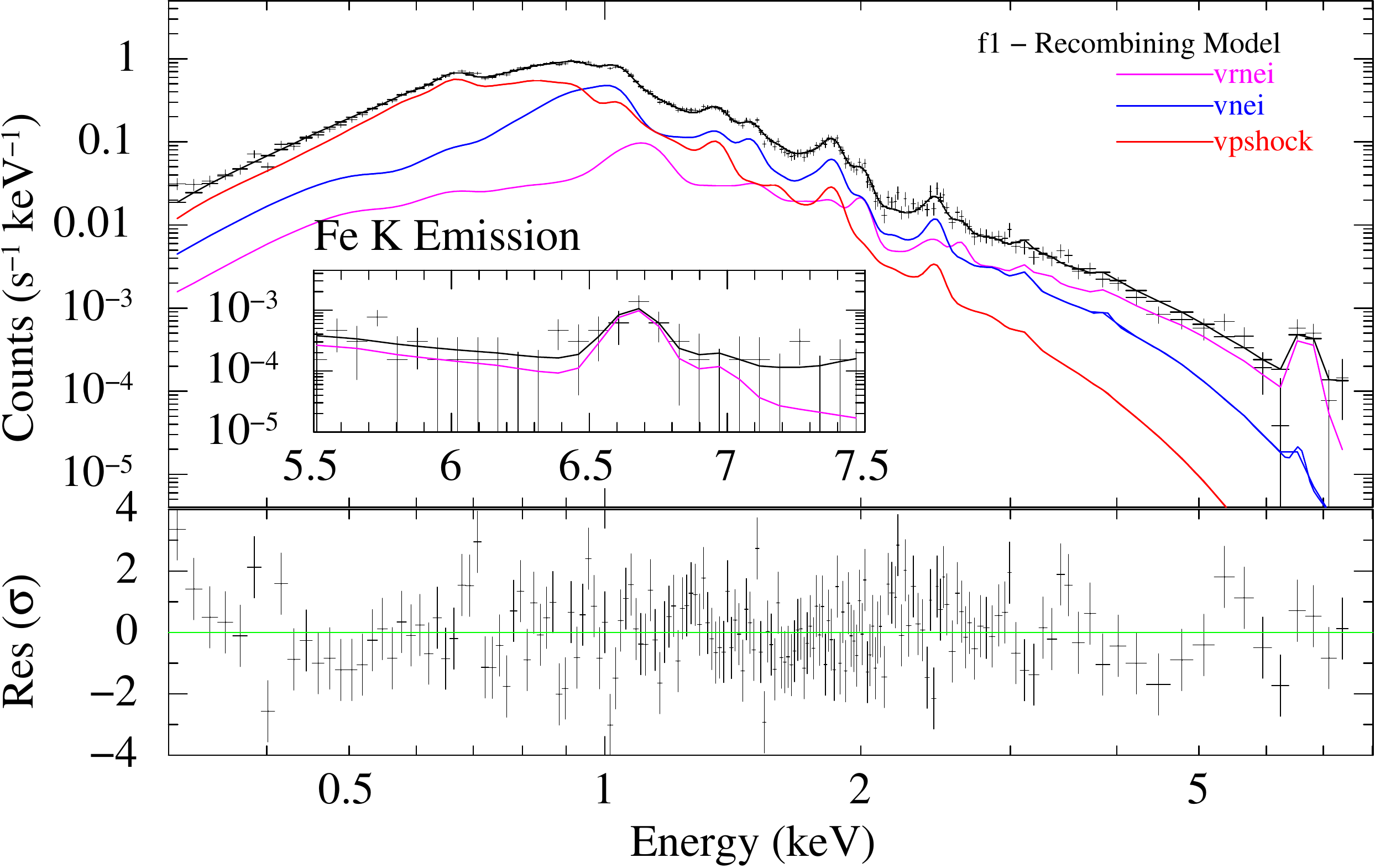}}
\caption{Fits and residuals for the region $\mathrm{f1}$, showing the weak Fe K emission feature present at $\sim6.7\,\kev$. Inset zooms into the energy range $5.5-7.5\,\kev$ that contains this emission. The fit parameters for both the ionizing and recombining models are presented in Table \ref{tab:fekregionss}. With the current number of counts, it is difficult to distinguish between the origin and the nature of the plasma since the spectral fits of both the models are statistically same. Note that the spectral counts have been rebinned differently in the main plot and the inset for display purposes. Spectral plots for other Fe K regions are present in Appendix \ref{s:app_rimfits}.}
\label{fig:fek_f1}
\end{figure}

We show the best-fit parameters for the ionizing and recombining models for regions $\mathrm{f1-f6}$ in Table \ref{tab:fekregionss}, and the corresponding spectral plots in Figures \ref{fig:fek_f1}, \ref{fig:fek_f2}, \ref{fig:fek_f3}, \ref{fig:fek_f4}, \ref{fig:fek_f5}, and \ref{fig:fek_f6}, respectively. Note that the total emission (top black histogram and curve) includes the background model, which is why it levels off at a higher level than the magenta curves in the inset in these Figures. There are several features of these fits that should be highlighted. Firstly, the quality of the fit in terms of the fit statistic \texttt{cstat(O)/dof} is indistinguishable for the \textit{ionizing} and \textit{recombining} plasma models in all the regions. Based on these results we can not conclude that one model is preferred over the other. We also note that the initial electron temperature (\texttt{kT\_init}) for the \texttt{vrnei} component in the recombining models is highly degenerate and gives similar results for temperatures higher than $10\,\kev$ (see also, \citealt{2018ApJ...854...71B}); hence, we freeze it at this value (see, for example, \citealt{2017ApJ...847..121A,2018PASJ...70..110K}). Further, we find that both the model fits for regions $\mathrm{f1,\,f2,\,f4}$ and $\mathrm{f6}$ are acceptable according to our criteria whereas those for region $\mathrm{f3}$ and $\mathrm{f5}$ are marginally inconsistent with our chosen acceptability criteria because for the latter two, $\mathtt{cstat (O)} \approx \mathtt{cstat (E)} + 3.1\,\sigma_{\mathrm{E}}$. This indicates that overall the fits are good, however, there are details which the models fail to reproduce. It also highlights the trade-off between using large regions that encompass sufficient Fe K emission and the existence of multiple plasma conditions within them that complicate the spectral modeling.

The electron temperatures and ionization timescales for the shell and the cooler X-ray emission model components are identical in both the models in all the six regions; they fall in the partial non-equilibration category as defined by \citet[see their Section 5.3]{2012A&ARv..20...49V}. Additionally, both the best-fit ionizing and recombining plasma models result in abundances for Si and Fe (in the hotter component) that are significantly enhanced compared to the expected abundances for the LMC in 4 out of the 6 regions. The enhanced abundance of Fe in these regions distinguishes them from the regions at the rim. The regions interior to the remnant contain plasma with a sufficiently high temperature and Fe abundance to produce the observed Fe K emission. We also let the abundances of O, Ne and Mg vary and find that both the best-fit models have enriched Ne and Mg in the cooler model component in regions $\mathrm{f1}$ and $\mathrm{f6}$, and in the recombining plasma model in region $\mathrm{f3}$. 

Although there are several similarities in the two models that lead us to conclude they cannot be distinguished with the current data, there are subtle differences that provide some understanding of the plasma conditions in these regions. For example, \texttt{cstat(O)} is slightly lesser for the recombining plasma models in regions $\mathrm{f1,\,f2,\,f3}$ and $\mathrm{f6}$; the ionization timescale for the hotter NEI component in the ionizing model (\texttt{vnei}1) approaches equilibrium within the uncertainties in regions $\mathrm{f3,\,f4}$ and $\mathrm{f6}$ whereas that for the recombining model essentially represents non-equilibrium plasma (except perhaps for region $\mathrm{f6}$). This could imply that the plasma is in fact evolving through recombination post equilibrium. The recombining models also exhibit typical LMC abundance of S in the \texttt{vrnei} component, except for region $\mathrm{f4}$. The constraints on best-fit abundances are tighter in the recombining models, and there are no values that might be nonphysical and simply a result of the complex nature of the fit; for example, the abundance of Si in the hotter model component in region $\mathrm{f2}$ in the ionizing model, which is $\sim7\times$ the solar value. The hottest plasma component in the recombining model contributes more to the emission at lower energies than the ionizing model, while both components explain the Fe K emission. A similar observation can also be made by realizing that the emission measures (\texttt{norm}) of the hotter component in the recombining model is larger than that in the corresponding ionizing model, except for region $\mathrm{f6}$. This hints at the possibility of different origins of the hot plasma in different parts of the remnant, as has been shown for the SNR G290.1-0.8 \citep{2015PASJ...67...16K}, as well as in simulations of an SNR shock interacting with a distribution of clouds in the ISM \citep{2019ApJ...875...81Z}, however, no definitive conclusions on the origin of the hot component can be drawn from the current data\footnote{Note that both \cite{2015PASJ...67...16K} and \cite{2019ApJ...875...81Z} work with mixed-morphology remnants whereas N132D has not been classified as one so far.}. 

Thus, we establish from this analysis that: 1. Fe K emission in N132D is not contained in a single ejecta clump or discrete feature, rather, it is largely spread across the southern half of the remnant, and 2. the plasma that leads to the production of Fe K is either hot with a surprisingly large value of the ionization timescale or undergoing recombination (with slight indications in the favor of the latter). In either case this plasma is physically distinct from the plasma that produces the soft X-ray emission. We further discuss its implications in Section \ref{s:fek}.

\begin{deluxetable*}{|l |c c c c c c r | l| c c c c c c r|}
\rotate
\tabletypesize{\scriptsize}
\tablenum{7}
\tablecolumns{16}
\label{tab:fekregionss}
\tablecaption{As in Table \ref{tab:rimfit1}, for ionizing and recombining plasma model fits for regions $\mathrm{f1, f2, f3, f4, f5}$ and $\mathrm{f6}$, which have weak signature of Fe K emission (see Figures \ref{fig:fek_f1}, \ref{fig:fek_f2}, \ref{fig:fek_f3}, \ref{fig:fek_f4}, \ref{fig:fek_f5} and \ref{fig:fek_f6}). Elemental abundances of respective model components not shown here are all frozen to average local LMC values (as in Table \ref{tab:scatter}).}
\tablehead{\colhead{Ionizing} & \colhead{} & \colhead{} & \colhead{} & \colhead{} & \colhead{} & \colhead{} & \colhead{} \vline & \colhead{Recombining} & \colhead{} & \colhead{} & \colhead{} & \colhead{} & \colhead{} & \colhead{} & \colhead{} \vline \\
\colhead{Model} & \colhead{} & \colhead{} & \colhead{} & \colhead{} & \colhead{} & \colhead{} & \colhead{} \vline & \colhead{Model} & \colhead{} & \colhead{} & \colhead{} & \colhead{} & \colhead{} & \colhead{} & \colhead{} \vline
}
\startdata
Parameter & Units & Region $\mathrm{f1}$ & Region $\mathrm{f2}$ & Region $\mathrm{f3}$ & Region $\mathrm{f4}$ & Region $\mathrm{f5}$ & Region $\mathrm{f6}$ & Parameter & Units & Region $\mathrm{f1}$ & Region $\mathrm{f2}$ & Region $\mathrm{f3}$ & Region $\mathrm{f4}$ & Region $\mathrm{f5}$ & Region $\mathrm{f6}$ \\
\hline
\nh&$10^{22}\,\mathrm{cm}^{-2}$&$0.09^{+0.01}_{-0.01}$&$0.16^{+0.01}_{-0.01}$&$0.16^{+0.01}_{-0.01}$&$0.14^{+0.01}_{-0.01}$ & $0.18^{+0.01}_{-0.01}$&$0.15^{+0.01}_{-0.01}$& \nh&$10^{22}\,\mathrm{cm}^{-2}$&$0.09^{+0.01}_{-0.01}$&$0.15^{+0.01}_{-0.01}$&$0.16^{+0.01}_{-0.01}$&$0.15^{+0.01}_{-0.01}$&$0.18^{+0.01}_{-0.01}$ & $0.15^{+0.01}_{-0.01}$\\
{\kt \texttt{vnei} 1}&$\kev$&$2.66^{+0.32}_{-0.29}$&$2.70^{+0.11}_{-0.14}$&$1.88^{+0.11}_{-0.13}$& $2.43^{+0.14}_{-0.32}$&$1.75^{+0.03}_{-0.14}$ &$1.45^{+0.06}_{-0.04}$&{\kt \texttt{vrnei}}&$\kev$&$1.97^{+0.21}_{-0.21}$&$1.33^{+0.16}_{-0.17}$&$1.27^{+0.08}_{-0.07}$&$2.06^{+0.45}_{-0.39}$&$1.54^{+0.19}_{-0.13}$ & $1.45^{+0.04}_{-0.07}$\\
...&...&...&...&...&...&...&...&\texttt{kT\_init} \texttt{vrnei}&$\kev$&10&10&10&10&10&10\\
Si&...&$3.27^{+2.41}_{-1.14}$&$6.95^{+3.45}_{-0.76}$&$1.67^{+0.68}_{-0.27}$&$2.30^{+1.76}_{-0.79}$ &$1.75^{+0.56}_{-0.45}$ &$1.02^{+0.16}_{-0.15}$&Si&...&$3.17^{+0.94}_{-0.82}$&$2.64^{+1.42}_{-0.95}$&$0.91^{+0.16}_{-0.10}$&$3.05^{+1.64}_{-0.88}$&$1.71^{+0.29}_{-0.34}$ & $1.03^{+0.18}_{-0.14}$\\
S&...&$1.69^{+0.81}_{-0.62}$&$1.50^{+0.69}_{-0.42}$&$0.73^{+0.29}_{-0.23}$& $1.56^{+0.92}_{-0.65}$&$0.79^{+0.43}_{-0.35}$ &$0.73^{+0.15}_{-0.14}$&S&...&$1.66^{+0.80}_{-0.66}$&$0.39^{+0.64}_{-0.04}$&$0.39^{+0.15}_{-0.15}$&$2.72^{+1.49}_{-0.86}$&$0.91^{+0.30}_{-0.28}$ & $0.56^{+0.17}_{-0.16}$\\
Fe&...&$2.59^{+0.79}_{-0.70}$&$1.87^{+0.37}_{-0.27}$&$0.90^{+0.17}_{-0.15}$& $1.37^{+0.57}_{-0.33}$&$0.61^{+0.15}_{-0.18}$ &$0.47^{+0.07}_{-0.11}$&Fe&...&$2.11^{+0.50}_{-0.42}$&$0.78^{+0.38}_{-0.29}$&$0.49^{+0.13}_{-0.06}$&$1.67^{+0.72}_{-0.19}$&$0.56^{+0.13}_{-0.12}$ & $0.43^{+0.06}_{-0.07}$\\
$\tau$ \texttt{vnei} 1&$10^{11}\,\mathrm{cm}^{-3}\,\mathrm{s}$&$4.84^{+1.54}_{-1.19}$&$5.91^{+3.66}_{-1.94}$&$7.11^{+5.93}_{-1.94}$&$6.21^{+10.58}_{-2.74}$ &$5.50^{+3.14}_{-1.52}$ &$6.27^{+5.13}_{-2.64}$&\texttt{$\tau$ vrnei}&$10^{11}\,\mathrm{cm}^{-3}\,\mathrm{s}$&$5.93^{+6.40}_{-2.12}$&$2.20^{+0.45}_{-0.49}$&$4.52^{+0.64}_{-0.44}$&$4.97^{+5.34}_{-2.09}$&$6.43^{+3.25}_{-2.15}$ & $7.71^{+3.01}_{-1.04}$\\
\texttt{norm vnei} 1&$10^{-4}\,\mathrm{cm}^{-5}$&$1.67^{+0.37}_{-0.26}$&$2.25^{+0.51}_{-0.49}$&$12.44^{+2.03}_{-1.12}$& $2.73^{+0.93}_{-0.93}$&$7.49^{+1.36}_{-1.01}$ &$16.49^{+1.42}_{-1.62}$&\texttt{norm vrnei}&$10^{-4}\,\mathrm{cm}^{-5}$&$2.73^{+0.34}_{-0.30}$&$8.60^{+3.83}_{-3.16}$&$34.31^{+8.59}_{-5.09}$&$2.84^{+1.28}_{-0.52}$&$9.33^{+1.88}_{-1.38}$ & $13.48^{+1.86}_{-1.23}$\\
{\kt \texttt{vnei} 2}&$\kev$&$1.04^{+0.03}_{-0.04}$&$0.74^{+0.01}_{-0.01}$&$0.92^{+0.02}_{-0.03}$& $0.89^{+0.04}_{-0.03}$& $0.87^{+0.04}_{-0.03}$&$0.88^{+0.02}_{-0.02}$&{\kt \texttt{vnei}}&$\kev$&$1.02^{+0.03}_{-0.02}$&$0.74^{+0.02}_{-0.01}$&$0.90^{+0.04}_{-0.03}$&$0.89^{+0.03}_{-0.03}$&$0.86^{+0.03}_{-0.03}$ & $0.89^{+0.04}_{-0.03}$\\
Ne&...&$1.65^{+0.69}_{-0.75}$&$0.87^{+0.10}_{-0.06}$&$0.89^{+0.15}_{-0.13}$& $0.77^{+0.14}_{-0.13}$& $0.83^{+0.18}_{-0.08}$&$1.31^{+0.22}_{-0.20}$&Ne&...&$1.91^{+0.85}_{-0.60}$&$1.00^{+0.24}_{-0.12}$&$1.40^{+1.65}_{-0.28}$&$0.74^{+0.09}_{-0.08}$&$0.83^{+0.08}_{-0.09}$ & $1.36^{+0.14}_{-0.24}$\\
Mg&...&$0.98^{+0.26}_{-0.27}$&$0.71^{+0.06}_{-0.02}$&$0.68^{+0.06}_{-0.07}$& $0.58^{+0.10}_{-0.04}$& $0.70^{+0.07}_{-0.08}$&$0.98^{+0.08}_{-0.10}$&Mg&...&$1.15^{+0.25}_{-0.15}$&$0.90^{+0.23}_{-0.12}$&$1.21^{+2.27}_{-0.30}$&$0.54^{+0.04}_{-0.02}$&$0.71^{+0.05}_{-0.07}$ & $0.99^{+0.23}_{-0.13}$\\
Si&...&$0.85^{+0.25}_{-0.16}$&$0.64^{+0.05}_{-0.03}$&$0.71^{+0.07}_{-0.06}$& $0.60^{+0.09}_{-0.08}$& $0.64^{+0.05}_{-0.09}$&$0.82^{+0.13}_{-0.04}$&Si&...&$0.98^{+0.12}_{-0.10}$&$0.81^{+0.21}_{-0.11}$&$1.41^{+4.32}_{-0.43}$&$0.57^{+0.10}_{-0.09}$&$0.67^{+0.06}_{-0.11}$ & $0.88^{+0.09}_{-0.13}$\\
S&...&$0.52^{+0.22}_{-0.17}$&$0.68^{+0.09}_{-0.07}$&$0.65^{+0.11}_{-0.11}$& $0.47^{+0.09}_{-0.10}$& $0.47^{+0.14}_{-0.17}$&$0.62^{+0.37}_{-0.30}$&S&...&$0.67^{+0.18}_{-0.17}$&$0.94^{+0.10}_{-0.13}$&$1.52^{+0.47}_{-0.20}$&$0.49^{+0.12}_{-0.05}$&$0.54^{+0.13}_{-0.08}$ & $0.97^{+0.11}_{-0.16}$\\
Fe&...&$0.45^{+0.10}_{-0.06}$&$0.37^{+0.02}_{-0.01}$&$0.38^{+0.03}_{-0.03}$& $0.34^{+0.04}_{-0.03}$& $0.40^{+0.06}_{-0.04}$&$0.41^{+0.07}_{-0.03}$&Fe&...&$0.51^{+0.04}_{-0.04}$&$0.50^{+0.08}_{-0.07}$&$0.76^{+3.64}_{-0.28}$&$0.32^{+0.05}_{-0.03}$&$0.42^{+0.02}_{-0.04}$ & $0.55^{+0.11}_{-0.06}$\\
$\tau$ \texttt{vnei} 2&$10^{11}\,\mathrm{cm}^{-3}\,\mathrm{s}$&$3.47^{+0.35}_{-0.30}$&$3.78^{+0.27}_{-0.25}$&$2.51^{+0.28}_{-0.14}$&$2.19^{+0.39}_{-0.32}$ &$2.88^{+0.31}_{-0.37}$ &$2.7^{+0.19}_{-0.17}$&$\tau$ \texttt{vnei}&$10^{11}\,\mathrm{cm}^{-3}\,\mathrm{s}$&$3.49^{+1.11}_{-0.70}$&$3.39^{+0.37}_{-0.33}$&$2.10^{+0.24}_{-0.26}$&$2.16^{+0.40}_{-0.29}$&$2.79^{+0.11}_{-0.11}$ & $2.88^{+0.43}_{-0.36}$\\
\texttt{norm vnei} 2&$10^{-4}\,\mathrm{cm}^{-5}$&$8.16^{+0.73}_{-0.33}$&$34.72^{+1.61}_{-0.49}$&$40.16^{+3.22}_{-2.64}$&$20.04^{+2.74}_{-2.51}$ &$22.63^{+3.79}_{-4.10}$ &$18.21^{+1.08}_{-0.49}$&\texttt{norm vnei}&$10^{-4}\,\mathrm{cm}^{-5}$&$7.14^{+0.31}_{-0.45}$&$25.64^{+4.48}_{-4.79}$&$18.43^{+8.17}_{-10.22}$&$21.31^{+0.59}_{-2.21}$&$21.57^{+2.22}_{-2.05}$ & $18.63^{+3.26}_{-2.88}$\\
{\kt \texttt{vpshock}}&$\kev$&$0.59^{+0.03}_{-0.04}$&$0.38^{+0.02}_{-0.07}$&$0.49^{+0.03}_{-0.03}$& $0.47^{+0.04}_{-0.06}$& $0.53^{+0.05}_{-0.05}$&$0.51^{+0.05}_{-0.04}$&{\kt \texttt{vpshock}}&$\kev$&$0.59^{+0.01}_{-0.01}$&$0.38^{+0.03}_{-0.08}$&$0.49^{+0.04}_{-0.06}$&$0.45^{+0.05}_{-0.03}$&$0.52^{+0.05}_{-0.06}$ & $0.52^{+0.06}_{-0.03}$\\
$\tau$ \texttt{vpshock}&$10^{11}\,\mathrm{cm}^{-3}\,\mathrm{s}$&$4.14^{+0.26}_{-0.26}$&$6.33^{+2.97}_{-1.22}$&$4.90^{+0.55}_{-0.42}$&$4.75^{+1.46}_{-0.62}$ & $5.27^{+0.38}_{-0.38}$&$3.10^{+0.46}_{-0.36}$&$\tau$ \texttt{vpshock}&$10^{11}\,\mathrm{cm}^{-3}\,\mathrm{s}$&$4.09^{+0.30}_{-0.25}$&$6.37^{+1.90}_{-1.69}$&$4.85^{+1.04}_{-0.79}$&$4.98^{+1.50}_{-1.22}$&$5.26^{+0.77}_{-0.59}$ & $3.29^{+0.57}_{-0.43}$\\
\texttt{norm vpshock}&$10^{-4}\,\mathrm{cm}^{-5}$&$12.09^{+0.40}_{-0.67}$&$21.10^{+4.31}_{-2.47}$&$31.76^{+1.36}_{-1.46}$& $23.88^{+1.56}_{-0.78}$& $19.63^{+1.72}_{-2.90}$&$16.20^{+1.69}_{-0.78}$&\texttt{norm vpshock}&$10^{-4}\,\mathrm{cm}^{-5}$&$11.72^{+0.73}_{-0.69}$&$21.74^{+2.36}_{-1.40}$&$31.16^{+1.94}_{-1.81}$&$23.65^{+1.61}_{-1.69}$&$19.18^{+2.48}_{-1.87}$ & $17.02^{+1.76}_{-1.40}$\\
\texttt{cstat(O)}/\texttt{dof} &...& 1126 / 966 & 1048 / 966 & 1190 / 966 & 1110 / 966 & 1173 / 966 & 1135 / 966 &\texttt{cstat(O)}/\texttt{dof} &...&1110 / 966 & 1042 / 966 & 1182 / 966 & 1110 / 966 & 1171 / 966 & 1126 / 966 \\
\texttt{cstat(E)}$\pm\,\sigma_{\mathrm{E}}$&...&$1018\pm45$&$1018\pm46$&$1039\pm49$&$1017\pm45$&$1025\pm48$ &$1018\pm45$&\texttt{cstat(E)}$\pm\,\sigma_{\mathrm{E}}$&...&$1021\pm45$&$1010\pm46$&$1038\pm47$ & $1017\pm45$ & $1025\pm47$ & $1014\pm45$\\
\enddata
\end{deluxetable*}

\section{Discussion}
\label{s:discuss}

\subsection{LMC Abundances and Their Variations}
\label{s:discuss_lmc_abundances}
Analysis of the rim regions provides a means to estimate the LMC abundances local to N132D that can inform us about the metallicity of the circumstellar medium (CSM) prior to the explosion. Table \ref{tab:lmcavg_compare} shows the mean abundances for the elements we include in this study, along with measurements from previous works. Before we can meaningfully compare these abundances, it is important to remark on several characteristics that influence these measurements and should be kept in mind. The first row of Table \ref{tab:lmcavg_compare} lists the number of regions used by different authors to determine the mean local abundances; `W' denotes that certain studies derived the mean abundances from fits to the integrated spectrum of the whole remnant. Further, note that some studies used a combination of regions on the rim and the interior of the remnant; 2 out of 8 regions used by \cite{2016AJ....151..161S} and 2 out of the 4 regions used by \cite{2018ApJS..237...10D} are in the interior. For the results from \cite{2018ApJS..237...10D}/\cite{2002A&A...385..143K} that are derived from observations with the ANU Wide-Field Spectrograph (WiFeS, \citealt{2010Ap&SS.327..245D}), we average over the four brightest clouds (in the optical) that were used to determine the abundances. There is no uncertainty on the abundance of Mg local to N132D because it could not be constrained from the data used and was fixed to half of the solar value. For Fe, spectral fits to the four brightest clouds returned the same value. The measurements from \cite{2019AJ....157...50D} are from the same regions as in \cite{2018ApJS..237...10D}, but with an improved shock modeling code that takes into account the emission from the photoionization region ahead of the forward shock. The quoted abundances of Mg, Si and Fe from \cite{2019AJ....157...50D} are the values assumed by the authors in the model.

The values in parenthesis that we quote from \cite{1998ApJ...505..732H} and \cite{2016AJ....151..161S} represent abundances averaged over multiple SNRs in the LMC, the ones we take from \cite{1992ApJ...384..508R} are averaged over SNRs as well as supergiants, and those from \cite{2018ApJS..237...10D}/\cite{2002A&A...385..143K} are determined from N132D and B-stars in NGC 2004. The ones we report from \cite{2019AJ....157...50D} in the parenthesis are averaged over SNRs, B stars, F-supergiants and \ion{H}{2} regions. Different studies have also used different observations to compute the mean abundances local to N132D, as we show in Table \ref{tab:lmcavg_compare}. The effect of dust depletion is only accounted for by \cite{2018ApJS..237...10D} and \cite{2019AJ....157...50D}, although the effects of depletion in different phases and shock conditions are largely unknown (see, for example, \citealt{2006ApJ...652L..33W,2019A&A...631A.127M}). For our work, we find the error and scatter of the fitted abundances using the method of Multiple Imputations \citep{2011ApJ...731..126L} which takes into account the statistical as well as systematic uncertainties on the parameters (see Appendix \ref{s:append_errors} for further details). We suspect that the abundance of S is not well constrained due to the low number of S counts on the rim; this is also apparent in its relatively large $1\sigma$ uncertainty. 

Our measurements that are local to N132D match closely with \cite{2000ApJ...537..667B}, except for Mg and S, however, the abundance of Mg reported in \cite{2000ApJ...537..667B} is classified as a lower limit by the authors whereas that of S is within the uncertainty. Our measurements of Ne and Mg are higher and lower, respectively, than the measurements of \cite{2018ApJS..237...10D}/\cite{2002A&A...385..143K}, however they are in good agreement with those of \cite{2019AJ....157...50D}. As compared to \cite{2016A&A...585A.162M}, we measure consistent (within the uncertainties) abundances of Ne, Si and Fe. There is a significant discrepancy of $\sim\,1.0\,\mathrm{dex}$ between the abundance of O measured by \cite{2016A&A...585A.162M} and other works including ours. We note that we have adopted the best-fit abundance values for the CSM/ISM model component for N132D from Table E.2 of \cite{2016A&A...585A.162M} whereas the majority of the O emission is fitted by the hotter, O-rich model component in their model.

Similarly, we can compare our results with that of \cite{2016AJ....151..161S}, especially because the same archival \chandra data have been used in both the studies. It is worth noting that our measurements of the abundances of O, Ne and Mg are $0.4\,\mathrm{dex}$ higher whereas those of Si and Fe are in excellent agreement with that of \cite{2016AJ....151..161S}. We speculate that the reason for this discrepancy can be that our measurements are derived from fitting the entire rim whereas those of \cite{2016AJ....151..161S} come from fitting certain regions located on different parts of the rim as well as some regions in the interior. In fact, we find that some of the regions on the rim analyzed by \cite{2016AJ....151..161S} have systematically lower abundances than the average in our fits to the same regions (regions $\mathrm{r13-r15}$).

\begin{deluxetable*}{|l c c c c c c c c r|}
\tabletypesize{\scriptsize}
\tablenum{8}
\tablecolumns{10}
\tablecaption{Comparison of mean abundances local to N132D with other works. Values in parenthesis denote those averaged over many SNRs in the LMC. The first row denotes the number of regions within the remnant used for the analysis; `W' denotes that the abundances were derived from the fit to the integrated spectrum of the entire remnant. The second row denotes the instrument/observatory source of the measurements local to N132D.}
\tablehead{\colhead{Element} & \colhead{Solar\tablenotemark{a}}& \colhead{Russell and Dopita\tablenotemark{b}} & \colhead{Hughes\tablenotemark{c}} & \colhead{Blair\tablenotemark{d}} & \colhead{Schenck\tablenotemark{e}} & \colhead{Maggi\tablenotemark{f}} & \colhead{Dopita/Korn\tablenotemark{g}} & \colhead{Dopita19\tablenotemark{h}} & \colhead{This Work}}
\startdata
...&...&W&W&1&8&W&4&4&19\\
...&...&...&\textit{ASCA}&\textit{HST}&\chandra&\textit{XMM-Newton}&WiFeS&WiFeS&\chandra\\
O&8.69&[8.35$\pm$0.06]&8.14$\pm$0.06 [8.21$\pm$0.07]&$8.45\pm0.10$&7.97$\pm$0.09 [8.04$\pm$0.04]&7.39$^{+0.17}_{-0.09}$ [8.01$^{+0.14}_{-0.21}$]&8.31$^{+0.01}_{-0.03}$ [8.32$\pm$0.06]&8.32$\pm$0.04 [8.40$\pm$0.05]&$8.39\pm0.19$\\
Ne&7.94&[7.61$\pm$0.05]&7.56$\pm$0.06 [7.55$\pm$0.08]&$7.64\pm0.10$&7.29$\pm$0.06 [7.39$\pm$0.06]&7.60$\pm$0.02 [7.39$^{+0.11}_{-0.15}$]&7.44$^{+0.01}_{-0.03}$ [7.52$\pm$0.09]&7.62$\pm$0.04 [7.70$\pm$0.09]&$7.71\pm0.12$\\
Mg&7.40&[7.47$\pm$0.13]&7.08$\pm$0.07 [7.08$\pm$0.07]&$6.75\pm0.10$&6.73$\pm$0.07 [6.88$\pm$0.06]&6.68$\pm$0.02 [6.92$^{+0.20}_{-0.37}$]&7.47 [7.37$\pm$0.06]&7.19 [7.19$\pm$0.09]&$7.08\pm0.15$\\
Si&7.27&[7.81]&7.08$\pm$0.13 [7.04$\pm$0.08]&$7.00\pm0.10$&7.00$\pm$0.07 [6.99$\pm$0.11]&6.86$\pm$0.03 [7.11$^{+0.20}_{-0.41}$]& [7.10$\pm$0.07]&7.11 [7.11$\pm$0.04]&$7.03\pm0.18$\\
S&7.09&[6.70$\pm$0.09]&6.73$\pm$0.06 [6.71]&$6.63\pm0.10$&...&...&7.01$^{+0.09}_{-0.06}$ [7.00$\pm$0.15]&7.10$\pm$0.07 [6.93$\pm$0.05]&$6.95\pm0.31$\\
Fe&7.43&[7.23$\pm$0.14]&7.08$\pm$0.06 [7.01$\pm$0.11]&$6.85\pm0.10$&6.97$\pm$0.07 [6.84$\pm$0.05]&6.88$\pm$0.02 [6.97$^{+0.13}_{-0.18}$]&7.23 [7.33$\pm$0.03]&7.33 [7.33]&$6.96\pm0.15$\\
\enddata
\let\amp=&
\catcode`\&=12
\tablecomments{References: (a) \citealt{2000ApJ...542..914W}; (b)\citealt{1992ApJ...384..508R}; (c) \citealt{1998ApJ...505..732H}; (d) \citealt{2000ApJ...537..667B}; (e) \citealt{2016AJ....151..161S}; (f) \citealt{2016A&A...585A.162M}; (g) \citealt{2018ApJS..237...10D,2002A&A...385..143K}; (h) \citealt{2019AJ....157...50D}.}
\label{tab:lmcavg_compare}
\end{deluxetable*}

\begin{deluxetable}{|l c c r|}
\tablenum{9}
\label{tab:ne}
\tablecolumns{4}
\tablecaption{Shock velocity ($v_{\mathrm{s}}$), electron density ($n_{\mathrm{e}}$) and shock age ($t$) in rim regions, with $1\sigma$ errors. $v_{\mathrm{s}}$ is estimated from equation \ref{eq:shockvelocity}, and $n_{\mathrm{e}}$ from the \texttt{norm} and geometrical approximations described in Appendix \ref{s:append_geometry}. Ionization timescales and electron density estimates are then used to calculate the age of the forward shock as $t = \tau/$\nee. The mean shock velocity is $\langle v_{\mathrm{s}} \rangle = 855\pm100\,\mathrm{km}\,\mathrm{s^{-1}}$, electron density is $\langle n_{\mathrm{e}}\rangle = (6\pm2)\,f^{-1/2}\,\mathrm{cm^{-3}}$ and shock age is $\langle t \rangle = (1200\pm270)\,f^{-1/2}\,\mathrm{yr}$, where $f$ represents the volume filling factor.}
\tablehead{\colhead{Region ID} & \colhead{$\mathrm{v}_s$ ($\kms$)} & \colhead{\nee$\,$(\textit{f}$^{-1/2}\,\mathrm{cm}^{-3}$)} & \colhead{$t$ (\textit{f}$^{-1/2}\,\mathrm{yr}$)}}
\startdata
\hline
$\mathrm{r1}$&$1029\pm46$&$3\pm1$&$680\pm110$\\
$\mathrm{r2}$&$893\pm34$&$4\pm1$&$1490\pm380$\\
$\mathrm{r3}$&$745\pm17$&$7\pm3$&$2620\pm1020$\\
$\mathrm{r4}$&$843\pm15$&$5\pm1$&$1130\pm130$\\
$\mathrm{r5}$&$784\pm11$&$8\pm2$&$1380\pm160$\\
$\mathrm{r6}$&$843\pm57$&$7\pm3$&$850\pm300$\\
$\mathrm{r7}$&$833\pm21$&$7\pm2$&$1220\pm190$\\
$\mathrm{r8}$&$774\pm17$&$7\pm1$&$760\pm180$\\
$\mathrm{r9}$&$779\pm22$&$4\pm3$&$1300\pm260$\\
$\mathrm{r10}$&$812\pm43$&$5\pm3$&$1990\pm530$\\
$\mathrm{r11}$&$950\pm55$&$5\pm2$&$990\pm280$\\
$\mathrm{r12}$&$917\pm19$&$4\pm1$&$1400\pm190$\\
$\mathrm{r13}$&$941\pm32$&$4\pm1$&$800\pm290$\\
$\mathrm{r14}$&$817\pm21$&$6\pm1$&$770\pm170$\\
$\mathrm{r15}$&$908\pm48$&$5\pm2$&$610\pm180$\\
$\mathrm{r16}$&$828\pm21$&$6\pm2$&$790\pm90$\\
$\mathrm{r17}$&$950\pm37$&$3\pm1$&$920\pm190$\\
$\mathrm{r18}$&$812\pm16$&$8\pm2$&$870\pm100$\\
$\mathrm{r19}$&$785\pm11$&$7\pm1$&$2400\pm280$\\
\hline
\enddata
\end{deluxetable}

\subsection{Shock Velocity and Electron Density}
\label{s:discuss_ne}
We calculate the forward shock velocity and an estimate of the shock age along the rim using average physical conditions of the plasma (temperature and ionization timescale), geometry of the region and its \texttt{norm}. Using the Rankine-Hugoniot conditions which predict mass proportional heating for electrons and ions \citep{1975ctf..book.....L,1999ApJ...526..385B,2012A&ARv..20...49V} and assuming no energy losses (for example, due to cosmic rays), we can relate the electron temperature to the shock velocity as
\begin{equation}
kT_\mathit{e} \approx \frac{3}{16} m_{\mu} v_{\mathrm{s}}^2\,,
\label{eq:shockvelocity}
\end{equation}
where $m_\mu$ is the mean mass per free particle and $v_{\mathrm{s}}$ is the shock velocity. Assuming the majority of the electrons are contributed by H and He, $m_\mu\,\sim\,0.59\,\mathrm{m}_\mathrm{H}$. The mean electron temperature we find from our analysis of the rim regions is $\langle$\kt$\rangle = 0.85\pm0.20\,\kev$ (see Table \ref{tab:scatter}). Then the mean shock velocity of the blast wave is $\langle v_{\mathrm{s}} \rangle = 855\pm100\,\mathrm{km}\,\mathrm{s}^{-1}$. We estimate the average shock age (by finding the electron density \nee\ using the \texttt{norm} and the 3D geometrical approximation described in Appendix \ref{s:append_geometry}) to be $\langle \tau \rangle  = (1200\pm270)\,\textit{f}^{-1/2}\,\mathrm{yr}$ for the rim, where \textit{f} denotes the volume filling factor of the shell region. Filling factor refers to the fraction of emitting plasma filling a volume in the remnant and is a parameter to account for our lack of knowledge about the extent of the emitting volume that is filled with X-ray emitting plasma \citep{1980ApJ...239..867H}. Table \ref{tab:ne} lists the corresponding shock velocities, electron densities and shock ages we find for all the rim regions.

Through simulations of a blast wave evolving into a cavity and colliding with clouds, \cite{2016ApJS..227...28T} propose that the shock velocity is decreased by roughly $\sqrt{n_e}$ when the blast wave hits the clouds. For the mean shock velocity ($855\pm100\,\mathrm{km \,s}^{-1}$) and electron density ($6\pm2\,f^{-1/2}\,\mathrm{cm}^{-3}$) we derive, this implies a mean pre-collision blast wave velocity of $\sim\,2100\,\mathrm{km \,s}^{-1}$ (if $f\sim1$). This is in good agreement with the pre-collision velocity of $1900\,\mathrm{km \,s}^{-1}$ proposed by \cite{2003ApJ...595..227C} for N132D, where the authors use a semi-analytical thin shell model to study an SNR crossing a density jump (a condition that can prevail in SNRs expanding in a low density cavity). For N132D, \cite{2003ApJ...595..227C} conclude that the current shock has been interacting with denser material for $\sim\,700\,\mathrm{yr}$ when it was slowed down to $\sim\,700\,\mathrm{km \,s}^{-1}$ from its pre-collision value due to impact with the walls of the cavity in which the massive progenitor is thought to have exploded. Thus, the observations are consistent with a scenario in which this SNR exploded inside a cavity (in a denser surrounding medium) possibly created by the winds of its progenitor.

\subsection{Deduction of Progenitor Mass}
\label{s:ejecta}
We estimate the mass of the progenitor with three different methods given in the literature using the spectral results from the interior regions.

\subsubsection{Estimates from Nucleosynthesis Models}
\label{s:progenitor_rege1}
We can compare the yields obtained from the spectral fit of region $\mathrm{e1}$ to models of low-metallicity CCSNe nucleosynthesis \citep{2006NuPhA.777..424N,2006ApJ...653.1145K}\footnote{Same models and corresponding yields are reported in \cite{2006NuPhA.777..424N} and \cite{2006ApJ...653.1145K}.} in order to deduce the mass of the progenitor, as was investigated by \citet[][see their Table 10]{2000ApJ...537..667B}. As we point out in Section \ref{s:analysis}, this is the only region that shows enhanced abundances of 5/6 elements that we fit across the remnant. Since direct model yields for LMC metallicity ($Z_{\mathrm{LMC}} = 0.008$) are not available in \cite{2006NuPhA.777..424N} and \cite{2006ApJ...653.1145K}, we take a geometric mean of model yields at SMC metallicities ($Z_{\mathrm{SMC}} = 0.004$) and Milky Way ($Z_{\mathrm{MW}} = 0.02$) to imitate the LMC environment. \cite{2011MNRAS.414.3231K} updated the yields given by \cite{2006ApJ...653.1145K} for $Z = 0.004;\,18\,M_{\odot}$ and $Z = 0.02;\,25\,M_{\odot}$ models because the earlier models produced large amounts of $^{13}$C and N due to erroneous mixing of H into the He-burning layer, also affecting the yields of other elements. Accordingly, we use the updated yields for these two models in our calculations.

\begin{figure}
\centering
\includegraphics[width=1.0\linewidth]{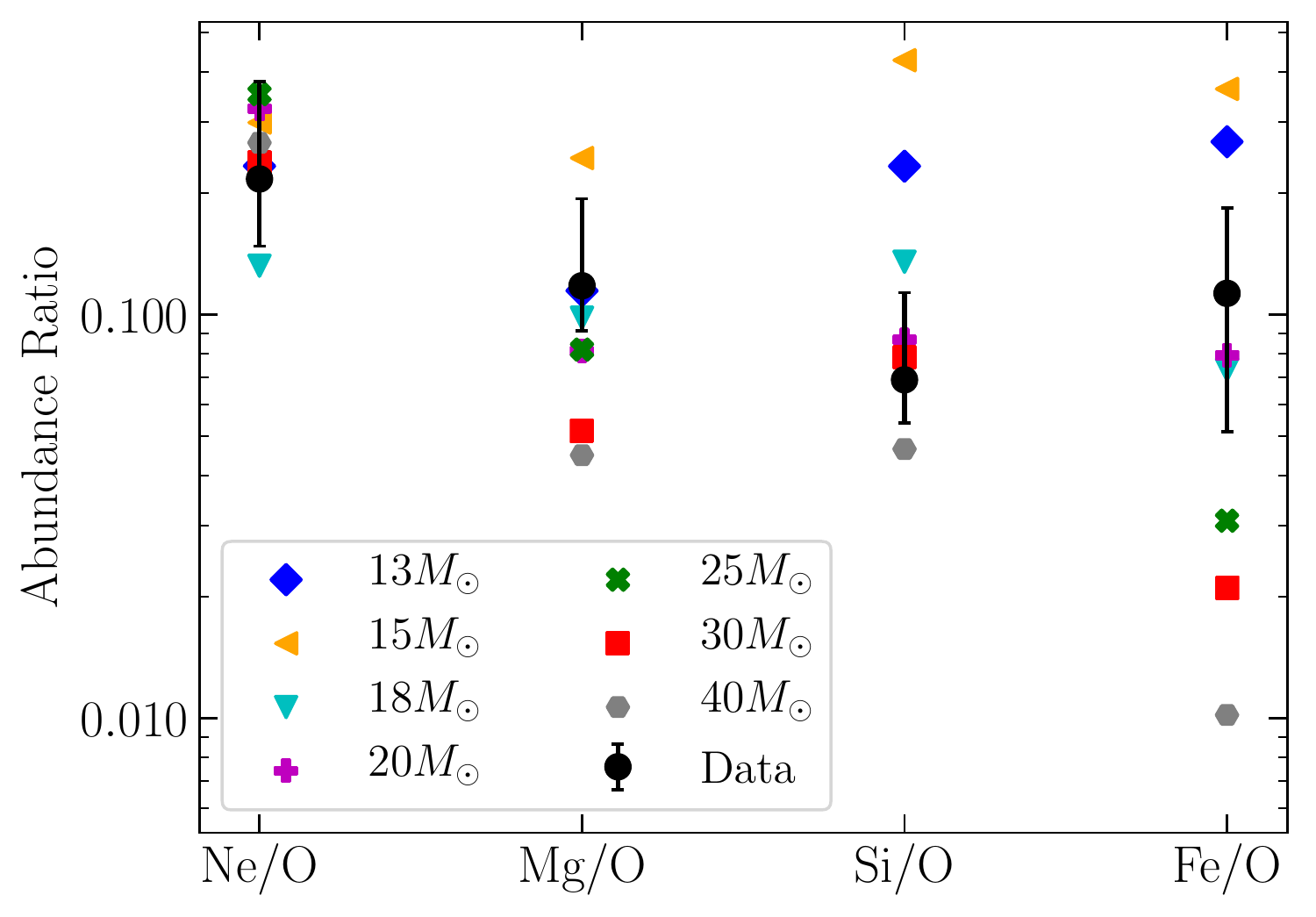}
\caption{Abundances of different elements (relative to O) from the ejecta component of the best-fit model to the spectrum of region $\mathrm{e1}$. Black markers denote the best-fit values and colored markers denote the nucleosynthesis yields for different progenitor masses from \cite{2006NuPhA.777..424N,2006ApJ...653.1145K,2011MNRAS.414.3231K}. The fit is present in Table \ref{tab:rimfit2} and the spectrum is shown in Figure \ref{fig:ejectafit}.}
\label{fig:yieldratios}
\end{figure}

\begin{figure}
\centering
\includegraphics[width=1.0\linewidth]{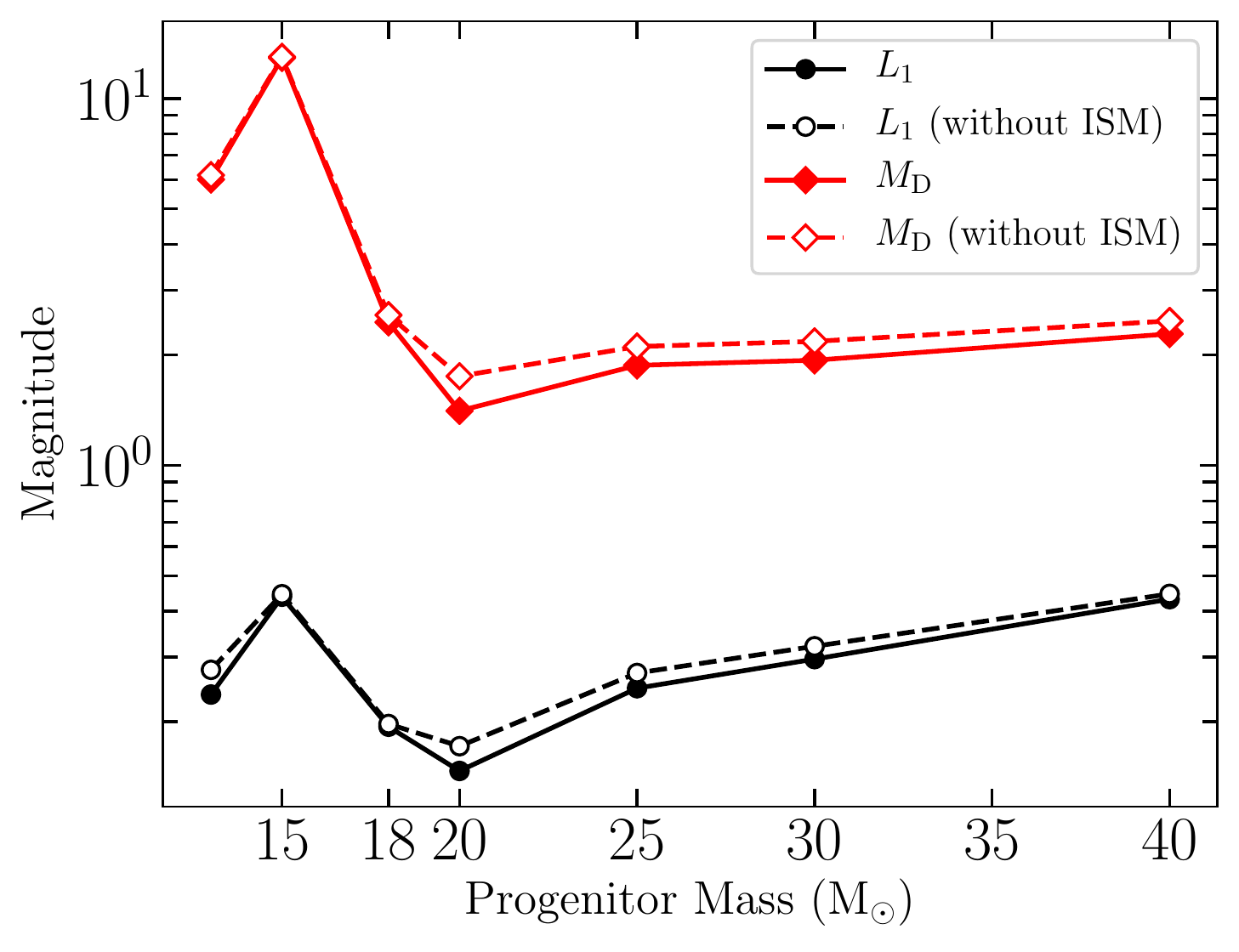}
\caption{Magnitudes of $L_1$ Norm and Mahalanobis Distance ($M_{\mathrm{D}}$) plotted against progenitor mass models, in order to deduce the progenitor mass of N132D by comparing its `ejecta' yields with that given by the low-metallicity CCNSe nucleosynthesis models of \cite{2006NuPhA.777..424N,2006ApJ...653.1145K,2011MNRAS.414.3231K}. The source model we use to find the best-fit parameters is the \texttt{vpshock+vnei} model; see Table \ref{tab:rimfit2}. Dashed curves use the residual of best-fit abundances after we subtract the local average ISM contribution, to check for swept-up ISM contamination.}
\label{fig:progenitormass}
\end{figure}

We use the Mahalanobis distance and $L_1$ norm methods to find the measure of closeness between our observed yields and the yields predicted by the models. The Mahalanobis distance ($M_{\mathrm{D}}$)
\begin{equation}
M_{\mathrm{D}} = \sqrt{\Sigma_{i=1}^n \frac{(x_i-y_i)^2}{s_i^2}}\,,
\end{equation}
is essentially an error ($s_i$) weighted Euclidean distance between the data (the test set $x_i$ representing the observed yields from our work) and the various nucleosynthesis models ($y_i$) \citep{zbMATH03023295}. The model yields are given in solar masses. Thus, to compare them with the best-fit values from the spectral fit to region $\mathrm{e1}$, we convert the latter to elemental yields by multiplying them with atomic mass, since the reference scale is defined for the number of atoms relative to H and not the atomic mass. The test set is best explained by that model set for which the Mahalanobis distance is a minimum. The $L_1$ norm method works on a similar principle of distance minimization; its logarithmic form is given by (for example, \citealt{2018ApJS..237...10D,2019AJ....157...50D}) \begin{equation}
L_1 = \frac{\sum_{j = 1}^{m}|\log_{10}\frac{x_i}{y_i}|}{m}\,,
\end{equation}
where $m$ = no. of elements. Figure \ref{fig:yieldratios} shows the abundance ratios relative to O for the different models and the data\footnote{We note that the results do not change if we take abundance ratios with respect to Si instead of O, as is often done in such comparisons (for example, \citealt{2014ApJ...781...41K,2015ApJ...810..113F,2019MNRAS.489.4444B}).}. Model abundance ratios with a large scatter across progenitor masses (Si/O and Fe/O) drive the mass estimate because they show larger differences than other ratios that have relatively less scatter (Ne/O and Mg/O). While the Mahalanobis distance and $L_1$ norm are guided by Si/O for lower progenitor masses, they are largely set by Fe/O for massive progenitors. We present the comparison of progenitor mass deduction using the two methods in Figure \ref{fig:progenitormass}. It can be noticed that both $M_{\mathrm{D}}$ and $L_1$ norm pass through the same global minimum as one moves from lower to higher mass progenitor models. Thus, we find a progenitor mass of $M_{\mathrm{p}} = 20\,M_{\odot}$ to be the closest to our observed yields. The model yields we use are calculated for an explosion energy of $10^{51}\,\mathrm{erg}$. If the explosion energy for N132D was higher \citep{2018ApJ...854...71B}, this will affect the comparison we make below because core collapse models are very sensitive to the production of $^{56}$Ni that correlates with the explosion energy (\citealt[see their Figure 20]{2015ApJ...801...90P}; \citealt[see their Figure 17]{2016ApJ...821...38S}). In case the explosion energy was $>\,10^{51}\,\mathrm{erg}$, it will produce more Fe \citep{2006ApJ...653.1145K,2013ARA&A..51..457N}. Further, these models are also sensitive to the rotation rate of the progenitor \citep{2000ARA&A..38..143M,2007ApJ...660..516T,2013ARA&A..51..457N}.

Note that our deduction assumes that the region contains pure ejecta, which is an ideal case. Nonetheless, the ratios are not particularly sensitive to contamination from swept-up ISM. We verify this by subtracting the local average ISM values for each abundance and finding that both methods still have global minima at $20\,M_{\odot}$ (see the dashed lines in Figure \ref{fig:progenitormass}). Although we utilize all possible elemental ratios to derive this estimate, we only use the X-ray heated ejecta from a single region (that represents a tiny fraction of the remnant in projection) to estimate the progenitor mass in this manner. Thus, the yields in this region may not be representative of the entire remnant. However, in order to compare these yields against the CCSNe model yields, we require sufficiently high abundances of more than 3 elements to remove the degeneracy between models of diverse progenitor masses. Given the depth of the existing data, we find $\mathrm{e1}$ to be the singular region which provides the most stringent constraints on these abundances. It is not surprising that we only find one eligible region for this analysis since such regions are difficult to extract because the remnant is dominated by swept-up ISM at the age of N132D. Given these caveats, it becomes clear why this technique is not sufficient to place robust constraints on the progenitor mass, and other avenues should be explored for the same.

\subsubsection{Estimates from Enriched Fe/Si}
\label{s:progenitor_katsudafesi}
\cite{2018ApJ...863..127K} point out that the estimates from elemental abundance ratios other than Fe/O or Fe/Si are not good tracers of progenitor mass because they are not sensitive to the CO core mass of the progenitor. Keeping this in mind, we also estimate the mass of the progenitor only from the Fe/Si ratio. As noted by \cite{2018ApJ...863..127K}, this technique cannot account for the unshocked ejecta in the SNR that can alter the measured Fe/Si ratio (see, however, \citealt{2012ApJ...746..130H} and \citealt{2014ApJ...785....7D} where it is proposed for SNR Cas A that up to 90 per cent of its ejecta has already been shocked). A major advantage of only using the Fe/Si in our case is that apart from region $\mathrm{e1}$, we can also use the Fe K regions since they show enhanced abundances of Fe/Si in the hottest model component that we assume comes largely from ejecta. Including these regions lets us cumulatively sample a large fraction of the remnant.

We use the best-fit relation provided by \cite{2018ApJ...863..127K} that the authors find by fitting revised progenitor mass estimates for several SNRs in the Milky Way and the Magellanic Clouds against Fe/Si measured from observations,
\begin{equation}
\frac{\mathrm{Fe/Si}}{\mathrm{Fe/Si}}_{\odot} = 1.13 \times \exp{\bigg(\frac{4.8 - M_{\mathrm{p}}}{10.6}\bigg)}\,.
\label{eq:katsudaequation}
\end{equation}
Table \ref{tab:fesikatsudaprogneitors} lists the progenitor mass estimates we derive from the spectral fits to regions $\mathrm{e1}$ and $\mathrm{f1 - f6}$. We utilize the best-fit abundances of the highest electron temperature component in the ionizing and recombining models for these regions (see Section \ref{s:results_fek} and Table \ref{tab:fekregionss}). The estimated masses from different regions lie between $10-20\,M_{\odot}$, with mean mass $\sim15\,M_{\odot}$. The variation in the deduced progenitor mass from region to region provides some insight into the importance of sampling as much of the ejecta as possible and the relatively large uncertainties on the deduced masses reflect the limitation imposed by the statistical precision of the current data.

\begin{deluxetable}{|l c r|}
\tablenum{10}
\label{tab:fesikatsudaprogneitors}
\tablecolumns{5}
\tablecaption{Estimates of the progenitor mass based on equation \ref{eq:katsudaequation} from \cite{2018ApJ...863..127K} that depends on the Fe/Si ratio measured in the ejecta component in spectral models for the interior regions. Region $\mathrm{e1}$ is only fit with a two-component (\texttt{vnei+vpshock}) model. Details of the estimation are present in Section \ref{s:progenitor_katsudafesi}.}
\tablehead{\colhead{Region ID} & \colhead{Ionizing Model} & \colhead{Recombining Model}}
\startdata
\hline
$\mathrm{e1}$&$12.15^{+7}_{-4}$&$...$\\
$\mathrm{f1}$&$9.65^{+7}_{-3}$&$10.67^{+3}_{-3}$\\
$\mathrm{f2}$&$20.01^{+6}_{-2}$&$19.02^{+8}_{-5}$\\
$\mathrm{f3}$&$12.65^{+5}_{-2}$&$12.66^{+5}_{-2}$\\
$\mathrm{f4}$&$11.59^{+9}_{-4}$&$12.48^{+7}_{-3}$\\
$\mathrm{f5}$&$16.70^{+4}_{-4}$&$17.93^{+3}_{-3}$\\
$\mathrm{f6}$&$14.31^{+3}_{-2}$&$15.35^{+2}_{-2}$\\
\hline
\enddata
\end{deluxetable}

\subsubsection{Estimates from Explosion in a Cavity Models}
\label{s:progenitor_cavity}
Finally, we also estimate the progenitor mass using the relation between the radius of the cavity ($R_{\mathrm{b}}$) and the progenitor mass ($M_{\mathrm{p}}$) proposed by \cite{2013ApJ...769L..16C} for SNRs evolving in cavities in or near giant molecular clouds
\begin{equation}
p^{1/3}_5R_{\mathrm{b}} = \Big[\alpha\Big(\frac{M_p}{M_{\odot}}\Big)-\beta\Big]\,\pc\,,
\label{eq:progenitormass_chen}
\end{equation}
where $p_5$ is the interclump pressure in units of $10^5\,\mathrm{cm}^{-3}\,\mathrm{K}$ (assumed to be unity, see \citealt{1993prpl.conf..125B,1999ApJ...511..798C,2009ApJ...699..850K}), and $\alpha = 1.22\pm0.05$ and $\beta = 9.16\pm1.77$ are derived from a linear regression. This assumes that the cavity was formed prior to the explosion by stellar winds of main sequence OB stars \citep{1987ApJ...314..103H} and does not take into account the effects of a Wolf-Rayet phase, if any, on the wind-blown bubble \citep{2005ApJ...619..839C}. Although such CCSNe undergo significant mass loss prior to the explosion \citep{2008MNRAS.389..113P,2012MNRAS.419.1515D,2013ApJ...767...71M,2016ApJ...818..111K}, its effect on the late-time dynamics when the blast wave interacts with the circumstellar shell has been shown to be of little importance \citep{2015ApJ...803..101P,2017ApJ...849..109P}. Thus, the predictions by explosion in a cavity models like this for SNRs older than a few centuries may not be affected by the pre-supernova mass loss \citep{2018ApJ...863..127K}. Since the shock has been interacting with the cloud in the south for the last few $100\,\mathrm{yr}$ \citep{2003ApJ...595..227C}, we assume that the radius of the cavity roughly equals the radius of X-ray emission. Adopting $R_{\mathrm{b}} = 12.5\,\pc$, we derive $M_{\mathrm{p}} = 17.8\pm3.8\,M_{\odot}$, in agreement with the progenitor mass we determine above.

Note that the interclump pressure in N132D will be more than the thermal pressure since additional pressure support can arise from turbulence and cosmic rays in dynamically active regions like supernova remnants \citep{2005ApJ...626..864M,2011ApJ...734...65J,2016ApJ...821..118W,2017ApJ...835..201H}. The average thermal pressure in the ISM of the LMC is estimated to be $p_{5,\mathrm{th}} = 0.1$ \citep{2016ApJ...821..118W}. If we use this value as a lower limit on the interclump pressure, the minimum progenitor mass we obtain for the same cavity size is $\sim 12\,M_{\odot}$, consistent with the results we summarize below in Section \ref{s:progenitor_summary}.

\subsubsection{Summary of Progenitor Mass Estimates}
\label{s:progenitor_summary}
Estimation from various different pathways (observational as well as theoretical) as we present above enables us to put a constraint on the progenitor mass. It is encouraging to find that the estimates of all three methods are within $2\sigma$ of each other. However, the results from nucleosynthesis yields and explosion in a cavity models favor a slightly more massive progenitor for N132D than the average estimated through the Fe/Si ratio. Nevertheless, our results suggest an intermediate mass ($M_{\mathrm{p}} < 25\,M_{\odot}$) progenitor for N132D, lower than the estimates of \cite{2000ApJ...537..667B} based on UV/optical data ($30-35\,M_{\odot}$) and by \cite{2009ApJ...707L..27F} based on \textit{Cosmic Origins Spectrograph} ($50^{+25}_{-15}\,M_{\odot}$) observations of N132D. 

Taking into account the uncertainties and systematic scatter in all the three methods listed above, we quote our estimate of the progenitor mass of N132D as $M_{\mathrm{p}} = 15\pm5\,M_{\odot}$, in line with the revised mass estimates of \cite{2018ApJ...863..127K}. This range of possible progenitor masses also overlaps with the suggested range of massive stars which can undergo a Wolf-Rayet phase in the LMC \citep{1999ApJ...511..798C}, as has been expected for N132D \citep{1995ApJ...439..365S}. Comparing with earlier predictions, we find that our progenitor mass estimate lies at the lower limit of \citet{1993ApJ...414..219H} where the authors used \textit{Einstein Observatory} data of N132D and nucleosynthesis models of \cite{1992eatc.conf...68T} to propose a progenitor mass of $20-25\,M_{\odot}$, whereas it is consistent with the estimate of slightly less than $20\,M_{\odot}$ given by \cite{1994ApJ...423..334B}.

If the mass of the progenitor was indeed within $10-20\,M_{\odot}$, this will have important implications on the explosion in a cavity scenario as well as the lifetime of the Wolf-Rayet phase, if any \citep{2012A&A...547A...3V}. A possible avenue to explore through simulations is to estimate the time and size of the creation of a cavity by pre-supernova winds for the estimated progenitor mass \citep{1991MNRAS.251..318T,1996A&A...316..133G,2007ApJ...667..226D,2017ApJ...849..109P}, however, it is beyond the scope of this work.

\subsection{High-Temperature Plasma and Fe K Emission}
\label{s:fek}
Emission in the $6.5-6.9\,\kev$ band can be mostly attributed to the presence of Fe He-like (Fe$\,$\ion{}{25}) line emission\footnote{Fe$\,$\ion{}{26} (Fe Li-like ion) also has certain line energies in the range $6.5-6.7\,\kev$, however, its emissivity is lower by at least an order of magnitude as compared to Fe$\,$\ion{}{25} and becomes comparable only at temperatures $\la\,1.3\,\kev$ while the component accounting for Fe K emission in both the ionizing and recombining models is $\ga\,1.5\,\kev$. Moreover, for there to be significant flux from Fe$\,$\ion{}{26}, a high ionization rate for the Li-like stage (high temperature) is required which cannot be possible near (optically thin coronal) equilibrium, not to mention that the process would anyway be unimportant in the recombining case. Thus, we can safely neglect the presence of significant flux from Fe Li like ions in this energy range.}. \textit{Suzaku} observations of N132D provide the centroid line energy of Fe K emission as $6656\pm9\,\mathrm{eV}$ \citep{2014ApJ...785L..27Y} whereas \textit{XMM-Newton} observations estimate it to be $6685^{+15}_{-14}\,\mathrm{eV}$ \citep{2016A&A...585A.162M}, proposed to be typical of middle-aged CCSNe (age > $2500\,\mathrm{yr}$) evolving in dense CSM with high ambient densities \citep{2014ApJ...785L..27Y,2015ApJ...803..101P}. The Fe \ion{}{25} He-like triplet spans about $64\,\mathrm{eV}$ between the Recombination and Forbidden lines. Thus, using the ionizing or recombining models affect the relative strengths of these lines in this complex, but with the limited statistics and spectral resolution of the ACIS data we are not sensitive to a shift in the centroid.

From both the ionizing (\texttt{vnei} + \texttt{vnei} + \texttt{vpshock}) and recombining (\texttt{vrnei} + \texttt{vnei} + \texttt{vpshock}) models we use to fit the six regions (see Table \ref{tab:fekregionss}) containing Fe K emission, we establish that a hotter ($\ga 1.5\,\kev$) plasma is needed to explain the Fe K emission in this remnant, while not over-producing the flux from Fe L at lower energies. A similar observation was also made by \cite{2016A&A...585A.162M}. However, both the models are able to explain this emission through hot NEI ionizing and recombining components, respectively. This degeneracy arises due to the low number of counts in the hard X-ray band and the complex nature of the fit with many free parameters. Our results are consistent with the conclusions of the \textit{Suzaku} + \textit{Hitomi} investigation by \cite{2018PASJ...70...16H} and the \textit{NuSTAR + Suzaku} analysis by \cite{2018ApJ...854...71B}. However, we are able to sample smaller and more compact regions with \chandra than these studies to show that the Fe K emission is distributed throughout the southern half of the remnant (not concentrated in a single feature) and the plasma history is most likely different for different regions. Together with the enhanced abundances we find in other regions, this provides some evidence for an asymmetric explosion. However, the current \chandra data for N132D are not deep enough to reconstruct the ejecta distribution with sufficient precision to conclude that the explosion was indeed asymmetric. An additional complication is the relatively large uncertainty in the explosion center for N132D, which is needed to constrain the ejecta distribution \citep{2007ApJ...670..635W,2017ApJ...844...84H,2018ApJ...856...18K}.

Although Fe K emission has been found in several SNRs in the LMC \citep[see their Table 2]{2016A&A...585A.162M}, the origin of a hotter plasma is not yet clear (see, for example, \citealt{2005ApJ...631..964P}). Applying the Rankine-Hugoniot strong shock conditions to the individual species leads to mass-proportional heating, so $T_\mathit{e} \ll T_\mathit{i}$ (by the ratio $m_\mathit{e}/m_\mathit{i}$; see, for example, \citealt{2000ApJ...543L..67S}). Coulomb equilibration (many small angle scatterings) would give a characteristic equilibration timescale of $\sim6000\,\mathrm{yr}$, too slow to account for electrons hot enough to excite X-ray emission lines. This argues for collisionless equilibration $-$ collective scattering of electrons with plasma magnetic field fluctuations. The available evidence is that this is effective for slow ($\la\,500 \,\mathrm{km}\,\mathrm{s}^{-1}$) shocks, but falls as roughly $\propto v_{\mathrm{s}}^{-2}$ and is much less effective for fast shocks. For example, for the estimated forward shock velocity of $855\,\kms$ the ratio of electron to proton temperature is $\approx0.2$ \citep{2007ApJ...654L..69G,2013SSRv..178..633G} indicating that the forward shock is unlikely to be the source of the Fe K emission. Presumably the reverse shock has a higher velocity into the ejecta with corresponding higher electron and ion temperatures than at the forward shock. For our fitted values of the temperature and ionization timescale, the RRC emission from Si and S would be weak. The Si and S RRCs would be stronger for lower plasma temperatures. We see no obvious RRC features for lower temperatures, so we can exclude that region of parameter space. Deeper observations may allow better constraints on anomalous line ratios for the He-like Fe, and RRCs for lighter ions. 

The high column densities of all the Fe K regions except region $\mathrm{f1}$ can be associated with the presence or absence of clumps of molecular clouds, respectively, as has been discussed for other SNRs interacting with molecular clouds \citep{1997ApJ...480..607B,2012ApJ...749...34L,2015SSRv..188..187S,2017ApJ...851...73M,2019ApJ...881...85S}. In fact, from Figure 2 of \cite{2018ApJS..237...10D}, we find that region $\mathrm{f1}$ does not contain any prominent shocked ISM clouds as observed in the optical, which is expected for its low column density. Thus, the origin of a recombining plasma in this region, if any, can be correlated with thermal conduction only if we assume that the dense cloud(s) in this region have already been evaporated. If the recombining plasma is due to thermal conduction, one would expect it to be interacting with dense gas, which is likely the case for region $\mathrm{f2}$. This analysis informs us of the spatially as well as spectrally diverse signatures of the plasma present in these regions that has evolved differently over time largely based on the surrounding environment. However, the origin of hot plasma that gives rise to the Fe K emission cannot be established from the available data.

\section{Summary}
\label{s:summary}
In this work, we have presented spatially resolved X-ray spectroscopy of N132D, the brightest SNR in the LMC, based on archival \chandra observations. By fitting the spectra of the entire well-defined rim of the remnant with a plane-parallel shock model, we calculate the mean local abundances of O, Ne, Mg, Si, S and Fe (Table \ref{tab:lmcavg_compare}) and find that Ne, Mg, Si and Fe show no excess or depletion on the rim around their mean and the associated total (statistical+systematic) uncertainty, whereas we find evidence of enhanced O and S on the north-western and north-eastern rim, respectively. A faint blob protruding outside the western rim shows enhanced abundance of O, however, extended X-ray observations are needed to ascertain if this blob is in fact an O-rich ejecta clump moving ahead of the blast wave. 

Using information from the rim regions, we derive a mean forward shock velocity $\langle v_{\mathrm{s}} \rangle = 855\pm100\,\mathrm{km}\,\mathrm{s}^{-1}$ and electron density $\langle n_{\mathrm{e}}\rangle = (6\pm2)\,f^{-1/2}\,\mathrm{cm^{-3}}$ where $f$ is the volume filling factor. For $f \sim 1$, our findings agree with the conclusions of \cite{2003ApJ...595..227C} where the authors propose that the shock collided with the cavity wall (inside which the progenitor exploded) $\sim 700\,\mathrm{yr}$ ago when it was slowed down from its pre-collision value of $\sim 1900\,\mathrm{km}\,\mathrm{s^{-1}}$. This is in line with the proposed explosion in a cavity scenario for this remnant, which partly comes from CO observations of molecular clouds in its surroundings \citep{1997ApJ...480..607B,2015ASPC..499..257S}.

We follow a mix of observational and theoretical approaches to estimate the mass of the progenitor of the remnant: 1. through comparison of best-fit ejecta abundances from region $\mathrm{e1}$ with the nucleosynthesis model yields from \cite{2006NuPhA.777..424N,2006ApJ...653.1145K,2011MNRAS.414.3231K}, 2. from Fe/Si ratio measured in the ejecta components of multiple regions in the interior \citep{2018ApJ...863..127K}, and 3. predictions from theoretical model of a core-collapse explosion in a cavity within a molecular cloud complex \citep{1999ApJ...511..798C,2013ApJ...769L..16C}. Our estimated progenitor mass of $15\pm5\,M_{\odot}$ is significantly lower than estimates based on optical data \citep{2000ApJ...537..667B,2009ApJ...707L..27F}, but consistent with those of \cite{2018ApJ...863..127K}.     

The presence of Fe K emission in N132D is well known \citep{2001A&A...365L.242B,2008AdSpR..41..416X,2014ApJ...785L..27Y,2016A&A...585A.162M,2018ApJ...854...71B,2018PASJ...70...16H}. With the spatial resolution of \chandra, we find that the Fe K complex emission is distributed largely across its southern half and is not located in a single feature. We fit the spectra of this emission in six regions using two different models that have three components each. These two models have two components in common, which account for the shell emission (plane-parallel shock) and cooler, soft X-ray emitting plasma (non-ionization equilibrium). The third component which accounts for the hotter, hard X-ray emitting plasma and Fe K emission is a non-equilibrium ionizing plasma in one model and a recombining plasma in the other model. In both the models, we find that a hot plasma ($\ga 1.5\,\kev$) is needed to explain the Fe K feature, and that this plasma is distinct from the soft x-ray emitting plasma. While our fits cannot distinguish between the ionizing and recombining plasma models for these regions because they result in similar fit statistics, we confirm the existence of such a hot plasma, in agreement with the findings of \cite{2018ApJ...854...71B}. A deeper observation and/or an observation with higher spectral resolution will help break the degeneracy between the two models and possibly shed light on the origin of the hot plasma and its interactions with molecular clouds in the region.

Thus, our analysis leads us to conclude that SNR N132D probably resulted from the core-collapse of an intermediate mass progenitor, in a cavity in the CSM created by pre-supernova winds. The exact type of the explosion, the possibility of a Wolf-Rayet phase prior to it, and the nature of the hot Fe K emitting plasma are some of the pertinent questions that still remain unanswered. Deeper observations with existing instruments and future observations with new instruments with enhanced capabilities will be required to address these questions.

\acknowledgments
We thank the anonymous referee for their comments that significantly improved the analysis and presentation of the results. We thank Chiaki Kobayashi for discussions on nucleosynthesis modeling of massive stars, Dominique Meyer for discussions on massive runaway stars, Katie Jameson, Nigel Maxted, Nickolas Pingel and Tony Wong for discussions on atomic and molecular gas in N132D, and Michael Dopita for discussions on collisionless shocks. We are grateful to Keith Arnaud and Craig Gordon for providing support with the X-ray fitting package \xspec. PS acknowledges the Birla Institute of Technology and Science Pilani Alumni Association (BITSAA) undergraduate summer research scholarship and the Australian Government Research Training Program Scholarship (AGRTP). TJG, VLK, and PPP acknowledge support under NASA contract NAS8-03060 with the \chandra X-ray Center. The scientific results reported in this article are based on data obtained from the \chandra Data Archive and the \chandra Source Catalog, and observations made by the \chandra X-ray Observatory. This research has also made use of software provided by the \chandra X-ray Center (CXC) in the application package CIAO, and the NASA's Astrophysics Data System (ADS).

\facility{CXO \citep{2000SPIE.4012....2W}}
\software{CIAO \citep{2006SPIE.6270E..1VF}, CALDB \citep{2007ChNew..14...33G}, Xspec \citep{1996ASPC..101...17A}, SAO ds9 \citep{2003ASPC..295..489J}, PINTofALE \citep{2000BASI...28..475K}, FTOOLS \citep{1995ASPC...77..367B}, Matplotlib \citep{Hunter:2007}, Numpy \citep{oliphant2006guide}.}

\bibliographystyle{yahapj}
\bibliography{references}

\begin{thebibliography}{}
\providecommand\natexlab[1]{#1}
\providecommand\JournalTitle[1]{#1}

\bibitem[{{Acero} {et~al.}(2016){Acero}, {Ackermann}, {Ajello}, {Baldini},
  {Ballet}, {Barbiellini}, {Bastieri}, {Bellazzini}, {Bissaldi}, {Blandford},
  {Bloom}, {Bonino}, {Bottacini}, {Brandt}, {Bregeon}, {Bruel}, {Buehler},
  {Buson}, {Caliandro}, {Cameron}, {Caputo}, {Caragiulo}, {Caraveo},
  {Casandjian}, {Cavazzuti}, {Cecchi}, {Chekhtman}, {Chiang}, {Chiaro},
  {Ciprini}, {Claus}, {Cohen}, {Cohen-Tanugi}, {Cominsky}, {Condon}, {Conrad},
  {Cutini}, {D'Ammando}, {de Angelis}, {de Palma}, {Desiante}, {Digel}, {Di
  Venere}, {Drell}, {Drlica-Wagner}, {Favuzzi}, {Ferrara}, {Franckowiak},
  {Fukazawa}, {Funk}, {Fusco}, {Gargano}, {Gasparrini}, {Giglietto}, {Giommi},
  {Giordano}, {Giroletti}, {Glanzman}, {Godfrey}, {Gomez-Vargas}, {Grenier},
  {Grondin}, {Guillemot}, {Guiriec}, {Gustafsson}, {Hadasch}, {Harding},
  {Hayashida}, {Hays}, {Hewitt}, {Hill}, {Horan}, {Hou}, {Iafrate}, {Jogler},
  {J{\'o}hannesson}, {Johnson}, {Kamae}, {Katagiri}, {Kataoka}, {Katsuta},
  {Kerr}, {Kn{\"o}dlseder}, {Kocevski}, {Kuss}, {Laffon}, {Lande}, {Larsson},
  {Latronico}, {Lemoine-Goumard}, {Li}, {Li}, {Longo}, {Loparco}, {Lovellette},
  {Lubrano}, {Magill}, {Maldera}, {Marelli}, {Mayer}, {Mazziotta}, {Michelson},
  {Mitthumsiri}, {Mizuno}, {Moiseev}, {Monzani}, {Moretti}, {Morselli},
  {Moskalenko}, {Murgia}, {Nemmen}, {Nuss}, {Ohsugi}, {Omodei}, {Orienti},
  {Orlando}, {Ormes}, {Paneque}, {Perkins}, {Pesce-Rollins}, {Petrosian},
  {Piron}, {Pivato}, {Porter}, {Rain{\`o}}, {Rando}, {Razzano}, {Razzaque},
  {Reimer}, {Reimer}, {Renaud}, {Reposeur}, {Rousseau}, {Saz Parkinson},
  {Schmid}, {Schulz}, {Sgr{\`o}}, {Siskind}, {Spada}, {Spandre}, {Spinelli},
  {Strong}, {Suson}, {Tajima}, {Takahashi}, {Tanaka}, {Thayer}, {Thompson},
  {Tibaldo}, {Tibolla}, {Torres}, {Tosti}, {Troja}, {Uchiyama}, {Vianello},
  {Wells}, {Wood}, {Wood}, {Yassine}, {den Hartog}, \&
  {Zimmer}}]{2016ApJS..224....8A}
{Acero}, F., {Ackermann}, M., {Ajello}, M., {et~al.} 2016,
  \href{http://dx.doi.org/10.3847/0067-0049/224/1/8}{\JournalTitle{\apjs}, 224,
  8}

\bibitem[{{Ackermann} {et~al.}(2016){Ackermann}, {Albert}, {Atwood}, {Baldini},
  {Ballet}, {Barbiellini}, {Bastieri}, {Bellazzini}, {Bissaldi}, {Bloom},
  {Bonino}, {Brandt}, {Bregeon}, {Bruel}, {Buehler}, {Caliandro}, {Cameron},
  {Caragiulo}, {Caraveo}, {Cavazzuti}, {Cecchi}, {Charles}, {Chekhtman},
  {Chiang}, {Chiaro}, {Ciprini}, {Cohen-Tanugi}, {Cutini}, {D'Ammando}, {de
  Angelis}, {de Palma}, {Desiante}, {Digel}, {Drell}, {Favuzzi}, {Ferrara},
  {Focke}, {Franckowiak}, {Fusco}, {Gargano}, {Gasparrini}, {Giglietto},
  {Giordano}, {Godfrey}, {Grenier}, {Grondin}, {Guillemot}, {Guiriec},
  {Harding}, {Hill}, {Horan}, {J{\'o}hannesson}, {Kn{\"o}dlseder}, {Kuss},
  {Larsson}, {Latronico}, {Li}, {Li}, {Longo}, {Loparco}, {Lubrano}, {Maldera},
  {Martin}, {Mayer}, {Mazziotta}, {Michelson}, {Mizuno}, {Monzani}, {Morselli},
  {Murgia}, {Nuss}, {Ohsugi}, {Orienti}, {Orlando}, {Ormes}, {Paneque},
  {Pesce-Rollins}, {Piron}, {Pivato}, {Porter}, {Rain{\`o}}, {Rando},
  {Razzano}, {Reimer}, {Reimer}, {Romani}, {S{\'a}nchez-Conde}, {Schulz},
  {Sgr{\`o}}, {Siskind}, {Smith}, {Spada}, {Spandre}, {Spinelli}, {Suson},
  {Takahashi}, {Thayer}, {Tibaldo}, {Torres}, {Tosti}, {Troja}, {Vianello},
  {Wood}, \& {Zimmer}}]{2016A&A...586A..71A}
{Ackermann}, M., {Albert}, A., {Atwood}, W.~B., {et~al.} 2016,
  \href{http://dx.doi.org/10.1051/0004-6361/201526920}{\JournalTitle{\aap},
  586, A71}

\bibitem[{{Arnaud}(1996)}]{1996ASPC..101...17A}
{Arnaud}, K.~A. 1996, in Astronomical Society of the Pacific Conference Series,
  Vol. 101, Astronomical Data Analysis Software and Systems V, ed. G.~H.
  {Jacoby} \& J.~{Barnes}, 17

\bibitem[{{Auchettl} {et~al.}(2017){Auchettl}, {Ng}, {Wong}, {Lopez}, \&
  {Slane}}]{2017ApJ...847..121A}
{Auchettl}, K., {Ng}, C.-Y., {Wong}, B.~T.~T., {Lopez}, L., \& {Slane}, P.
  2017, \href{http://dx.doi.org/10.3847/1538-4357/aa830e}{\JournalTitle{\apj},
  847, 121}

\bibitem[{{Ballet}(1999)}]{1999A&AS..135..371B}
{Ballet}, J. 1999,
  \href{http://dx.doi.org/10.1051/aas:1999179}{\JournalTitle{\aaps}, 135, 371}

\bibitem[{{Bamba} {et~al.}(2018){Bamba}, {Ohira}, {Yamazaki}, {Sawada},
  {Terada}, {Koyama}, {Miller}, {Yamaguchi}, {Katsuda}, {Nobukawa}, \&
  {Nobukawa}}]{2018ApJ...854...71B}
{Bamba}, A., {Ohira}, Y., {Yamazaki}, R., {et~al.} 2018,
  \href{http://dx.doi.org/10.3847/1538-4357/aaa5a0}{\JournalTitle{\apj}, 854,
  71}

\bibitem[{{Banas} {et~al.}(1997){Banas}, {Hughes}, {Bronfman}, \&
  {Nyman}}]{1997ApJ...480..607B}
{Banas}, K.~R., {Hughes}, J.~P., {Bronfman}, L., \& {Nyman}, L.-{\AA}. 1997,
  \href{http://dx.doi.org/10.1086/303989}{\JournalTitle{\apj}, 480, 607}

\bibitem[{{Bartalucci} {et~al.}(2014){Bartalucci}, {Mazzotta}, {Bourdin}, \&
  {Vikhlinin}}]{2014A&A...566A..25B}
{Bartalucci}, I., {Mazzotta}, P., {Bourdin}, H., \& {Vikhlinin}, A. 2014,
  \href{http://dx.doi.org/10.1051/0004-6361/201423443}{\JournalTitle{\aap},
  566, A25}

\bibitem[{{Bautz} {et~al.}(1998){Bautz}, {Pivovaroff}, {Baganoff}, {Isobe},
  {Jones}, {Kissel}, {Lamarr}, {Manning}, {Prigozhin}, {Ricker}, {Nousek},
  {Grant}, {Nishikida}, {Scholze}, {Thornagel}, \& {Ulm}}]{1998SPIE.3444..210B}
{Bautz}, M.~W., {Pivovaroff}, M., {Baganoff}, F., {et~al.} 1998, Society of
  Photo-Optical Instrumentation Engineers (SPIE) Conference Series, Vol. 3444,
  {X-ray CCD calibration for the AXAF CCD Imaging Spectrometer}, ed. R.~B.
  {Hoover} \& A.~B. {Walker}, 210

\bibitem[{{Bearden}(1967)}]{1967RvMP...39...78B}
{Bearden}, J.~A. 1967,
  \href{http://dx.doi.org/10.1103/RevModPhys.39.78}{\JournalTitle{Reviews of
  Modern Physics}, 39, 78}

\bibitem[{{Behar} {et~al.}(2001){Behar}, {Rasmussen}, {Griffiths}, {Dennerl},
  {Audard}, {Aschenbach}, \& {Brinkman}}]{2001A&A...365L.242B}
{Behar}, E., {Rasmussen}, A.~P., {Griffiths}, R.~G., {et~al.} 2001,
  \href{http://dx.doi.org/10.1051/0004-6361:20000082}{\JournalTitle{\aap}, 365,
  L242}

\bibitem[{{Berezhko} \& {Ellison}(1999)}]{1999ApJ...526..385B}
{Berezhko}, E.~G., \& {Ellison}, D.~C. 1999,
  \href{http://dx.doi.org/10.1086/307993}{\JournalTitle{\apj}, 526, 385}

\bibitem[{{Blackburn}(1995)}]{1995ASPC...77..367B}
{Blackburn}, J.~K. 1995, Astronomical Society of the Pacific Conference Series,
  Vol.~77, {FTOOLS: A FITS Data Processing and Analysis Software Package}, ed.
  R.~A. {Shaw}, H.~E. {Payne}, \& J.~J.~E. {Hayes}, 367

\bibitem[{{Blair} {et~al.}(1994){Blair}, {Raymond}, \&
  {Long}}]{1994ApJ...423..334B}
{Blair}, W.~P., {Raymond}, J.~C., \& {Long}, K.~S. 1994,
  \href{http://dx.doi.org/10.1086/173811}{\JournalTitle{\apj}, 423, 334}

\bibitem[{{Blair} {et~al.}(2000){Blair}, {Morse}, {Raymond}, {Kirshner},
  {Hughes}, {Dopita}, {Sutherland}, {Long}, \& {Winkler}}]{2000ApJ...537..667B}
{Blair}, W.~P., {Morse}, J.~A., {Raymond}, J.~C., {et~al.} 2000,
  \href{http://dx.doi.org/10.1086/309077}{\JournalTitle{\apj}, 537, 667}

\bibitem[{{Blitz}(1993)}]{1993prpl.conf..125B}
{Blitz}, L. 1993, in Protostars and Planets III, ed. E.~H. {Levy} \& J.~I.
  {Lunine}, 125

\bibitem[{{Borkowski} {et~al.}(2007){Borkowski}, {Hendrick}, \&
  {Reynolds}}]{2007ApJ...671L..45B}
{Borkowski}, K.~J., {Hendrick}, S.~P., \& {Reynolds}, S.~P. 2007,
  \href{http://dx.doi.org/10.1086/524733}{\JournalTitle{\apjl}, 671, L45}

\bibitem[{{Borkowski} {et~al.}(2001){Borkowski}, {Lyerly}, \&
  {Reynolds}}]{2001ApJ...548..820B}
{Borkowski}, K.~J., {Lyerly}, W.~J., \& {Reynolds}, S.~P. 2001,
  \href{http://dx.doi.org/10.1086/319011}{\JournalTitle{\apj}, 548, 820}

\bibitem[{{Braun} {et~al.}(2019){Braun}, {Safi-Harb}, \&
  {Fryer}}]{2019MNRAS.489.4444B}
{Braun}, C., {Safi-Harb}, S., \& {Fryer}, C.~L. 2019,
  \href{http://dx.doi.org/10.1093/mnras/stz2437}{\JournalTitle{\mnras}, 489,
  4444}

\bibitem[{{Cash}(1979)}]{1979ApJ...228..939C}
{Cash}, W. 1979, \href{http://dx.doi.org/10.1086/156922}{\JournalTitle{\apj},
  228, 939}

\bibitem[{{Chen} {et~al.}(1997){Chen}, {Fabian}, \&
  {Gendreau}}]{1997MNRAS.285..449C}
{Chen}, L.~W., {Fabian}, A.~C., \& {Gendreau}, K.~C. 1997,
  \href{http://dx.doi.org/10.1093/mnras/285.3.449}{\JournalTitle{\mnras}, 285,
  449}

\bibitem[{{Chen} {et~al.}(2003){Chen}, {Zhang}, {Williams}, \&
  {Wang}}]{2003ApJ...595..227C}
{Chen}, Y., {Zhang}, F., {Williams}, R.~M., \& {Wang}, Q.~D. 2003,
  \href{http://dx.doi.org/10.1086/377353}{\JournalTitle{\apj}, 595, 227}

\bibitem[{{Chen} {et~al.}(2013){Chen}, {Zhou}, \& {Chu}}]{2013ApJ...769L..16C}
{Chen}, Y., {Zhou}, P., \& {Chu}, Y.-H. 2013,
  \href{http://dx.doi.org/10.1088/2041-8205/769/1/L16}{\JournalTitle{\apjl},
  769, L16}

\bibitem[{{Chevalier}(1999)}]{1999ApJ...511..798C}
{Chevalier}, R.~A. 1999,
  \href{http://dx.doi.org/10.1086/306710}{\JournalTitle{\apj}, 511, 798}

\bibitem[{{Chevalier}(2005)}]{2005ApJ...619..839C}
---. 2005, \href{http://dx.doi.org/10.1086/426584}{\JournalTitle{\apj}, 619,
  839}

\bibitem[{{Clementini} {et~al.}(2003){Clementini}, {Gratton}, {Bragaglia},
  {Carretta}, {Di Fabrizio}, \& {Maio}}]{2003AJ....125.1309C}
{Clementini}, G., {Gratton}, R., {Bragaglia}, A., {et~al.} 2003,
  \href{http://dx.doi.org/10.1086/367773}{\JournalTitle{\aj}, 125, 1309}

\bibitem[{{Danziger} \& {Dennefeld}(1976)}]{1976ApJ...207..394D}
{Danziger}, I.~J., \& {Dennefeld}, M. 1976,
  \href{http://dx.doi.org/10.1086/154507}{\JournalTitle{\apj}, 207, 394}

\bibitem[{{Davis}(2001)}]{2001ApJ...562..575D}
{Davis}, J.~E. 2001,
  \href{http://dx.doi.org/10.1086/323488}{\JournalTitle{\apj}, 562, 575}

\bibitem[{{DeLaney} {et~al.}(2014){DeLaney}, {Kassim}, {Rudnick}, \&
  {Perley}}]{2014ApJ...785....7D}
{DeLaney}, T., {Kassim}, N.~E., {Rudnick}, L., \& {Perley}, R.~A. 2014,
  \href{http://dx.doi.org/10.1088/0004-637X/785/1/7}{\JournalTitle{\apj}, 785,
  7}

\bibitem[{{Desai} {et~al.}(2010){Desai}, {Chu}, {Gruendl}, {Dluger}, {Katz},
  {Wong}, {Chen}, {Looney}, {Hughes}, {Muller}, {Ott}, \&
  {Pineda}}]{2010AJ....140..584D}
{Desai}, K.~M., {Chu}, Y.-H., {Gruendl}, R.~A., {et~al.} 2010,
  \href{http://dx.doi.org/10.1088/0004-6256/140/2/584}{\JournalTitle{\aj}, 140,
  584}

\bibitem[{{Dickel} \& {Milne}(1995)}]{1995AJ....109..200D}
{Dickel}, J.~R., \& {Milne}, D.~K. 1995,
  \href{http://dx.doi.org/10.1086/117266}{\JournalTitle{\aj}, 109, 200}

\bibitem[{{Dickey} \& {Lockman}(1990)}]{1990ARA&A..28..215D}
{Dickey}, J.~M., \& {Lockman}, F.~J. 1990,
  \href{http://dx.doi.org/10.1146/annurev.aa.28.090190.001243}{\JournalTitle{\araa},
  28, 215}

\bibitem[{{Dopita} {et~al.}(2010){Dopita}, {Rhee}, {Farage}, {McGregor},
  {Bloxham}, {Green}, {Roberts}, {Neilson}, {Wilson}, {Young}, {Firth},
  {Busarello}, \& {Merluzzi}}]{2010Ap&SS.327..245D}
{Dopita}, M., {Rhee}, J., {Farage}, C., {et~al.} 2010,
  \href{http://dx.doi.org/10.1007/s10509-010-0335-9}{\JournalTitle{\apss}, 327,
  245}

\bibitem[{{Dopita} {et~al.}(2019){Dopita}, {Seitenzahl}, {Sutherland },
  {Nicholls}, {Vogt}, {Ghavamian}, \& {Ruiter}}]{2019AJ....157...50D}
{Dopita}, M.~A., {Seitenzahl}, I.~R., {Sutherland }, R.~S., {et~al.} 2019,
  \href{http://dx.doi.org/10.3847/1538-3881/aaf235}{\JournalTitle{\aj}, 157,
  50}

\bibitem[{{Dopita} {et~al.}(2018){Dopita}, {Vogt}, {Sutherland}, {Seitenzahl},
  {Ruiter}, \& {Ghavamian}}]{2018ApJS..237...10D}
{Dopita}, M.~A., {Vogt}, F.~P.~A., {Sutherland}, R.~S., {et~al.} 2018,
  \href{http://dx.doi.org/10.3847/1538-4365/aac837}{\JournalTitle{\apjs}, 237,
  10}

\bibitem[{{Dufour} {et~al.}(1982){Dufour}, {Shields}, \&
  {Talbot}}]{1982ApJ...252..461D}
{Dufour}, R.~J., {Shields}, G.~A., \& {Talbot}, Jr., R.~J. 1982,
  \href{http://dx.doi.org/10.1086/159574}{\JournalTitle{\apj}, 252, 461}

\bibitem[{{Dwarkadas}(2007)}]{2007ApJ...667..226D}
{Dwarkadas}, V.~V. 2007,
  \href{http://dx.doi.org/10.1086/520670}{\JournalTitle{\apj}, 667, 226}

\bibitem[{{Dwarkadas} \& {Gruszko}(2012)}]{2012MNRAS.419.1515D}
{Dwarkadas}, V.~V., \& {Gruszko}, J. 2012,
  \href{http://dx.doi.org/10.1111/j.1365-2966.2011.19808.x}{\JournalTitle{\mnras},
  419, 1515}

\bibitem[{{Ellison} {et~al.}(2007){Ellison}, {Patnaude}, {Slane}, {Blasi}, \&
  {Gabici}}]{2007ApJ...661..879E}
{Ellison}, D.~C., {Patnaude}, D.~J., {Slane}, P., {Blasi}, P., \& {Gabici}, S.
  2007, \href{http://dx.doi.org/10.1086/517518}{\JournalTitle{\apj}, 661, 879}

\bibitem[{{Favata} {et~al.}(1997){Favata}, {Vink}, {Parmar}, {Kaastra}, \&
  {Mineo}}]{1997A&A...324L..45F}
{Favata}, F., {Vink}, J., {Parmar}, A.~N., {Kaastra}, J.~S., \& {Mineo}, T.
  1997, \JournalTitle{\aap}, 324, L45

\bibitem[{{Foster} {et~al.}(2013){Foster}, {Ji}, {Yamaguchi}, {Smith}, \&
  {Brickhouse}}]{2013AIPC.1545..252F}
{Foster}, A.~R., {Ji}, L., {Yamaguchi}, H., {Smith}, R.~K., \& {Brickhouse},
  N.~S. 2013, \href{http://dx.doi.org/10.1063/1.4815861}{in American Institute
  of Physics Conference Series, Vol. 1545, American Institute of Physics
  Conference Series, ed. J.~D. {Gillaspy}, W.~L. {Wiese}, \& Y.~A. {Podpaly}},
  252

\bibitem[{{France} {et~al.}(2009){France}, {Beasley}, {Keeney}, {Danforth},
  {Froning}, {Green}, \& {Shull}}]{2009ApJ...707L..27F}
{France}, K., {Beasley}, M., {Keeney}, B.~A., {et~al.} 2009,
  \href{http://dx.doi.org/10.1088/0004-637X/707/1/L27}{\JournalTitle{\apjl},
  707, L27}

\bibitem[{{Frank} {et~al.}(2015){Frank}, {Burrows}, \&
  {Park}}]{2015ApJ...810..113F}
{Frank}, K.~A., {Burrows}, D.~N., \& {Park}, S. 2015,
  \href{http://dx.doi.org/10.1088/0004-637X/810/2/113}{\JournalTitle{\apj},
  810, 113}

\bibitem[{{Fruscione} {et~al.}(2006){Fruscione}, {McDowell}, {Allen},
  {Brickhouse}, {Burke}, {Davis}, {Durham}, {Elvis}, {Galle}, {Harris},
  {Huenemoerder}, {Houck}, {Ishibashi}, {Karovska}, {Nicastro}, {Noble},
  {Nowak}, {Primini}, {Siemiginowska}, {Smith}, \&
  {Wise}}]{2006SPIE.6270E..1VF}
{Fruscione}, A., {McDowell}, J.~C., {Allen}, G.~E., {et~al.} 2006,
  \href{http://dx.doi.org/10.1117/12.671760}{in \procspie, Vol. 6270, Society
  of Photo-Optical Instrumentation Engineers (SPIE) Conference Series}, 62701V

\bibitem[{{Fukui} {et~al.}(2008){Fukui}, {Kawamura}, {Minamidani}, {Mizuno},
  {Kanai}, {Mizuno}, {Onishi}, {Yonekura}, {Mizuno}, {Ogawa}, \&
  {Rubio}}]{2008ApJS..178...56F}
{Fukui}, Y., {Kawamura}, A., {Minamidani}, T., {et~al.} 2008,
  \href{http://dx.doi.org/10.1086/589833}{\JournalTitle{\apjs}, 178, 56}

\bibitem[{{Gaetz}(1990)}]{1990ApJ...353..245G}
{Gaetz}, T.~J. 1990,
  \href{http://dx.doi.org/10.1086/168611}{\JournalTitle{\apj}, 353, 245}

\bibitem[{{Garcia-Segura} {et~al.}(1996){Garcia-Segura}, {Langer}, \& {Mac
  Low}}]{1996A&A...316..133G}
{Garcia-Segura}, G., {Langer}, N., \& {Mac Low}, M.-M. 1996,
  \JournalTitle{\aap}, 316, 133

\bibitem[{{Garofali} {et~al.}(2017){Garofali}, {Williams}, {Plucinsky},
  {Gaetz}, {Wold}, {Haberl}, {Long}, {Blair}, {Pannuti}, {Winkler}, \&
  {Gross}}]{2017MNRAS.472..308G}
{Garofali}, K., {Williams}, B.~F., {Plucinsky}, P.~P., {et~al.} 2017,
  \href{http://dx.doi.org/10.1093/mnras/stx1905}{\JournalTitle{\mnras}, 472,
  308}

\bibitem[{{Ghavamian} {et~al.}(2005){Ghavamian}, {Hughes}, \&
  {Williams}}]{2005ApJ...635..365G}
{Ghavamian}, P., {Hughes}, J.~P., \& {Williams}, T.~B. 2005,
  \href{http://dx.doi.org/10.1086/497283}{\JournalTitle{\apj}, 635, 365}

\bibitem[{{Ghavamian} {et~al.}(2007){Ghavamian}, {Laming}, \&
  {Rakowski}}]{2007ApJ...654L..69G}
{Ghavamian}, P., {Laming}, J.~M., \& {Rakowski}, C.~E. 2007,
  \href{http://dx.doi.org/10.1086/510740}{\JournalTitle{\apjl}, 654, L69}

\bibitem[{{Ghavamian} {et~al.}(2013){Ghavamian}, {Schwartz}, {Mitchell},
  {Masters}, \& {Laming}}]{2013SSRv..178..633G}
{Ghavamian}, P., {Schwartz}, S.~J., {Mitchell}, J., {Masters}, A., \& {Laming},
  J.~M. 2013,
  \href{http://dx.doi.org/10.1007/s11214-013-9999-0}{\JournalTitle{\ssr}, 178,
  633}

\bibitem[{{Graessle} {et~al.}(2007){Graessle}, {Evans}, {Glotfelty}, {He},
  {Evans}, {Rots}, {Fabbiano}, \& {Brissenden}}]{2007ChNew..14...33G}
{Graessle}, D.~E., {Evans}, I.~N., {Glotfelty}, K., {et~al.} 2007,
  \JournalTitle{Chandra News}, 14, 33

\bibitem[{{Hamilton} {et~al.}(1983){Hamilton}, {Sarazin}, \&
  {Chevalier}}]{1983ApJS...51..115H}
{Hamilton}, A.~J.~S., {Sarazin}, C.~L., \& {Chevalier}, R.~A. 1983,
  \href{http://dx.doi.org/10.1086/190841}{\JournalTitle{\apjs}, 51, 115}

\bibitem[{{Henize}(1956)}]{1956ApJS....2..315H}
{Henize}, K.~G. 1956,
  \href{http://dx.doi.org/10.1086/190025}{\JournalTitle{\apjs}, 2, 315}

\bibitem[{{Herrera-Camus} {et~al.}(2017){Herrera-Camus}, {Bolatto}, {Wolfire},
  {Ostriker}, {Draine}, {Leroy}, {Sandstrom}, {Hunt}, {Kennicutt}, {Calzetti},
  {Smith}, {Croxall}, {Galametz}, {de Looze}, {Dale}, {Crocker}, \&
  {Groves}}]{2017ApJ...835..201H}
{Herrera-Camus}, R., {Bolatto}, A., {Wolfire}, M., {et~al.} 2017,
  \href{http://dx.doi.org/10.3847/1538-4357/835/2/201}{\JournalTitle{\apj},
  835, 201}

\bibitem[{{H.E.S.S.~Collaboration} {et~al.}(2015){H.E.S.S.~Collaboration},
  {Abramowski}, {Aharonian}, {Ait Benkhali}, {Akhperjanian}, {Ang{\"u}ner},
  {Backes}, {Balenderan}, {Balzer}, {Barnacka}, \&
  et~al.}]{2015Sci...347..406H}
{H.E.S.S.~Collaboration}, {Abramowski}, A., {Aharonian}, F., {et~al.} 2015,
  \href{http://dx.doi.org/10.1126/science.1261313}{\JournalTitle{Science}, 347,
  406}

\bibitem[{{Higdon} \& {Lingenfelter}(1980)}]{1980ApJ...239..867H}
{Higdon}, J.~C., \& {Lingenfelter}, R.~E. 1980,
  \href{http://dx.doi.org/10.1086/158171}{\JournalTitle{\apj}, 239, 867}

\bibitem[{{Hitomi Collaboration} {et~al.}(2018){Hitomi Collaboration},
  {Aharonian}, {Akamatsu}, {Akimoto}, {Allen}, {Angelini}, {Audard}, {Awaki},
  {Axelsson}, {Bamba}, {Bautz}, {Blandford}, {Brenneman}, {Brown}, {Bulbul},
  {Cackett}, {Chernyakova}, {Chiao}, {Coppi}, {Costantini}, {de Plaa}, {de
  Vries}, {den Herder}, {Done}, {Dotani}, {Ebisawa}, {Eckart}, {Enoto}, {Ezoe},
  {Fabian}, {Ferrigno}, {Foster}, {Fujimoto}, {Fukazawa}, {Furuzawa},
  {Galeazzi}, {Gallo}, {Gandhi}, {Giustini}, {Goldwurm}, {Gu}, {Guainazzi},
  {Haba}, {Hagino}, {Hamaguchi}, {Harrus}, {Hatsukade}, {Hayashi}, {Hayashi},
  {Hayashida}, {Hiraga}, {Hornschemeier}, {Hoshino}, {Hughes}, {Ichinohe},
  {Iizuka}, {Inoue}, {Inoue}, {Ishida}, {Ishikawa}, {Ishisaki}, {Iwai},
  {Kaastra}, {Kallman}, {Kamae}, {Kataoka}, {Katsuda}, {Kawai}, {Kelley},
  {Kilbourne}, {Kitaguchi}, {Kitamoto}, {Kitayama}, {Kohmura}, {Kokubun},
  {Koyama}, {Koyama}, {Kretschmar}, {Krimm}, {Kubota}, {Kunieda}, {Laurent},
  {Lee}, {Leutenegger}, {Limousin}, {Loewenstein}, {Long}, {Lumb}, {Madejski},
  {Maeda}, {Maier}, {Makishima}, {Markevitch}, {Matsumoto}, {Matsushita},
  {McCammon}, {McNamara}, {Mehdipour}, {Miller}, {Miller}, {Mineshige},
  {Mitsuda}, {Mitsuishi}, {Miyazawa}, {Mizuno}, {Mori}, {Mori}, {Mukai},
  {Murakami}, {Mushotzky}, {Nakagawa}, {Nakajima}, {Nakamori}, {Nakashima},
  {Nakazawa}, {Nobukawa}, {Nobukawa}, {Noda}, {Odaka}, {Ohashi}, {Ohno},
  {Okajima}, {Ota}, {Ozaki}, {Paerels}, {Paltani}, {Petre}, {Pinto}, {Porter},
  {Pottschmidt}, {Reynolds}, {Safi-Harb}, {Saito}, {Sakai}, {Sasaki}, {Sato},
  {Sato}, {Sato}, {Sato}, {Sawada}, {Schartel}, {Serlemtsos}, {Seta},
  {Shidatsu}, {Simionescu}, {Smith}, {Soong}, {Stawarz}, {Sugawara}, {Sugita},
  {Szymkowiak}, {Tajima}, {Takahashi}, {Takahashi}, {Takeda}, {Takei},
  {Tamagawa}, {Tamura}, {Tanaka}, {Tanaka}, {Tanaka}, {Tashiro}, {Tawara},
  {Terada}, {Terashima}, {Tombesi}, {Tomida}, {Tsuboi}, {Tsujimoto}, {Tsunemi},
  {Tsuru}, {Uchida}, {Uchiyama}, {Uchiyama}, {Ueda}, {Ueda}, {Uno}, {Urry},
  {Ursino}, {Watanabe}, {Werner}, {Wilkins}, {Williams}, {Yamada}, {Yamaguchi},
  {Yamaoka}, {Yamasaki}, {Yamauchi}, {Yamauchi}, {Yaqoob}, {Yatsu}, {Yonetoku},
  {Zhuravleva}, \& {Zoghbi}}]{2018PASJ...70...16H}
{Hitomi Collaboration}, {Aharonian}, F., {Akamatsu}, H., {et~al.} 2018,
  \href{http://dx.doi.org/10.1093/pasj/psx151}{\JournalTitle{\pasj}, 70, 16}

\bibitem[{{Holland-Ashford} {et~al.}(2017){Holland-Ashford}, {Lopez},
  {Auchettl}, {Temim}, \& {Ramirez-Ruiz}}]{2017ApJ...844...84H}
{Holland-Ashford}, T., {Lopez}, L.~A., {Auchettl}, K., {Temim}, T., \&
  {Ramirez-Ruiz}, E. 2017,
  \href{http://dx.doi.org/10.3847/1538-4357/aa7a5c}{\JournalTitle{\apj}, 844,
  84}

\bibitem[{{Hughes}(1987)}]{1987ApJ...314..103H}
{Hughes}, J.~P. 1987,
  \href{http://dx.doi.org/10.1086/165043}{\JournalTitle{\apj}, 314, 103}

\bibitem[{{Hughes} {et~al.}(1998){Hughes}, {Hayashi}, \&
  {Koyama}}]{1998ApJ...505..732H}
{Hughes}, J.~P., {Hayashi}, I., \& {Koyama}, K. 1998,
  \href{http://dx.doi.org/10.1086/306202}{\JournalTitle{\apj}, 505, 732}

\bibitem[{Hunter(2007)}]{Hunter:2007}
Hunter, J.~D. 2007,
  \href{http://dx.doi.org/10.1109/MCSE.2007.55}{\JournalTitle{Computing in
  Science \& Engineering}, 9, 90}

\bibitem[{{Hwang} {et~al.}(1993){Hwang}, {Hughes}, {Canizares}, \&
  {Markert}}]{1993ApJ...414..219H}
{Hwang}, U., {Hughes}, J.~P., {Canizares}, C.~R., \& {Markert}, T.~H. 1993,
  \href{http://dx.doi.org/10.1086/173070}{\JournalTitle{\apj}, 414, 219}

\bibitem[{{Hwang} \& {Laming}(2012)}]{2012ApJ...746..130H}
{Hwang}, U., \& {Laming}, J.~M. 2012,
  \href{http://dx.doi.org/10.1088/0004-637X/746/2/130}{\JournalTitle{\apj},
  746, 130}

\bibitem[{{Itoh}(1977)}]{1977PASJ...29..813I}
{Itoh}, H. 1977, \JournalTitle{\pasj}, 29, 813

\bibitem[{{Itoh} \& {Masai}(1989)}]{1989MNRAS.236..885I}
{Itoh}, H., \& {Masai}, K. 1989,
  \href{http://dx.doi.org/10.1093/mnras/236.4.885}{\JournalTitle{\mnras}, 236,
  885}

\bibitem[{{Jenkins} \& {Tripp}(2011)}]{2011ApJ...734...65J}
{Jenkins}, E.~B., \& {Tripp}, T.~M. 2011,
  \href{http://dx.doi.org/10.1088/0004-637X/734/1/65}{\JournalTitle{\apj}, 734,
  65}

\bibitem[{{Jones} \& {Ellison}(1991)}]{1991SSRv...58..259J}
{Jones}, F.~C., \& {Ellison}, D.~C. 1991,
  \href{http://dx.doi.org/10.1007/BF01206003}{\JournalTitle{\ssr}, 58, 259}

\bibitem[{{Joye} \& {Mandel}(2003)}]{2003ASPC..295..489J}
{Joye}, W.~A., \& {Mandel}, E. 2003, in Astronomical Society of the Pacific
  Conference Series, Vol. 295, Astronomical Data Analysis Software and Systems
  XII, ed. H.~E. {Payne}, R.~I. {Jedrzejewski}, \& R.~N. {Hook}, 489

\bibitem[{{Kaastra}(2017)}]{2017A&A...605A..51K}
{Kaastra}, J.~S. 2017,
  \href{http://dx.doi.org/10.1051/0004-6361/201629319}{\JournalTitle{\aap},
  605, A51}

\bibitem[{{Kamble} {et~al.}(2016){Kamble}, {Margutti}, {Soderberg},
  {Chakraborti}, {Fransson}, {Chevalier}, {Powell}, {Milisavljevic}, {Parrent},
  \& {Bietenholz}}]{2016ApJ...818..111K}
{Kamble}, A., {Margutti}, R., {Soderberg}, A.~M., {et~al.} 2016,
  \href{http://dx.doi.org/10.3847/0004-637X/818/2/111}{\JournalTitle{\apj},
  818, 111}

\bibitem[{{Kamitsukasa} {et~al.}(2015){Kamitsukasa}, {Koyama}, {Uchida},
  {Nakajima}, {Hayashida}, {Mori}, {Katsuda}, \&
  {Tsunemi}}]{2015PASJ...67...16K}
{Kamitsukasa}, F., {Koyama}, K., {Uchida}, H., {et~al.} 2015,
  \href{http://dx.doi.org/10.1093/pasj/psu149}{\JournalTitle{\pasj}, 67, 16}

\bibitem[{{Kashyap} \& {Drake}(2000)}]{2000BASI...28..475K}
{Kashyap}, V., \& {Drake}, J.~J. 2000, \JournalTitle{Bulletin of the
  Astronomical Society of India}, 28, 475

\bibitem[{{Katsuda} {et~al.}(2018{\natexlab{a}}){Katsuda}, {Takiwaki},
  {Tominaga}, {Moriya}, \& {Nakamura}}]{2018ApJ...863..127K}
{Katsuda}, S., {Takiwaki}, T., {Tominaga}, N., {Moriya}, T.~J., \& {Nakamura},
  K. 2018{\natexlab{a}},
  \href{http://dx.doi.org/10.3847/1538-4357/aad2d8}{\JournalTitle{\apj}, 863,
  127}

\bibitem[{{Katsuda} {et~al.}(2018{\natexlab{b}}){Katsuda}, {Morii}, {Janka},
  {Wongwathanarat}, {Nakamura}, {Kotake}, {Mori}, {M{\"u}ller}, {Takiwaki},
  {Tanaka}, {Tominaga}, \& {Tsunemi}}]{2018ApJ...856...18K}
{Katsuda}, S., {Morii}, M., {Janka}, H.-T., {et~al.} 2018{\natexlab{b}},
  \href{http://dx.doi.org/10.3847/1538-4357/aab092}{\JournalTitle{\apj}, 856,
  18}

\bibitem[{{Katsuragawa} {et~al.}(2018){Katsuragawa}, {Nakashima}, {Matsumura},
  {Tanaka}, {Uchida}, {Lee}, {Uchiyama}, {Arakawa}, \&
  {Takahashi}}]{2018PASJ...70..110K}
{Katsuragawa}, M., {Nakashima}, S., {Matsumura}, H., {et~al.} 2018,
  \href{http://dx.doi.org/10.1093/pasj/psy114}{\JournalTitle{\pasj}, 70, 110}

\bibitem[{{Kavanagh} {et~al.}(2019){Kavanagh}, {Sasaki}, {Breitschwerdt}, {de
  Avillez}, {Filipovic}, {Galvin}, {Haberl}, {Hatzidimitriou}, {Henze},
  {Plucinsky}, {Saeedi}, {Sokolovsky}, \& {Williams}}]{2019arXiv191012754K}
{Kavanagh}, P.~J., {Sasaki}, M., {Breitschwerdt}, D., {et~al.} 2019,
  \JournalTitle{arXiv e-prints}, arXiv:1910.12754

\bibitem[{{Kawasaki} {et~al.}(2002){Kawasaki}, {Ozaki}, {Nagase}, {Masai},
  {Ishida}, \& {Petre}}]{2002ApJ...572..897K}
{Kawasaki}, M.~T., {Ozaki}, M., {Nagase}, F., {et~al.} 2002,
  \href{http://dx.doi.org/10.1086/340383}{\JournalTitle{\apj}, 572, 897}

\bibitem[{{Kim} {et~al.}(2003){Kim}, {Staveley-Smith}, {Dopita}, {Sault},
  {Freeman}, {Lee}, \& {Chu}}]{2003ApJS..148..473K}
{Kim}, S., {Staveley-Smith}, L., {Dopita}, M.~A., {et~al.} 2003,
  \href{http://dx.doi.org/10.1086/376980}{\JournalTitle{\apjs}, 148, 473}

\bibitem[{{Kobayashi} {et~al.}(2011){Kobayashi}, {Karakas}, \&
  {Umeda}}]{2011MNRAS.414.3231K}
{Kobayashi}, C., {Karakas}, A.~I., \& {Umeda}, H. 2011,
  \href{http://dx.doi.org/10.1111/j.1365-2966.2011.18621.x}{\JournalTitle{\mnras},
  414, 3231}

\bibitem[{{Kobayashi} {et~al.}(2006){Kobayashi}, {Umeda}, {Nomoto}, {Tominaga},
  \& {Ohkubo}}]{2006ApJ...653.1145K}
{Kobayashi}, C., {Umeda}, H., {Nomoto}, K., {Tominaga}, N., \& {Ohkubo}, T.
  2006, \href{http://dx.doi.org/10.1086/508914}{\JournalTitle{\apj}, 653, 1145}

\bibitem[{{Korn} {et~al.}(2002){Korn}, {Keller}, {Kaufer}, {Langer},
  {Przybilla}, {Stahl}, \& {Wolf}}]{2002A&A...385..143K}
{Korn}, A.~J., {Keller}, S.~C., {Kaufer}, A., {et~al.} 2002,
  \href{http://dx.doi.org/10.1051/0004-6361:20020116}{\JournalTitle{\aap}, 385,
  143}

\bibitem[{{Krumholz} {et~al.}(2009){Krumholz}, {McKee}, \&
  {Tumlinson}}]{2009ApJ...699..850K}
{Krumholz}, M.~R., {McKee}, C.~F., \& {Tumlinson}, J. 2009,
  \href{http://dx.doi.org/10.1088/0004-637X/699/1/850}{\JournalTitle{\apj},
  699, 850}

\bibitem[{{Kumar} {et~al.}(2014){Kumar}, {Safi-Harb}, {Slane}, \&
  {Gotthelf}}]{2014ApJ...781...41K}
{Kumar}, H.~S., {Safi-Harb}, S., {Slane}, P.~O., \& {Gotthelf}, E.~V. 2014,
  \href{http://dx.doi.org/10.1088/0004-637X/781/1/41}{\JournalTitle{\apj}, 781,
  41}

\bibitem[{{Kuntz} \& {Snowden}(2001)}]{2001ApJ...554..684K}
{Kuntz}, K.~D., \& {Snowden}, S.~L. 2001,
  \href{http://dx.doi.org/10.1086/321421}{\JournalTitle{\apj}, 554, 684}

\bibitem[{{Kuntz} \& {Snowden}(2010)}]{2010ApJS..188...46K}
---. 2010,
  \href{http://dx.doi.org/10.1088/0067-0049/188/1/46}{\JournalTitle{\apjs},
  188, 46}

\bibitem[{{Laki{\'c}evi{\'c}} {et~al.}(2015){Laki{\'c}evi{\'c}}, {van Loon},
  {Meixner}, {Gordon}, {Bot}, {Roman-Duval}, {Babler}, {Bolatto},
  {Engelbracht}, {Filipovi{\'c}}, {Hony}, {Indebetouw}, {Misselt}, {Montiel},
  {Okumura}, {Panuzzo}, {Patat}, {Sauvage}, {Seale}, {Sonneborn}, {Temim},
  {Uro{\v{s}}evi{\'c}}, \& {Zanardo}}]{2015ApJ...799...50L}
{Laki{\'c}evi{\'c}}, M., {van Loon}, J.~T., {Meixner}, M., {et~al.} 2015,
  \href{http://dx.doi.org/10.1088/0004-637X/799/1/50}{\JournalTitle{\apj}, 799,
  50}

\bibitem[{{Landau} \& {Lifshitz}(1975)}]{1975ctf..book.....L}
{Landau}, L.~D., \& {Lifshitz}, E.~M. 1975, {The classical theory of fields}

\bibitem[{{Lasker}(1978)}]{1978ApJ...223..109L}
{Lasker}, B.~M. 1978,
  \href{http://dx.doi.org/10.1086/156241}{\JournalTitle{\apj}, 223, 109}

\bibitem[{{Lasker}(1980)}]{1980ApJ...237..765L}
---. 1980, \href{http://dx.doi.org/10.1086/157923}{\JournalTitle{\apj}, 237,
  765}

\bibitem[{{Law} {et~al.}(2020){Law}, {Milisavljevic}, {Patnaude}, {Plucinsky},
  {Gladders}, {Schmidt}, {Sravan}, {Banovetz}, {Sano}, {McGraw}, {Takahashi},
  \& {Orlando}}]{2020arXiv200400016L}
{Law}, C.~J., {Milisavljevic}, D., {Patnaude}, D.~J., {et~al.} 2020,
  \JournalTitle{arXiv e-prints}, arXiv:2004.00016

\bibitem[{{Leccardi} \& {Molendi}(2007)}]{2007A&A...472...21L}
{Leccardi}, A., \& {Molendi}, S. 2007,
  \href{http://dx.doi.org/10.1051/0004-6361:20077290}{\JournalTitle{\aap}, 472,
  21}

\bibitem[{{Lee} {et~al.}(2011){Lee}, {Kashyap}, {van Dyk}, {Connors}, {Drake},
  {Izem}, {Meng}, {Min}, {Park}, {Ratzlaff}, {Siemiginowska}, \&
  {Zezas}}]{2011ApJ...731..126L}
{Lee}, H., {Kashyap}, V.~L., {van Dyk}, D.~A., {et~al.} 2011,
  \href{http://dx.doi.org/10.1088/0004-637X/731/2/126}{\JournalTitle{\apj},
  731, 126}

\bibitem[{{Lee} {et~al.}(2012){Lee}, {Koo}, {Snell}, {Yun}, {Heyer}, \&
  {Burton}}]{2012ApJ...749...34L}
{Lee}, J.-J., {Koo}, B.-C., {Snell}, R.~L., {et~al.} 2012,
  \href{http://dx.doi.org/10.1088/0004-637X/749/1/34}{\JournalTitle{\apj}, 749,
  34}

\bibitem[{{Long} \& {Helfand}(1979)}]{1979ApJ...234L..77L}
{Long}, K.~S., \& {Helfand}, D.~J. 1979,
  \href{http://dx.doi.org/10.1086/183113}{\JournalTitle{\apj}, 234, L77}

\bibitem[{{Lopez} {et~al.}(2013){Lopez}, {Pearson}, {Ramirez-Ruiz}, {Castro},
  {Yamaguchi}, {Slane}, \& {Smith}}]{2013ApJ...777..145L}
{Lopez}, L.~A., {Pearson}, S., {Ramirez-Ruiz}, E., {et~al.} 2013,
  \href{http://dx.doi.org/10.1088/0004-637X/777/2/145}{\JournalTitle{\apj},
  777, 145}

\bibitem[{{Mac Low} {et~al.}(2005){Mac Low}, {Balsara}, {Kim}, \& {de
  Avillez}}]{2005ApJ...626..864M}
{Mac Low}, M.-M., {Balsara}, D.~S., {Kim}, J., \& {de Avillez}, M.~A. 2005,
  \href{http://dx.doi.org/10.1086/430122}{\JournalTitle{\apj}, 626, 864}

\bibitem[{{Maeder} \& {Meynet}(2000)}]{2000ARA&A..38..143M}
{Maeder}, A., \& {Meynet}, G. 2000,
  \href{http://dx.doi.org/10.1146/annurev.astro.38.1.143}{\JournalTitle{\araa},
  38, 143}

\bibitem[{{Maggi} {et~al.}(2016){Maggi}, {Haberl}, {Kavanagh}, {Sasaki},
  {Bozzetto}, {Filipovi{\'c}}, {Vasilopoulos}, {Pietsch}, {Points}, {Chu},
  {Dickel}, {Ehle}, {Williams}, \& {Greiner}}]{2016A&A...585A.162M}
{Maggi}, P., {Haberl}, F., {Kavanagh}, P.~J., {et~al.} 2016,
  \href{http://dx.doi.org/10.1051/0004-6361/201526932}{\JournalTitle{\aap},
  585, A162}

\bibitem[{{Maggi} {et~al.}(2019){Maggi}, {Filipovi{\'c}}, {Vukoti{\'c}},
  {Ballet}, {Haberl}, {Maitra}, {Kavanagh}, {Sasaki}, \&
  {Stupar}}]{2019A&A...631A.127M}
{Maggi}, P., {Filipovi{\'c}}, M.~D., {Vukoti{\'c}}, B., {et~al.} 2019,
  \href{http://dx.doi.org/10.1051/0004-6361/201936583}{\JournalTitle{\aap},
  631, A127}

\bibitem[{{Mahalanobis}(1936)}]{zbMATH03023295}
{Mahalanobis}, P.~C. 1936, \JournalTitle{{Proc. Natl. Inst. Sci. India}}, 2, 49

\bibitem[{{Masai}(1994)}]{1994ApJ...437..770M}
{Masai}, K. 1994, \href{http://dx.doi.org/10.1086/175037}{\JournalTitle{\apj},
  437, 770}

\bibitem[{{Matsumura} {et~al.}(2017){Matsumura}, {Tanaka}, {Uchida}, {Okon}, \&
  {Tsuru}}]{2017ApJ...851...73M}
{Matsumura}, H., {Tanaka}, T., {Uchida}, H., {Okon}, H., \& {Tsuru}, T.~G.
  2017, \href{http://dx.doi.org/10.3847/1538-4357/aa9bdf}{\JournalTitle{\apj},
  851, 73}

\bibitem[{{McCammon} {et~al.}(2002){McCammon}, {Almy}, {Apodaca}, {Bergmann
  Tiest}, {Cui}, {Deiker}, {Galeazzi}, {Juda}, {Lesser}, {Mihara},
  {Morgenthaler}, {Sanders}, {Zhang}, {Figueroa-Feliciano}, {Kelley},
  {Moseley}, {Mushotzky}, {Porter}, {Stahle}, \&
  {Szymkowiak}}]{2002ApJ...576..188M}
{McCammon}, D., {Almy}, R., {Apodaca}, E., {et~al.} 2002,
  \href{http://dx.doi.org/10.1086/341727}{\JournalTitle{\apj}, 576, 188}

\bibitem[{{McKee}(1974)}]{1974ApJ...188..335M}
{McKee}, C.~F. 1974,
  \href{http://dx.doi.org/10.1086/152721}{\JournalTitle{\apj}, 188, 335}

\bibitem[{{Mewe} \& {Gronenschild}(1981)}]{1981A&AS...45...11M}
{Mewe}, R., \& {Gronenschild}, E.~H.~B.~M. 1981, \JournalTitle{\aaps}, 45, 11

\bibitem[{{Milisavljevic} {et~al.}(2013){Milisavljevic}, {Margutti},
  {Soderberg}, {Pignata}, {Chomiuk}, {Fesen}, {Bufano}, {Sanders}, {Parrent},
  {Parker}, {Mazzali}, {Pian}, {Pickering}, {Buckley}, {Crawford}, {Gulbis},
  {Hettlage}, {Hooper}, {Nordsieck}, {O'Donoghue}, {Husser}, {Potter},
  {Kniazev}, {Kotze}, {Romero-Colmenero}, {Vaisanen}, {Wolf}, {Bietenholz},
  {Bartel}, {Fransson}, {Walker}, {Brunthaler}, {Chakraborti}, {Levesque},
  {MacFadyen}, {Drescher}, {Bock}, {Marples}, {Anderson}, {Benetti},
  {Reichart}, \& {Ivarsen}}]{2013ApJ...767...71M}
{Milisavljevic}, D., {Margutti}, R., {Soderberg}, A.~M., {et~al.} 2013,
  \href{http://dx.doi.org/10.1088/0004-637X/767/1/71}{\JournalTitle{\apj}, 767,
  71}

\bibitem[{{Morse} {et~al.}(1995){Morse}, {Winkler}, \&
  {Kirshner}}]{1995AJ....109.2104M}
{Morse}, J.~A., {Winkler}, P.~F., \& {Kirshner}, R.~P. 1995,
  \href{http://dx.doi.org/10.1086/117436}{\JournalTitle{\aj}, 109, 2104}

\bibitem[{{Morse} {et~al.}(1996){Morse}, {Blair}, {Dopita}, {Hughes},
  {Kirshner}, {Long}, {Raymond}, {Sutherland}, \&
  {Winkler}}]{1996AJ....112.2350M}
{Morse}, J.~A., {Blair}, W.~P., {Dopita}, M.~A., {et~al.} 1996,
  \href{http://dx.doi.org/10.1086/118189}{\JournalTitle{\aj}, 112, 2350}

\bibitem[{{Nomoto} {et~al.}(2013){Nomoto}, {Kobayashi}, \&
  {Tominaga}}]{2013ARA&A..51..457N}
{Nomoto}, K., {Kobayashi}, C., \& {Tominaga}, N. 2013,
  \href{http://dx.doi.org/10.1146/annurev-astro-082812-140956}{\JournalTitle{\araa},
  51, 457}

\bibitem[{{Nomoto} {et~al.}(2006){Nomoto}, {Tominaga}, {Umeda}, {Kobayashi}, \&
  {Maeda}}]{2006NuPhA.777..424N}
{Nomoto}, K., {Tominaga}, N., {Umeda}, H., {Kobayashi}, C., \& {Maeda}, K.
  2006,
  \href{http://dx.doi.org/10.1016/j.nuclphysa.2006.05.008}{\JournalTitle{Nuclear
  Physics A}, 777, 424}

\bibitem[{{Nousek} \& {Shue}(1989)}]{1989ApJ...342.1207N}
{Nousek}, J.~A., \& {Shue}, D.~R. 1989,
  \href{http://dx.doi.org/10.1086/167676}{\JournalTitle{\apj}, 342, 1207}

\bibitem[{{Okon} {et~al.}(2018){Okon}, {Uchida}, {Tanaka}, {Matsumura}, \&
  {Tsuru}}]{2018PASJ...70...35O}
{Okon}, H., {Uchida}, H., {Tanaka}, T., {Matsumura}, H., \& {Tsuru}, T.~G.
  2018, \href{http://dx.doi.org/10.1093/pasj/psy022}{\JournalTitle{\pasj}, 70,
  35}

\bibitem[{{Okon} {et~al.}(2020){Okon}, {Tanaka}, {Uchida}, {Yamaguchi},
  {Tsuru}, {Seta}, {Smith}, {Yoshiike}, {Orlando}, {Bocchino}, \&
  {Miceli}}]{2020ApJ...890...62O}
{Okon}, H., {Tanaka}, T., {Uchida}, H., {et~al.} 2020,
  \href{http://dx.doi.org/10.3847/1538-4357/ab6987}{\JournalTitle{\apj}, 890,
  62}

\bibitem[{Oliphant(2006)}]{oliphant2006guide}
Oliphant, T.~E. 2006, A guide to NumPy, Vol.~1 (Trelgol Publishing USA)

\bibitem[{{Ozawa} {et~al.}(2009){Ozawa}, {Koyama}, {Yamaguchi}, {Masai}, \&
  {Tamagawa}}]{2009ApJ...706L..71O}
{Ozawa}, M., {Koyama}, K., {Yamaguchi}, H., {Masai}, K., \& {Tamagawa}, T.
  2009,
  \href{http://dx.doi.org/10.1088/0004-637X/706/1/L71}{\JournalTitle{\apjl},
  706, L71}

\bibitem[{{Park} {et~al.}(2005){Park}, {Muno}, {Baganoff}, {Maeda}, {Morris},
  {Chartas}, {Sanwal}, {Burrows}, \& {Garmire}}]{2005ApJ...631..964P}
{Park}, S., {Muno}, M.~P., {Baganoff}, F.~K., {et~al.} 2005,
  \href{http://dx.doi.org/10.1086/432639}{\JournalTitle{\apj}, 631, 964}

\bibitem[{{Pastorello} {et~al.}(2008){Pastorello}, {Mattila}, {Zampieri},
  {Della Valle}, {Smartt}, {Valenti}, {Agnoletto}, {Benetti}, {Benn}, {Branch},
  {Cappellaro}, {Dennefeld}, {Eldridge}, {Gal-Yam}, {Harutyunyan}, {Hunter},
  {Kjeldsen}, {Lipkin}, {Mazzali}, {Milne}, {Navasardyan}, {Ofek}, {Pian},
  {Shemmer}, {Spiro}, {Stathakis}, {Taubenberger}, {Turatto}, \&
  {Yamaoka}}]{2008MNRAS.389..113P}
{Pastorello}, A., {Mattila}, S., {Zampieri}, L., {et~al.} 2008,
  \href{http://dx.doi.org/10.1111/j.1365-2966.2008.13602.x}{\JournalTitle{\mnras},
  389, 113}

\bibitem[{{Patnaude} \& {Fesen}(2014)}]{2014ApJ...789..138P}
{Patnaude}, D.~J., \& {Fesen}, R.~A. 2014,
  \href{http://dx.doi.org/10.1088/0004-637X/789/2/138}{\JournalTitle{\apj},
  789, 138}

\bibitem[{{Patnaude} {et~al.}(2015){Patnaude}, {Lee}, {Slane}, {Badenes},
  {Heger}, {Ellison}, \& {Nagataki}}]{2015ApJ...803..101P}
{Patnaude}, D.~J., {Lee}, S.-H., {Slane}, P.~O., {et~al.} 2015,
  \href{http://dx.doi.org/10.1088/0004-637X/803/2/101}{\JournalTitle{\apj},
  803, 101}

\bibitem[{{Patnaude} {et~al.}(2017){Patnaude}, {Lee}, {Slane}, {Badenes},
  {Nagataki}, {Ellison}, \& {Milisavljevic}}]{2017ApJ...849..109P}
---. 2017,
  \href{http://dx.doi.org/10.3847/1538-4357/aa9189}{\JournalTitle{\apj}, 849,
  109}

\bibitem[{{Pejcha} \& {Thompson}(2015)}]{2015ApJ...801...90P}
{Pejcha}, O., \& {Thompson}, T.~A. 2015,
  \href{http://dx.doi.org/10.1088/0004-637X/801/2/90}{\JournalTitle{\apj}, 801,
  90}

\bibitem[{{Pietrzy{\'n}ski} {et~al.}(2013){Pietrzy{\'n}ski}, {Graczyk},
  {Gieren}, {Thompson}, {Pilecki}, {Udalski}, {Soszy{\'n}ski}, {Koz{\l}owski},
  {Konorski}, {Suchomska}, {Bono}, {Moroni}, {Villanova}, {Nardetto},
  {Bresolin}, {Kudritzki}, {Storm}, {Gallenne}, {Smolec}, {Minniti}, {Kubiak},
  {Szyma{\'n}ski}, {Poleski}, {Wyrzykowski}, {Ulaczyk}, {Pietrukowicz},
  {G{\'o}rski}, \& {Karczmarek}}]{2013Natur.495...76P}
{Pietrzy{\'n}ski}, G., {Graczyk}, D., {Gieren}, W., {et~al.} 2013,
  \href{http://dx.doi.org/10.1038/nature11878}{\JournalTitle{\nat}, 495, 76}

\bibitem[{{Pietrzy{\'n}ski} {et~al.}(2019){Pietrzy{\'n}ski}, {Graczyk},
  {Gallenne}, {Gieren}, {Thompson}, {Pilecki}, {Karczmarek}, {G{\'o}rski},
  {Suchomska}, {Taormina}, {Zgirski}, {Wielg{\'o}rski}, {Ko{\l}aczkowski},
  {Konorski}, {Villanova}, {Nardetto}, {Kervella}, {Bresolin}, {Kudritzki},
  {Storm}, {Smolec}, \& {Narloch}}]{2019Natur.567..200P}
{Pietrzy{\'n}ski}, G., {Graczyk}, D., {Gallenne}, A., {et~al.} 2019,
  \href{http://dx.doi.org/10.1038/s41586-019-0999-4}{\JournalTitle{\nat}, 567,
  200}

\bibitem[{{Porquet} {et~al.}(2010){Porquet}, {Dubau}, \&
  {Grosso}}]{2010SSRv..157..103P}
{Porquet}, D., {Dubau}, J., \& {Grosso}, N. 2010,
  \href{http://dx.doi.org/10.1007/s11214-010-9731-2}{\JournalTitle{\ssr}, 157,
  103}

\bibitem[{{Rho} \& {Petre}(1998)}]{1998ApJ...503L.167R}
{Rho}, J., \& {Petre}, R. 1998,
  \href{http://dx.doi.org/10.1086/311538}{\JournalTitle{\apjl}, 503, L167}

\bibitem[{{Russell} \& {Dopita}(1992)}]{1992ApJ...384..508R}
{Russell}, S.~C., \& {Dopita}, M.~A. 1992,
  \href{http://dx.doi.org/10.1086/170893}{\JournalTitle{\apj}, 384, 508}

\bibitem[{{Sano}(2019)}]{2019asrc.confE.123S}
{Sano}, H. 2019, \href{http://dx.doi.org/10.5281/zenodo.3585447}{in ALMA2019:
  Science Results and Cross-Facility Synergies}, 123

\bibitem[{{Sano} {et~al.}(2015){Sano}, {Fukui}, {Yoshiike}, {Fukuda},
  {Tachihara}, {Inutsuka}, {Kawamura}, {Fujii}, {Mizuno}, {Inoue}, {Onishi},
  {Acero}, \& {Vink}}]{2015ASPC..499..257S}
{Sano}, H., {Fukui}, Y., {Yoshiike}, S., {et~al.} 2015, in Astronomical Society
  of the Pacific Conference Series, Vol. 499, Revolution in Astronomy with
  ALMA: The Third Year, ed. D.~{Iono}, K.~{Tatematsu}, A.~{Wootten}, \&
  L.~{Testi}, 257

\bibitem[{{Sano} {et~al.}(2019){Sano}, {Matsumura}, {Yamane}, {Maggi}, {Fujii},
  {Tsuge}, {Tokuda}, {Alsaberi}, {Filipovi{\'c}}, {Maxted}, {Rowell}, {Uchida},
  {Tanaka}, {Muraoka}, {Takekoshi}, {Onishi}, {Kawamura}, {Minamidani},
  {Mizuno}, {Yamamoto}, {Tachihara}, {Inoue}, {Inutsuka}, {Voisin}, {Tothill},
  {Sasaki}, {McClure-Griffiths}, \& {Fukui}}]{2019ApJ...881...85S}
{Sano}, H., {Matsumura}, H., {Yamane}, Y., {et~al.} 2019,
  \href{http://dx.doi.org/10.3847/1538-4357/ab2ade}{\JournalTitle{\apj}, 881,
  85}

\bibitem[{{Schenck} {et~al.}(2014){Schenck}, {Park}, {Burrows}, {Hughes},
  {Lee}, \& {Mori}}]{2014ApJ...791...50S}
{Schenck}, A., {Park}, S., {Burrows}, D.~N., {et~al.} 2014,
  \href{http://dx.doi.org/10.1088/0004-637X/791/1/50}{\JournalTitle{\apj}, 791,
  50}

\bibitem[{{Schenck} {et~al.}(2016){Schenck}, {Park}, \&
  {Post}}]{2016AJ....151..161S}
{Schenck}, A., {Park}, S., \& {Post}, S. 2016,
  \href{http://dx.doi.org/10.3847/0004-6256/151/6/161}{\JournalTitle{\aj}, 151,
  161}

\bibitem[{{Seok} {et~al.}(2013){Seok}, {Koo}, \& {Onaka}}]{2013ApJ...779..134S}
{Seok}, J.~Y., {Koo}, B.-C., \& {Onaka}, T. 2013,
  \href{http://dx.doi.org/10.1088/0004-637X/779/2/134}{\JournalTitle{\apj},
  779, 134}

\bibitem[{{Shimada} \& {Hoshino}(2000)}]{2000ApJ...543L..67S}
{Shimada}, N., \& {Hoshino}, M. 2000,
  \href{http://dx.doi.org/10.1086/318161}{\JournalTitle{\apjl}, 543, L67}

\bibitem[{{Shimizu} {et~al.}(2012){Shimizu}, {Masai}, \&
  {Koyama}}]{2012PASJ...64...24S}
{Shimizu}, T., {Masai}, K., \& {Koyama}, K. 2012,
  \href{http://dx.doi.org/10.1093/pasj/64.2.24}{\JournalTitle{\pasj}, 64, 24}

\bibitem[{{Slane} {et~al.}(2015){Slane}, {Bykov}, {Ellison}, {Dubner}, \&
  {Castro}}]{2015SSRv..188..187S}
{Slane}, P., {Bykov}, A., {Ellison}, D.~C., {Dubner}, G., \& {Castro}, D. 2015,
  \href{http://dx.doi.org/10.1007/s11214-014-0062-6}{\JournalTitle{\ssr}, 188,
  187}

\bibitem[{{Smith} \& {MCELS Team}(1999)}]{1999IAUS..190...28S}
{Smith}, R.~C., \& {MCELS Team}. 1999, in IAU Symposium, Vol. 190, New Views of
  the Magellanic Clouds, ed. Y.~H. {Chu}, N.~{Suntzeff}, J.~{Hesser}, \&
  D.~{Bohlender}, 28

\bibitem[{{Smith} {et~al.}(2004){Smith}, {Points}, {Aguilera}, {Leiton}, {Chu},
  {Winkler}, \& {MCELS}}]{2004AAS...20510108S}
{Smith}, R.~C., {Points}, S., {Aguilera}, C., {et~al.} 2004, in American
  Astronomical Society Meeting Abstracts, Vol. 205, American Astronomical
  Society Meeting Abstracts, 101.08

\bibitem[{{Snowden} {et~al.}(2004){Snowden}, {Valencic}, {Perry}, {Arida}, \&
  {Kuntz}}]{2004xmmg.rept.....S}
{Snowden}, S., {Valencic}, L., {Perry}, B., {Arida}, M., \& {Kuntz}, K.~D.
  2004, {The XMM-Newton ABC Guide: An Introduction to XMM-Newton Data
  Analysis}, Tech. rep.

\bibitem[{{Snowden} {et~al.}(1998){Snowden}, {Egger}, {Finkbeiner}, {Freyberg},
  \& {Plucinsky}}]{1998ApJ...493..715S}
{Snowden}, S.~L., {Egger}, R., {Finkbeiner}, D.~P., {Freyberg}, M.~J., \&
  {Plucinsky}, P.~P. 1998,
  \href{http://dx.doi.org/10.1086/305135}{\JournalTitle{\apj}, 493, 715}

\bibitem[{{Snowden} {et~al.}(2008){Snowden}, {Mushotzky}, {Kuntz}, \&
  {Davis}}]{2008A&A...478..615S}
{Snowden}, S.~L., {Mushotzky}, R.~F., {Kuntz}, K.~D., \& {Davis}, D.~S. 2008,
  \href{http://dx.doi.org/10.1051/0004-6361:20077930}{\JournalTitle{\aap}, 478,
  615}

\bibitem[{{Snowden} {et~al.}(1997){Snowden}, {Egger}, {Freyberg}, {McCammon},
  {Plucinsky}, {Sanders}, {Schmitt}, {Tr{\"u}mper}, \&
  {Voges}}]{1997ApJ...485..125S}
{Snowden}, S.~L., {Egger}, R., {Freyberg}, M.~J., {et~al.} 1997,
  \href{http://dx.doi.org/10.1086/304399}{\JournalTitle{\apj}, 485, 125}

\bibitem[{{Sukhbold} {et~al.}(2016){Sukhbold}, {Ertl}, {Woosley}, {Brown}, \&
  {Janka}}]{2016ApJ...821...38S}
{Sukhbold}, T., {Ertl}, T., {Woosley}, S.~E., {Brown}, J.~M., \& {Janka}, H.-T.
  2016,
  \href{http://dx.doi.org/10.3847/0004-637X/821/1/38}{\JournalTitle{\apj}, 821,
  38}

\bibitem[{{Sutherland} \& {Dopita}(1995)}]{1995ApJ...439..365S}
{Sutherland}, R.~S., \& {Dopita}, M.~A. 1995,
  \href{http://dx.doi.org/10.1086/175180}{\JournalTitle{\apj}, 439, 365}

\bibitem[{{Tang} {et~al.}(2016){Tang}, {Reynolds}, \&
  {Ressler}}]{2016ApJS..227...28T}
{Tang}, Z., {Reynolds}, S.~P., \& {Ressler}, S.~M. 2016,
  \href{http://dx.doi.org/10.3847/1538-4365/227/2/28}{\JournalTitle{\apjs},
  227, 28}

\bibitem[{{Tappe} {et~al.}(2012){Tappe}, {Rho}, {Boersma}, \&
  {Micelotta}}]{2012ApJ...754..132T}
{Tappe}, A., {Rho}, J., {Boersma}, C., \& {Micelotta}, E.~R. 2012,
  \href{http://dx.doi.org/10.1088/0004-637X/754/2/132}{\JournalTitle{\apj},
  754, 132}

\bibitem[{{Tappe} {et~al.}(2006){Tappe}, {Rho}, \&
  {Reach}}]{2006ApJ...653..267T}
{Tappe}, A., {Rho}, J., \& {Reach}, W.~T. 2006,
  \href{http://dx.doi.org/10.1086/508741}{\JournalTitle{\apj}, 653, 267}

\bibitem[{{Tenorio-Tagle} {et~al.}(1991){Tenorio-Tagle}, {Rozyczka}, {Franco},
  \& {Bodenheimer}}]{1991MNRAS.251..318T}
{Tenorio-Tagle}, G., {Rozyczka}, M., {Franco}, J., \& {Bodenheimer}, P. 1991,
  \href{http://dx.doi.org/10.1093/mnras/251.2.318}{\JournalTitle{\mnras}, 251,
  318}

\bibitem[{{Thielemann} {et~al.}(1992){Thielemann}, {Nomoto}, {Shigeyama},
  {Tsujimoto}, \& {Hashimoto}}]{1992eatc.conf...68T}
{Thielemann}, F.~K., {Nomoto}, K., {Shigeyama}, T., {Tsujimoto}, T., \&
  {Hashimoto}, M. 1992, in Elements and the Cosmos, ed. M.~G. {Edmunds} \&
  R.~{Terlevich}, 68

\bibitem[{{Tominaga} {et~al.}(2007){Tominaga}, {Umeda}, \&
  {Nomoto}}]{2007ApJ...660..516T}
{Tominaga}, N., {Umeda}, H., \& {Nomoto}, K. 2007,
  \href{http://dx.doi.org/10.1086/513063}{\JournalTitle{\apj}, 660, 516}

\bibitem[{{van Dyk} {et~al.}(2001){van Dyk}, {Connors}, {Kashyap}, \&
  {Siemiginowska}}]{2001ApJ...548..224V}
{van Dyk}, D.~A., {Connors}, A., {Kashyap}, V.~L., \& {Siemiginowska}, A. 2001,
  \href{http://dx.doi.org/10.1086/318656}{\JournalTitle{\apj}, 548, 224}

\bibitem[{{van Marle} \& {Keppens}(2012)}]{2012A&A...547A...3V}
{van Marle}, A.~J., \& {Keppens}, R. 2012,
  \href{http://dx.doi.org/10.1051/0004-6361/201218957}{\JournalTitle{\aap},
  547, A3}

\bibitem[{{Vink}(2012)}]{2012A&ARv..20...49V}
{Vink}, J. 2012,
  \href{http://dx.doi.org/10.1007/s00159-011-0049-1}{\JournalTitle{\aapr}, 20,
  49}

\bibitem[{{Vogt} \& {Dopita}(2011)}]{2011Ap&SS.331..521V}
{Vogt}, F., \& {Dopita}, M.~A. 2011,
  \href{http://dx.doi.org/10.1007/s10509-010-0479-7}{\JournalTitle{\apss}, 331,
  521}

\bibitem[{{Weisskopf} {et~al.}(2000){Weisskopf}, {Tananbaum}, {Van Speybroeck},
  \& {O'Dell}}]{2000SPIE.4012....2W}
{Weisskopf}, M.~C., {Tananbaum}, H.~D., {Van Speybroeck}, L.~P., \& {O'Dell},
  S.~L. 2000, Society of Photo-Optical Instrumentation Engineers (SPIE)
  Conference Series, Vol. 4012, {Chandra X-ray Observatory (CXO): overview},
  ed. J.~E. {Truemper} \& B.~{Aschenbach}, 2

\bibitem[{{Welty} {et~al.}(2016){Welty}, {Lauroesch}, {Wong}, \&
  {York}}]{2016ApJ...821..118W}
{Welty}, D.~E., {Lauroesch}, J.~T., {Wong}, T., \& {York}, D.~G. 2016,
  \href{http://dx.doi.org/10.3847/0004-637X/821/2/118}{\JournalTitle{\apj},
  821, 118}

\bibitem[{{Westerlund}(1997)}]{1997macl.book.....W}
{Westerlund}, B.~E. 1997, {The Magellanic Clouds}

\bibitem[{{Westerlund} \& {Mathewson}(1966)}]{1966MNRAS.131..371W}
{Westerlund}, B.~E., \& {Mathewson}, D.~S. 1966,
  \href{http://dx.doi.org/10.1093/mnras/131.3.371}{\JournalTitle{\mnras}, 131,
  371}

\bibitem[{{White} \& {Long}(1991)}]{1991ApJ...373..543W}
{White}, R.~L., \& {Long}, K.~S. 1991,
  \href{http://dx.doi.org/10.1086/170073}{\JournalTitle{\apj}, 373, 543}

\bibitem[{{Williams} {et~al.}(2006){Williams}, {Borkowski}, {Reynolds},
  {Blair}, {Ghavamian}, {Hendrick}, {Long}, {Points}, {Raymond}, {Sankrit},
  {Smith}, \& {Winkler}}]{2006ApJ...652L..33W}
{Williams}, B.~J., {Borkowski}, K.~J., {Reynolds}, S.~P., {et~al.} 2006,
  \href{http://dx.doi.org/10.1086/509876}{\JournalTitle{\apjl}, 652, L33}

\bibitem[{{Wilms} {et~al.}(2000){Wilms}, {Allen}, \&
  {McCray}}]{2000ApJ...542..914W}
{Wilms}, J., {Allen}, A., \& {McCray}, R. 2000,
  \href{http://dx.doi.org/10.1086/317016}{\JournalTitle{\apj}, 542, 914}

\bibitem[{{Winkler} \& {Petre}(2007)}]{2007ApJ...670..635W}
{Winkler}, P.~F., \& {Petre}, R. 2007,
  \href{http://dx.doi.org/10.1086/522101}{\JournalTitle{\apj}, 670, 635}

\bibitem[{{Wong} {et~al.}(2011){Wong}, {Hughes}, {Ott}, {Muller}, {Pineda},
  {Bernard}, {Chu}, {Fukui}, {Gruendl}, \& {Henkel}}]{2011ApJS..197...16W}
{Wong}, T., {Hughes}, A., {Ott}, J., {et~al.} 2011,
  \href{http://dx.doi.org/10.1088/0067-0049/197/2/16}{\JournalTitle{\apjs},
  197, 16}

\bibitem[{{Xiao} \& {Chen}(2008)}]{2008AdSpR..41..416X}
{Xiao}, X., \& {Chen}, Y. 2008,
  \href{http://dx.doi.org/10.1016/j.asr.2007.03.071}{\JournalTitle{Advances in
  Space Research}, 41, 416}

\bibitem[{{Yamaguchi} {et~al.}(2009){Yamaguchi}, {Ozawa}, {Koyama}, {Masai},
  {Hiraga}, {Ozaki}, \& {Yonetoku}}]{2009ApJ...705L...6Y}
{Yamaguchi}, H., {Ozawa}, M., {Koyama}, K., {et~al.} 2009,
  \href{http://dx.doi.org/10.1088/0004-637X/705/1/L6}{\JournalTitle{\apjl},
  705, L6}

\bibitem[{{Yamaguchi} {et~al.}(2014){Yamaguchi}, {Badenes}, {Petre}, {Nakano},
  {Castro}, {Enoto}, {Hiraga}, {Hughes}, {Maeda}, {Nobukawa}, {Safi-Harb},
  {Slane}, {Smith}, \& {Uchida}}]{2014ApJ...785L..27Y}
{Yamaguchi}, H., {Badenes}, C., {Petre}, R., {et~al.} 2014,
  \href{http://dx.doi.org/10.1088/2041-8205/785/2/L27}{\JournalTitle{\apjl},
  785, L27}

\bibitem[{{Yamaguchi} {et~al.}(2018){Yamaguchi}, {Tanaka}, {Wik}, {Rho},
  {Bamba}, {Castro}, {Smith}, {Foster}, {Uchida}, \&
  {Petre}}]{2018ApJ...868L..35Y}
{Yamaguchi}, H., {Tanaka}, T., {Wik}, D.~R., {et~al.} 2018,
  \href{http://dx.doi.org/10.3847/2041-8213/aaf055}{\JournalTitle{\apjl}, 868,
  L35}

\bibitem[{{Zhang} {et~al.}(2019){Zhang}, {Slavin}, {Foster}, {Smith}, {ZuHone},
  {Zhou}, \& {Chen}}]{2019ApJ...875...81Z}
{Zhang}, G.-Y., {Slavin}, J.~D., {Foster}, A., {et~al.} 2019,
  \href{http://dx.doi.org/10.3847/1538-4357/ab0f9a}{\JournalTitle{\apj}, 875,
  81}

\bibitem[{{Zhou} {et~al.}(2011){Zhou}, {Miceli}, {Bocchino}, {Orland o}, \&
  {Chen}}]{2011MNRAS.415..244Z}
{Zhou}, X., {Miceli}, M., {Bocchino}, F., {Orland o}, S., \& {Chen}, Y. 2011,
  \href{http://dx.doi.org/10.1111/j.1365-2966.2011.18695.x}{\JournalTitle{\mnras},
  415, 244}

\bibitem[{{Zhu} {et~al.}(2019){Zhu}, {Slane}, {Raymond}, \&
  {Tian}}]{2019ApJ...882..135Z}
{Zhu}, H., {Slane}, P., {Raymond}, J., \& {Tian}, W.~W. 2019,
  \href{http://dx.doi.org/10.3847/1538-4357/ab3226}{\JournalTitle{\apj}, 882,
  135}

\end{thebibliography}

\appendix
\counterwithin{figure}{section}

\section{Pileup Map and Narrow Band Images} 
\label{s:app_nbandandpileup}
\begin{figure}
\centering
\includegraphics[width=0.5\linewidth]{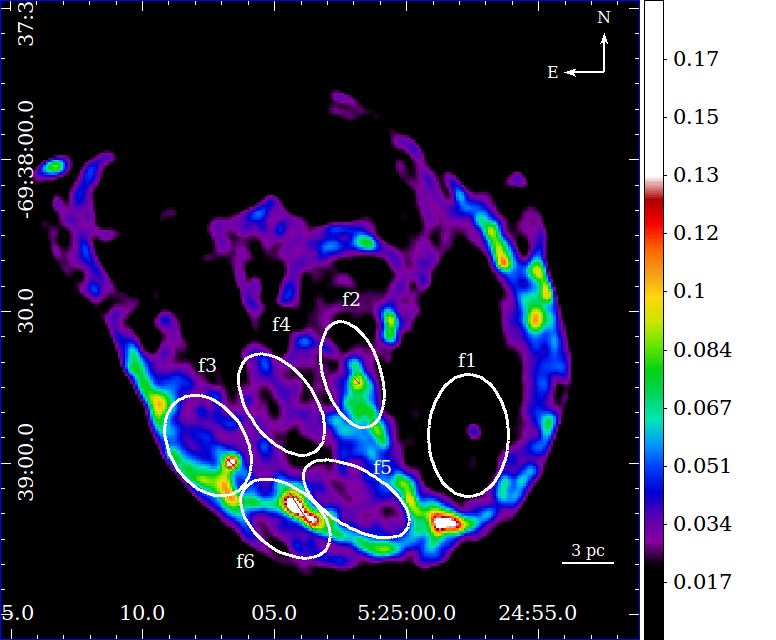}
\caption{Pileup map of SNR N132D from 2006 \chandra archival observations. The colorbar denotes the pileup fraction calculated in CIAO over all energies. The bright patches that are significantly affected by pileup in regions $\mathrm{f2,\,f3}$ and $\mathrm{f6}$ are excluded from the X-ray analysis.}
\label{fig:pileup}
\end{figure}

\begin{figure}
\centering
\includegraphics[width=0.3\linewidth]{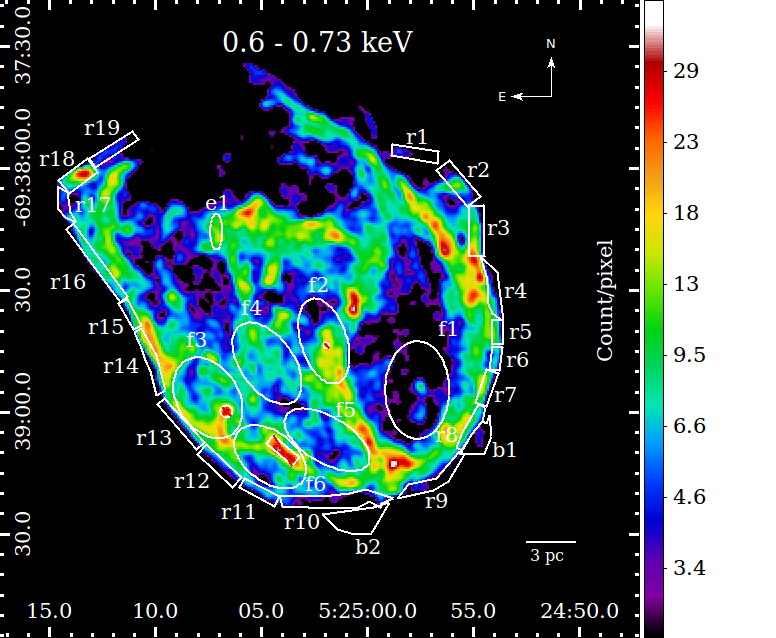}
\includegraphics[width=0.3\linewidth]{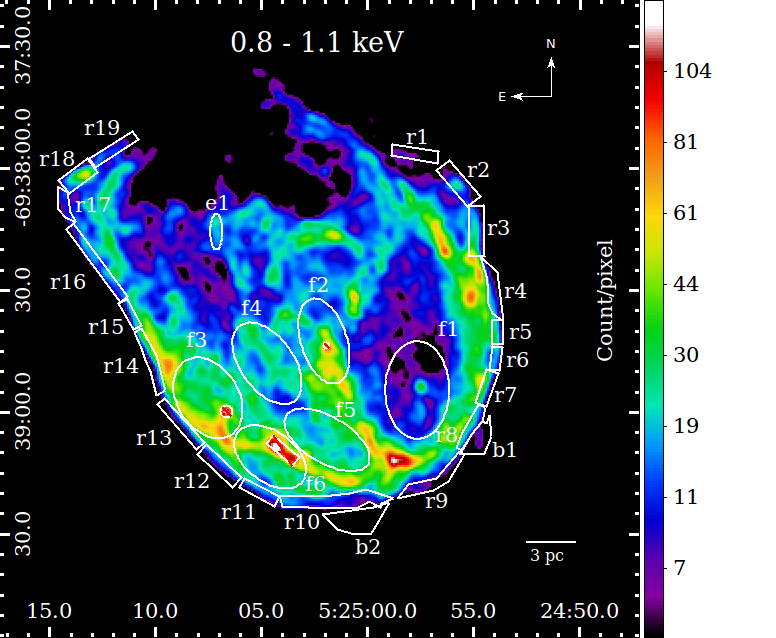}
\includegraphics[width=0.3\linewidth]{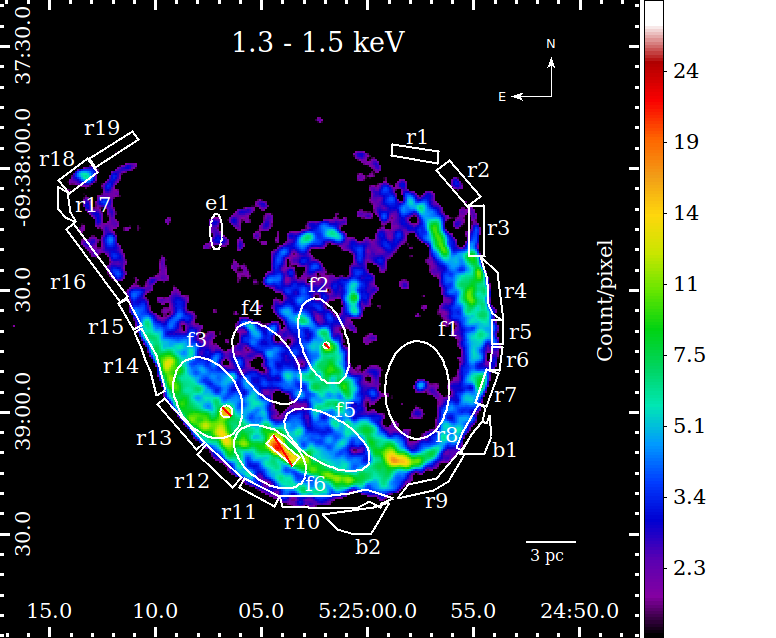}
\includegraphics[width=0.3\linewidth]{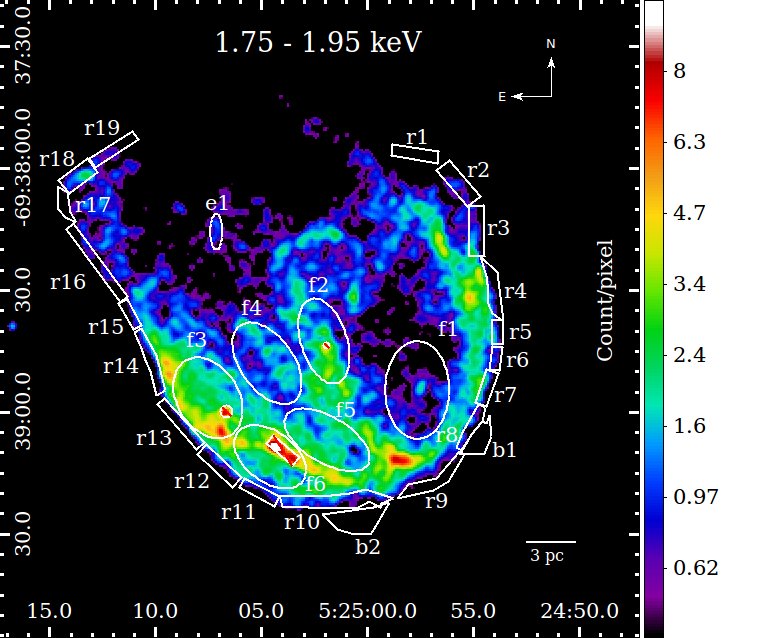}
\includegraphics[width=0.3\linewidth]{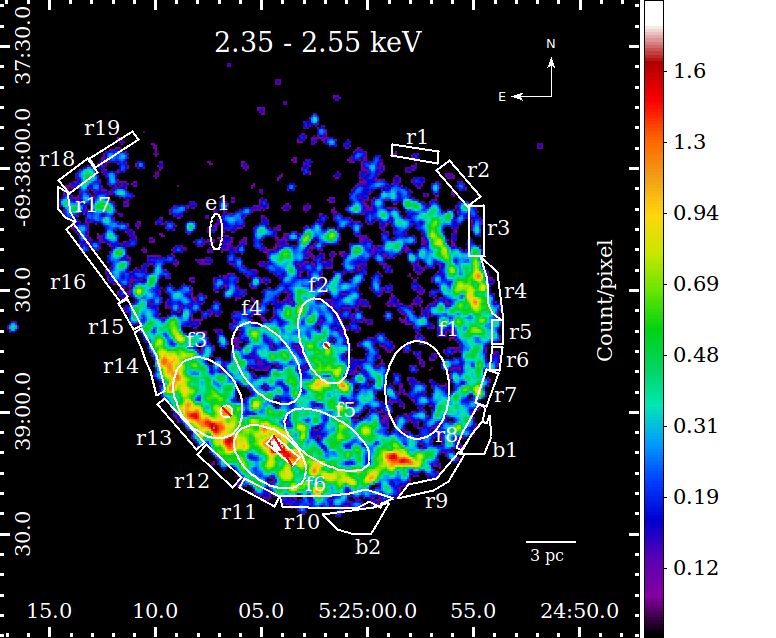}
\includegraphics[width=0.3\linewidth]{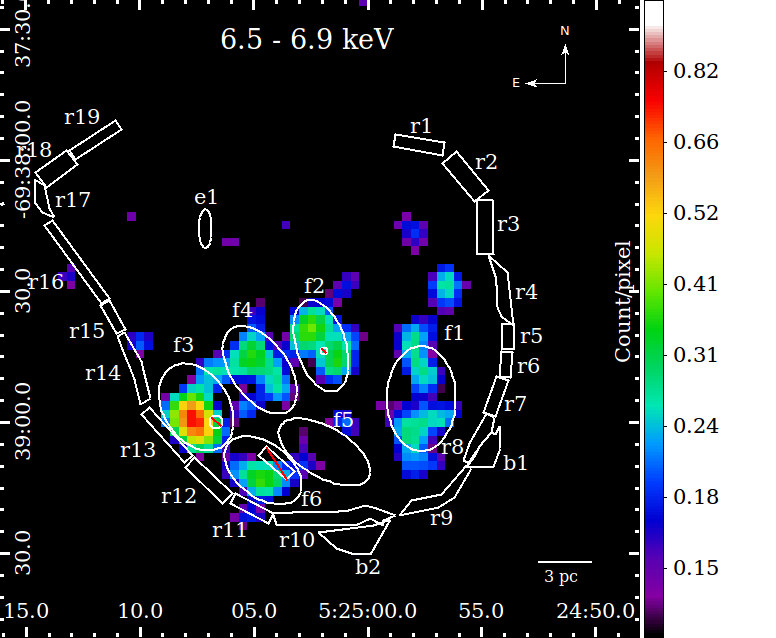}
\caption{Narrow band images of SNR N132D from \chandra 2006 archival observations, with bands covering (\textit{Top}, L to R) brightest line emissions of O, Ne, Mg, and (\textit{Bottom}, L to R) Si, S and Fe. All the images have been smoothed with a Gaussian kernel of radius 3, except for Fe where it has been binned by 4 and smoothed with a kernel of radius 3 owing to weak Fe K emission. Note that Fe also has numerous L-shell lines around $1\,\kev$.}
\label{fig:narrowbands}
\end{figure}

Figure \ref{fig:pileup} shows the pileup fraction in the remnant. Note that CIAO estimates the pileup in an image without filtering over energy. Figure \ref{fig:narrowbands} shows the narrow band images around various bright line emissions of O, Ne, Mg, Si, S and Fe in the remnant.

\section{Error and Scatter on the Mean Using Multiple Imputations}
\label{s:append_errors}
We use the method of Multiple Imputations \citep{2011ApJ...731..126L} to get an estimate of 68 per cent confidence intervals (similar to $1\sigma$ in the Gaussian case) for the mean values calculated for the blast wave spectral parameters across the rim of the remnant (viz., Table \ref{tab:scatter}). This method incorporates systematic uncertainties caused by scatter amongst the best-fit values in different regions along with standard statistical uncertainties in the estimates of each parameter. Thus, it is a better descriptor of the scatter present in the samples. Further, it also lets us quantify the systematic variations of a parameter around the rim. We estimate the combined statistical and systematic uncertainty by computing the weighted average \citep[see Section 3.1.2 of][]{2011ApJ...731..126L} of the so-called between variance ($B$; the variance of the best-fit values and a measure of the systematic scatter present in the data) and the within variance ($W$; the average of the individual variances in each measurement, and a measure of the statistical quality of the data) as,
\begin{equation}
V = \mathrm{W + \bigg(1+\frac{1}{M}\bigg)B}\,,
\label{eq:Appendixeq_1}
\end{equation}
where $\mathrm{M}$ are the number of regions, and $\sqrt{V}$ represents the width of a $t_\nu$-distribution with $\nu$ degrees of freedom,
 \begin{equation}
\nu =  \mathrm{(M-1)\Bigg(1+\frac{M{\cdot}W}{(M+1)B}\Bigg)^2}\,.
\label{eq:Appendixeq_2}
\end{equation}
The $t_\nu$-distribution has inherently heavier tails than the Gaussian distribution, but closely approximates the width of the latter for large $\nu$ ($\ga 7$). We compute a correction factor $C_\nu$ to map the 84$^{th}$-percentile quantile of the $t_\nu$-distribution to $\sqrt{V}$, and define a 1$\sigma$-equivalent error bar
\begin{equation}
\sigma_\mu = C_\nu \times \sqrt{V} \,.
\label{eq:Appendixeq_3}
\end{equation}
We find $\nu\in\{20,79\}$ for the various parameters considered. The correction factor $C_\nu{\rightarrow}1$ as $\nu{\rightarrow}\infty$, and is $\approx\,$2 per cent for $\nu=20$. These 1$\sigma$-equivalent error bars are reported in Table \ref{tab:scatter}.

The separation of the statistical (W) and systematic (B) variances also allows us to explore when systematic variations are large compared to the accuracy with which the parameters are measured. Large values of the scatter, $\frac{\sqrt{\mathrm{B}}}{\sqrt{\mathrm{W}}}$, shows where systematic variations overwhelm the statistical error (see Table \ref{tab:scatter}). When scatter $>\,1$, it signifies that there is more systematic than statistical uncertainty in the parameter value. We consider all abundance samples where this threshold is exceeded as showing localized enhancements.

\section{Spectral Fits}
\label{s:app_rimfits}
\begin{figure}
\includegraphics[width=0.3\linewidth, angle=270]{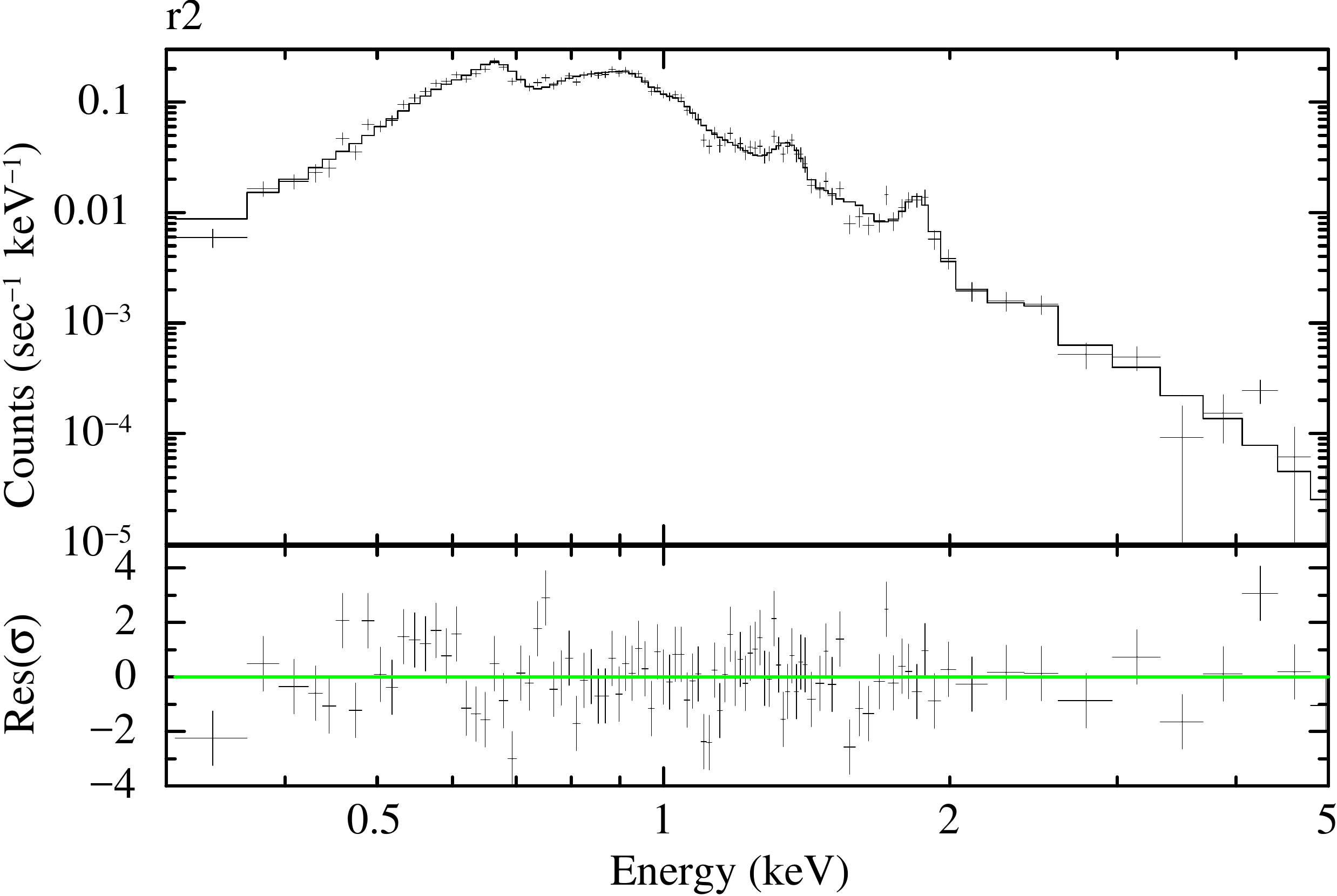}
\includegraphics[width=0.3\linewidth, angle=270]{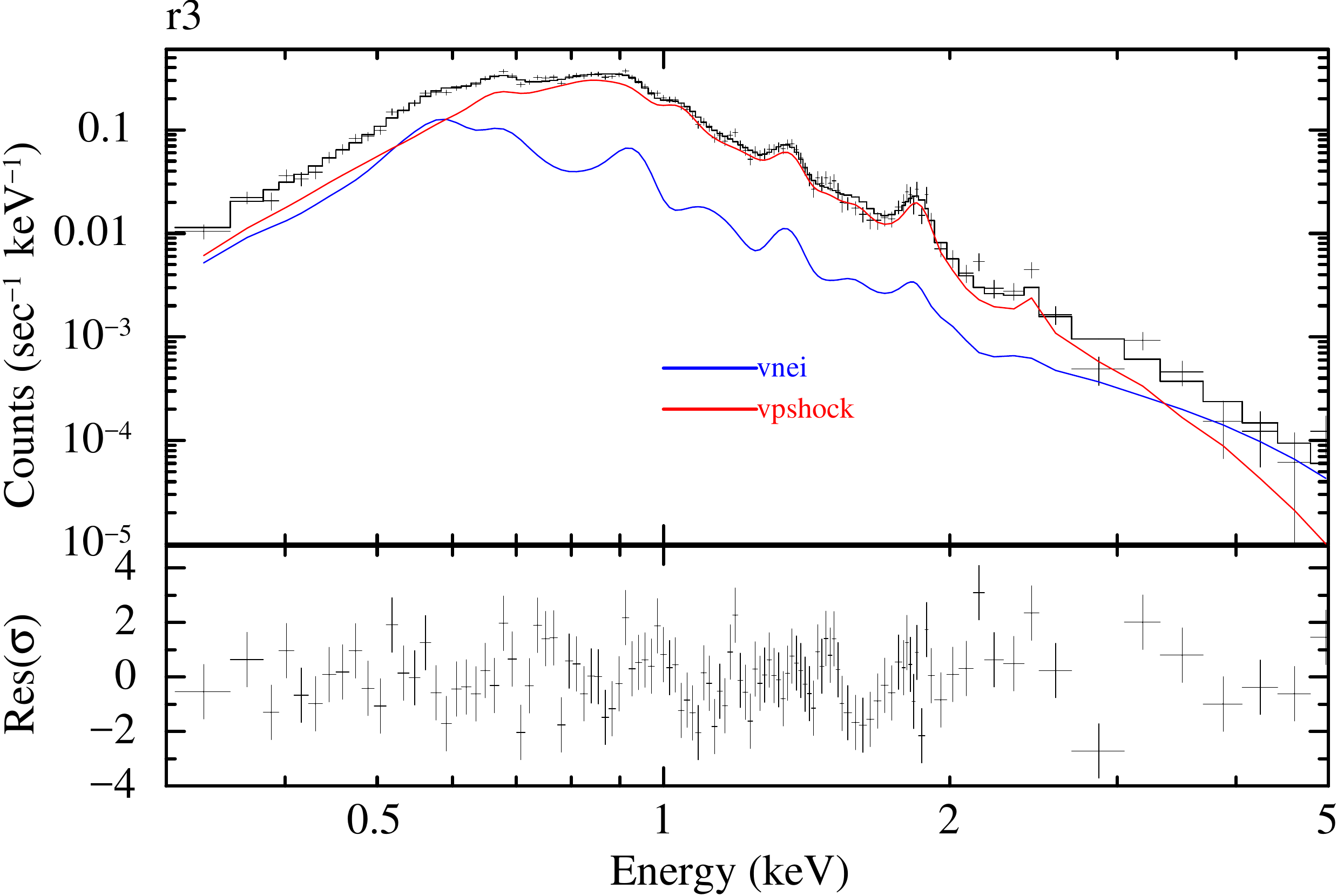}
\caption{Fits for regions $\mathrm{r2}$ and $\mathrm{r3}$ with the single \texttt{vpshock} and \texttt{vnei+vpshock} models, respectively.}
\label{fig:rims23}
\end{figure}

\begin{figure}
\includegraphics[width=0.3\linewidth, angle=270]{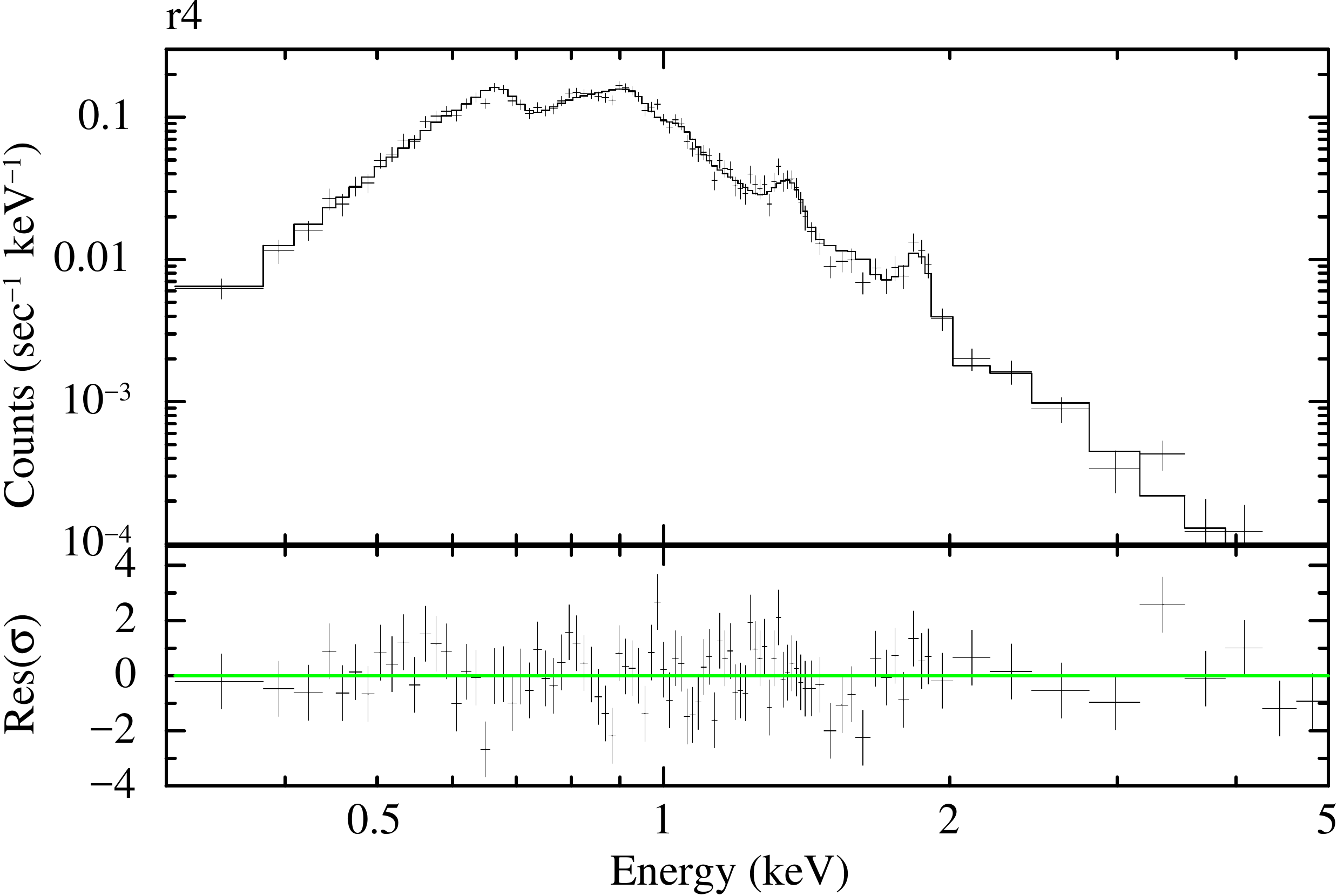}
\includegraphics[width=0.3\linewidth, angle=270]{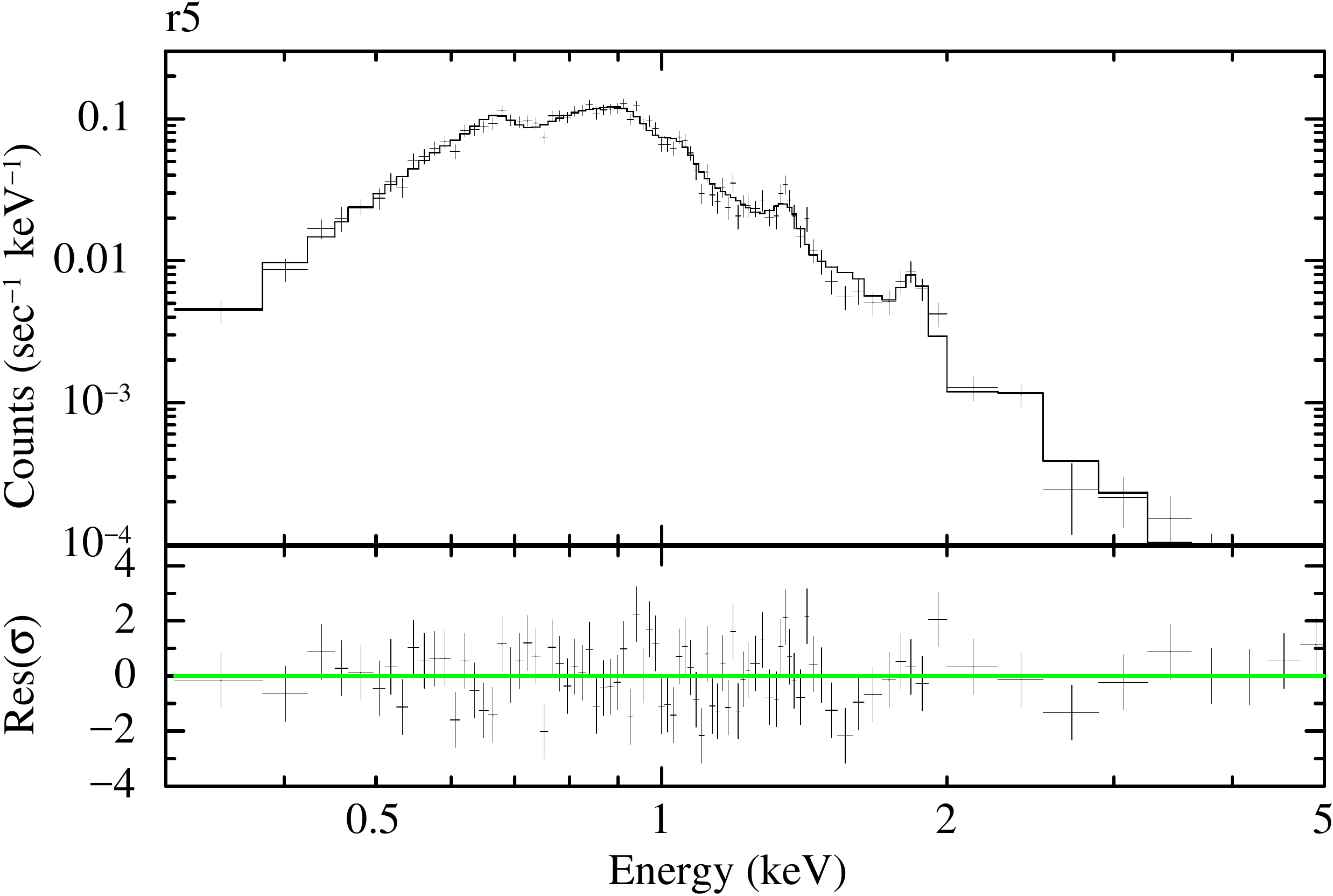}
\caption{Fits for regions $\mathrm{r4}$ and $\mathrm{r5}$ with the single \texttt{vpshock}.}
\label{fig:rims45}
\end{figure}

\begin{figure}
\includegraphics[width=0.3\linewidth, angle=270]{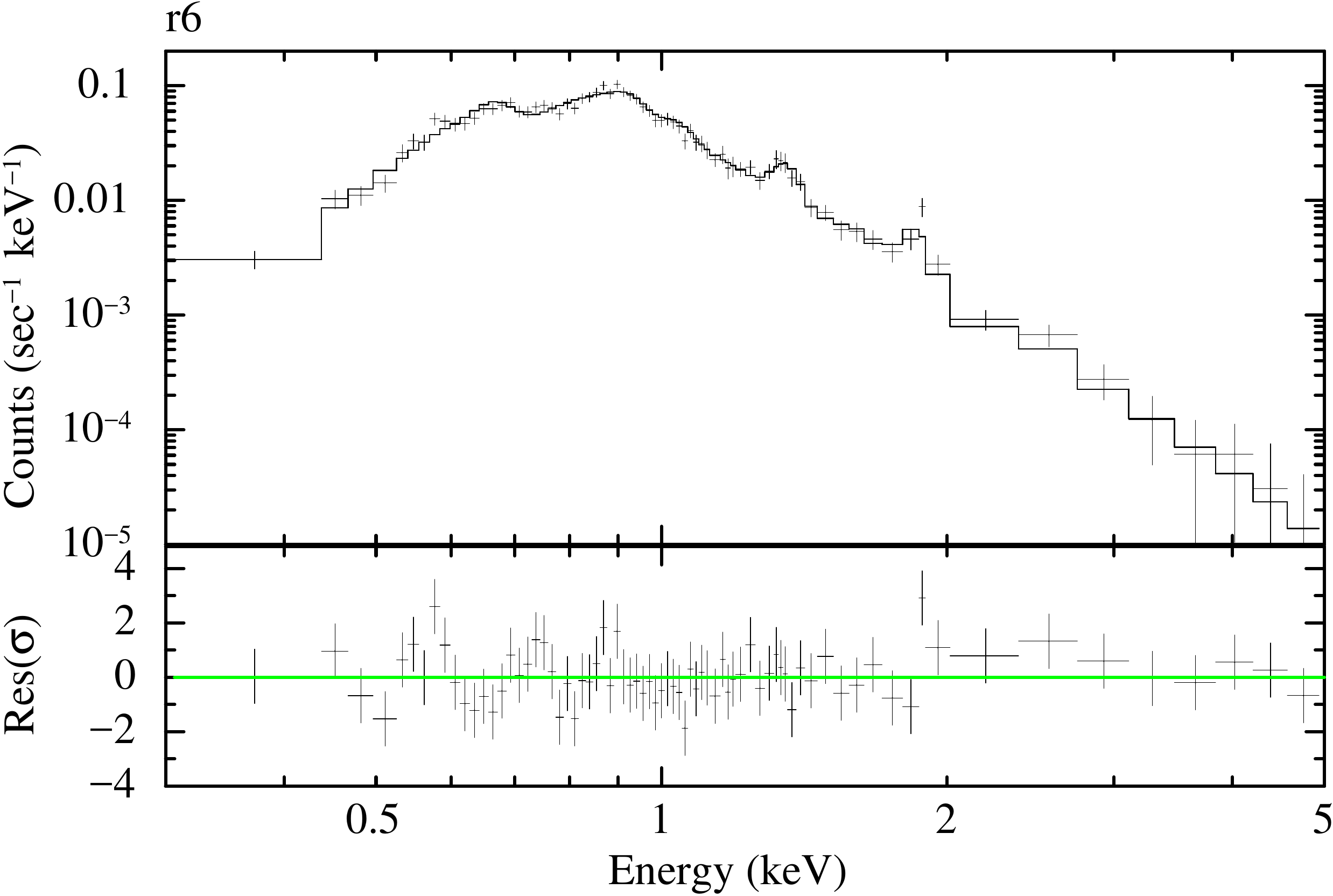}
\includegraphics[width=0.3\linewidth, angle=270]{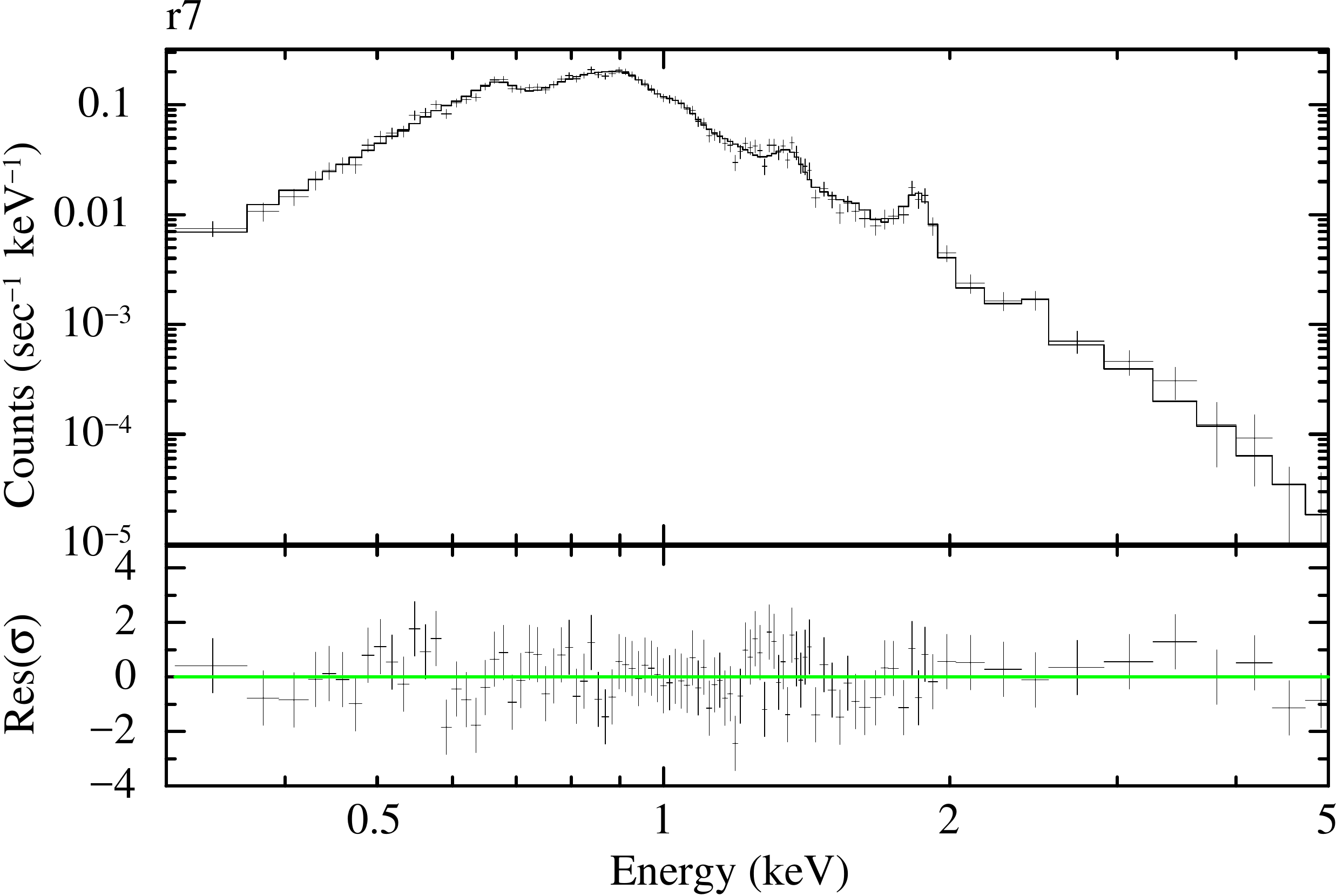}
\caption{Fits for regions $\mathrm{r6}$ and $\mathrm{r7}$ with the single \texttt{vpshock}.}
\label{fig:rims67}
\end{figure}

\begin{figure}
\includegraphics[width=0.3\linewidth, angle=270]{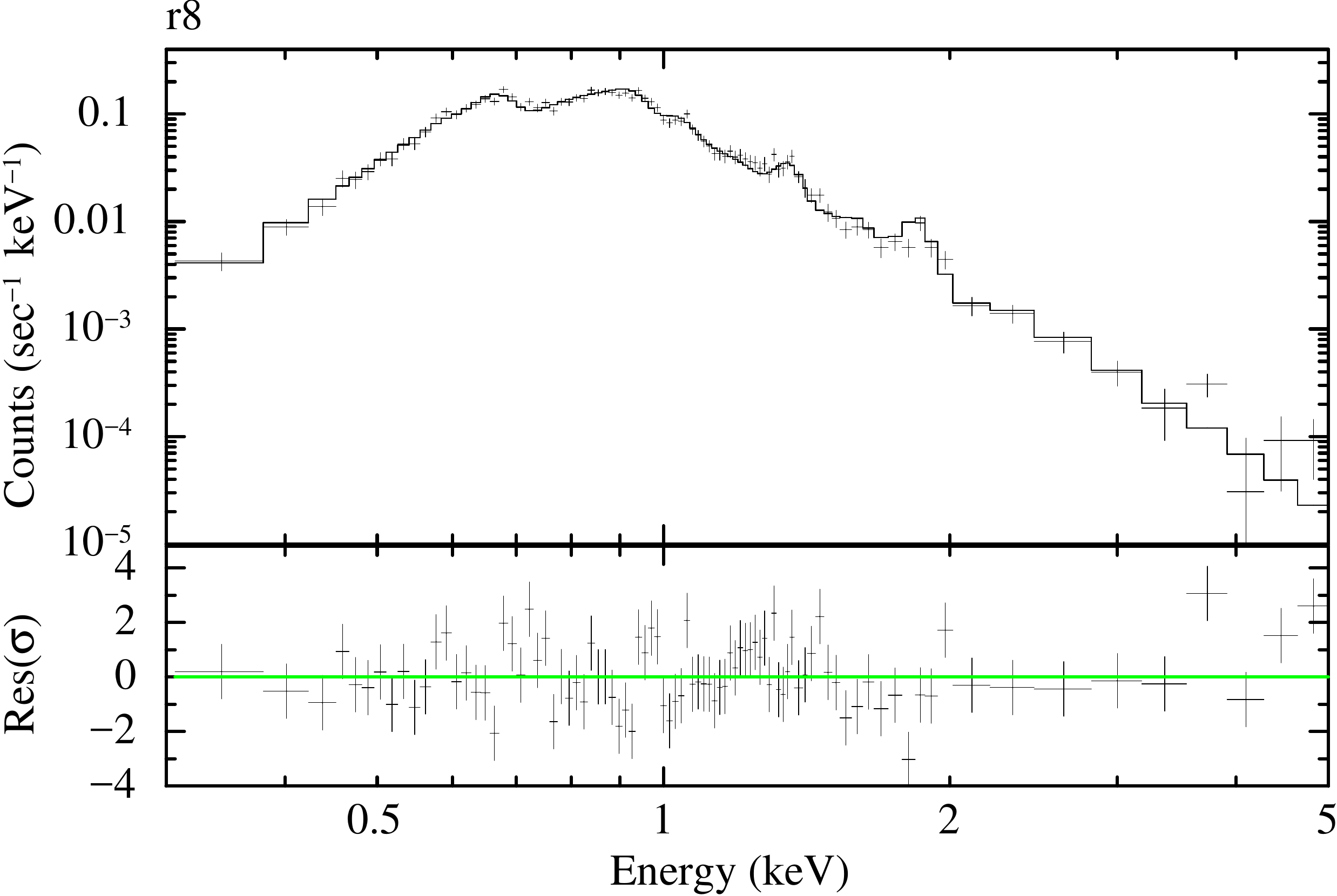}
\includegraphics[width=0.3\linewidth, angle=270]{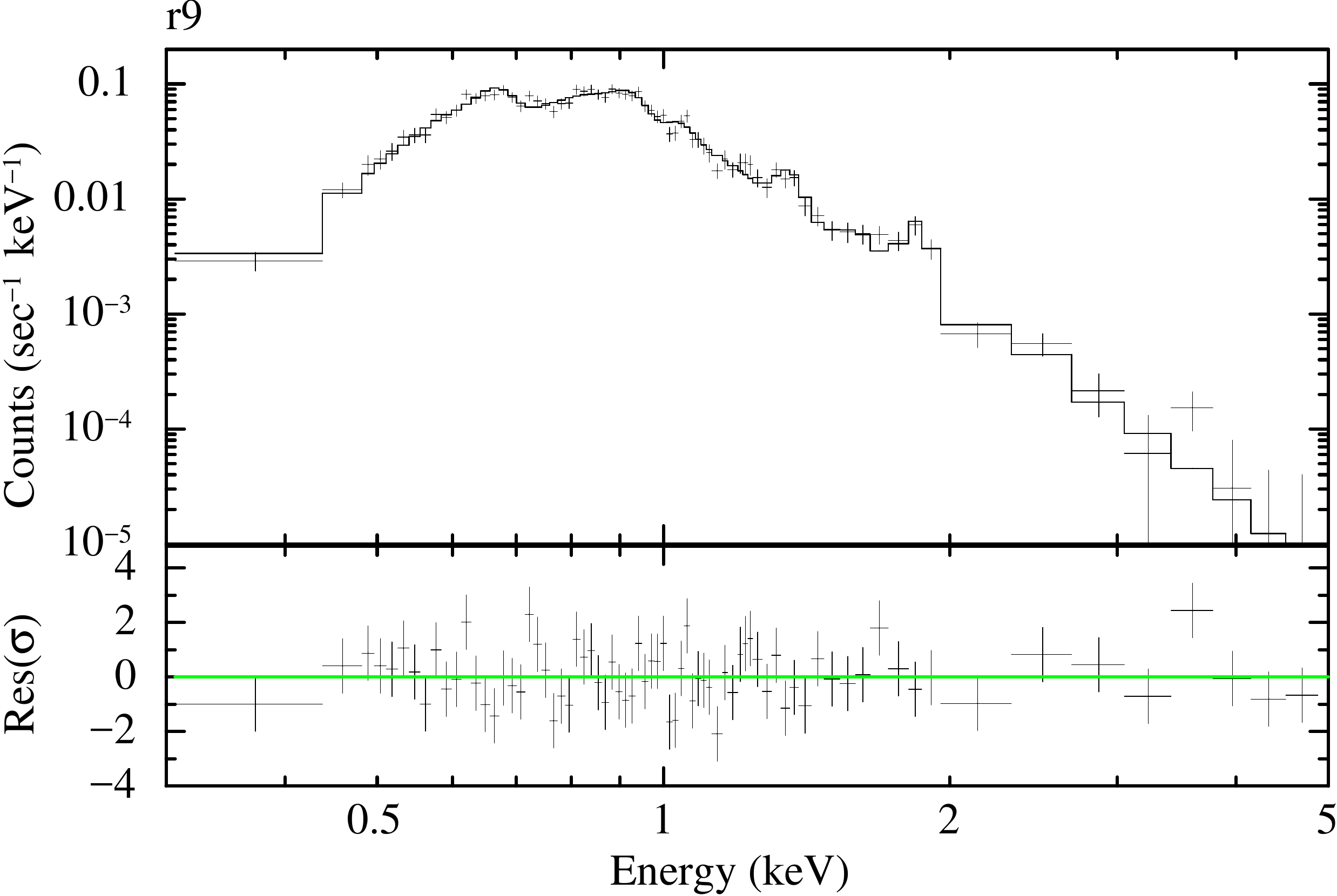}
\caption{Fits for regions $\mathrm{r8}$ and $\mathrm{r9}$ with the single \texttt{vpshock}.}
\label{fig:rims89}
\end{figure}

\begin{figure}
\includegraphics[width=0.3\linewidth, angle=270]{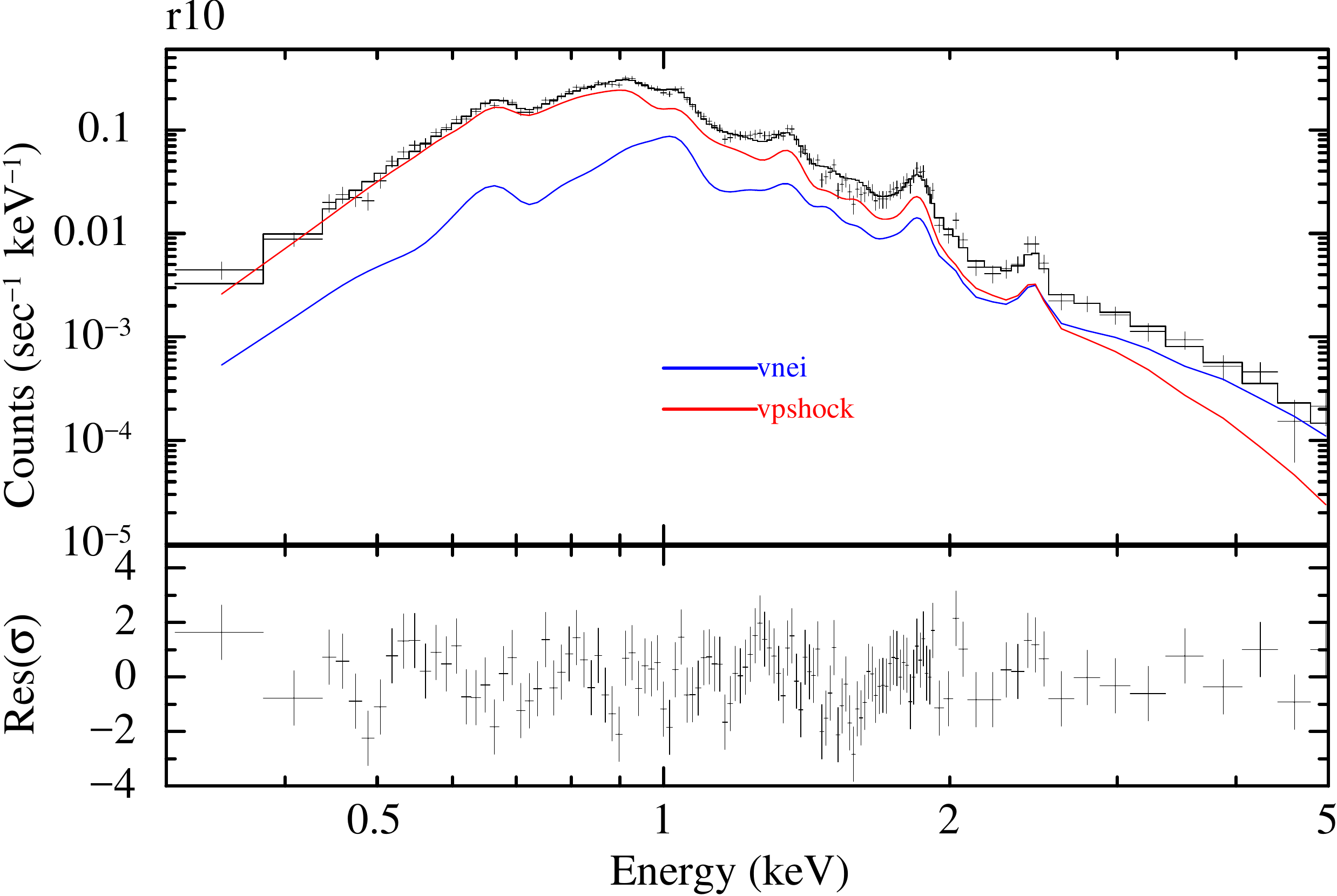}
\includegraphics[width=0.3\linewidth, angle=270]{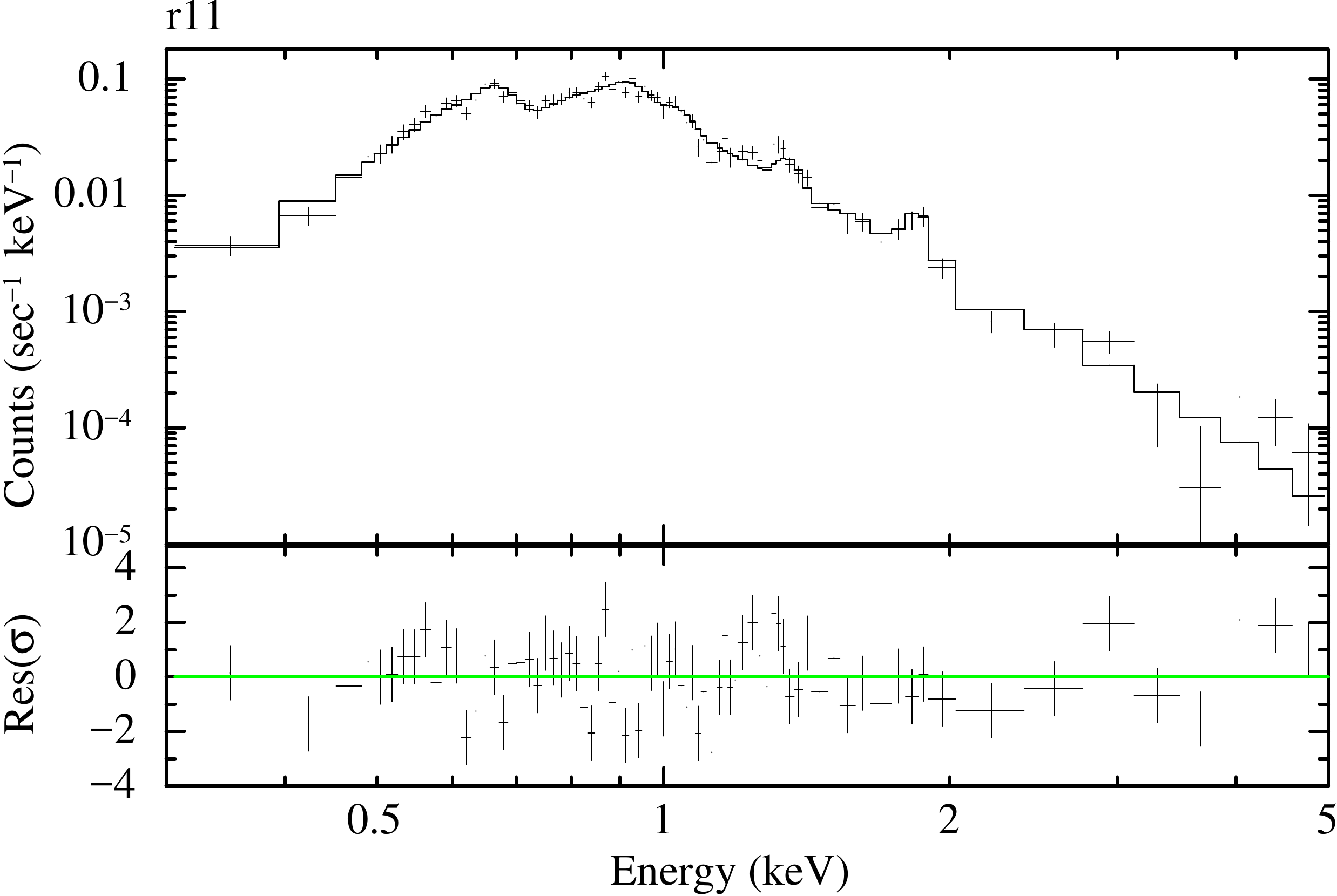}
\caption{Fits for regions $\mathrm{r10}$ and $\mathrm{r11}$ with the \texttt{vnei+vpshock} and the single \texttt{vpshock} models, respectively.}
\label{fig:rims1011}
\end{figure}

\begin{figure}
\includegraphics[width=0.3\linewidth, angle=270]{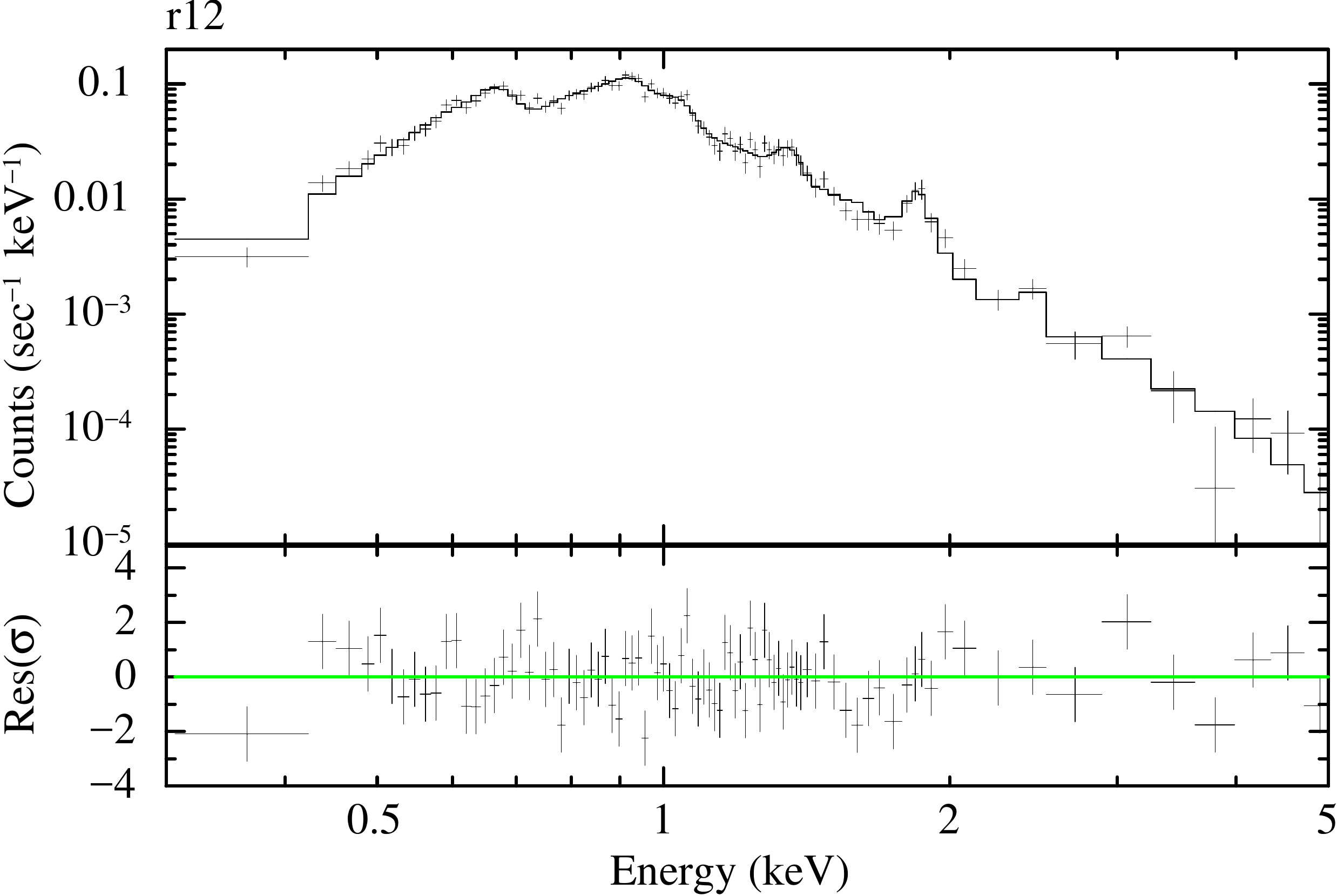}
\includegraphics[width=0.3\linewidth, angle=270]{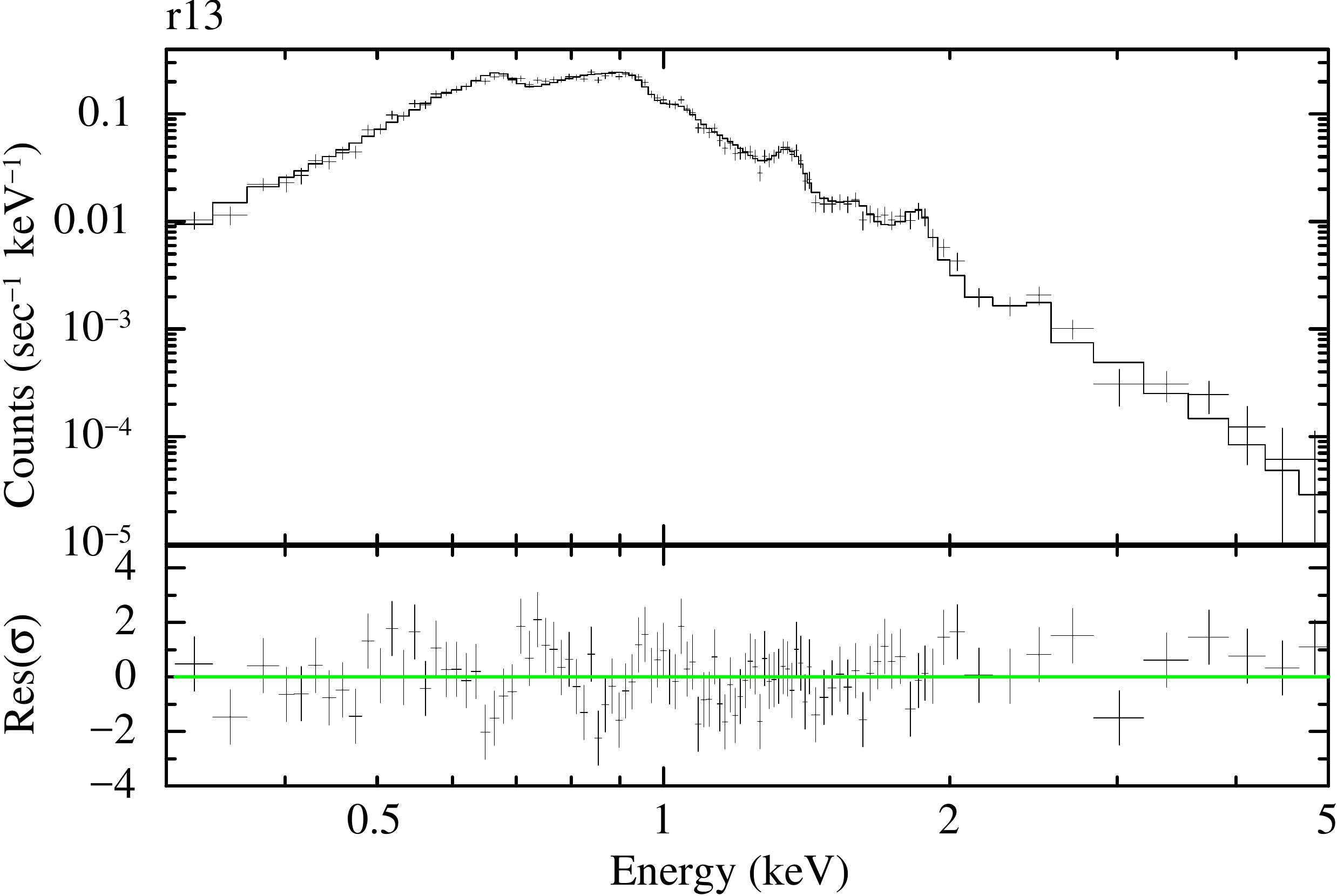}
\caption{Fits for regions $\mathrm{r12}$ and $\mathrm{r13}$ with the single \texttt{vpshock}.}
\label{fig:rims1213}
\end{figure}

\begin{figure}
\includegraphics[width=0.3\linewidth, angle=270]{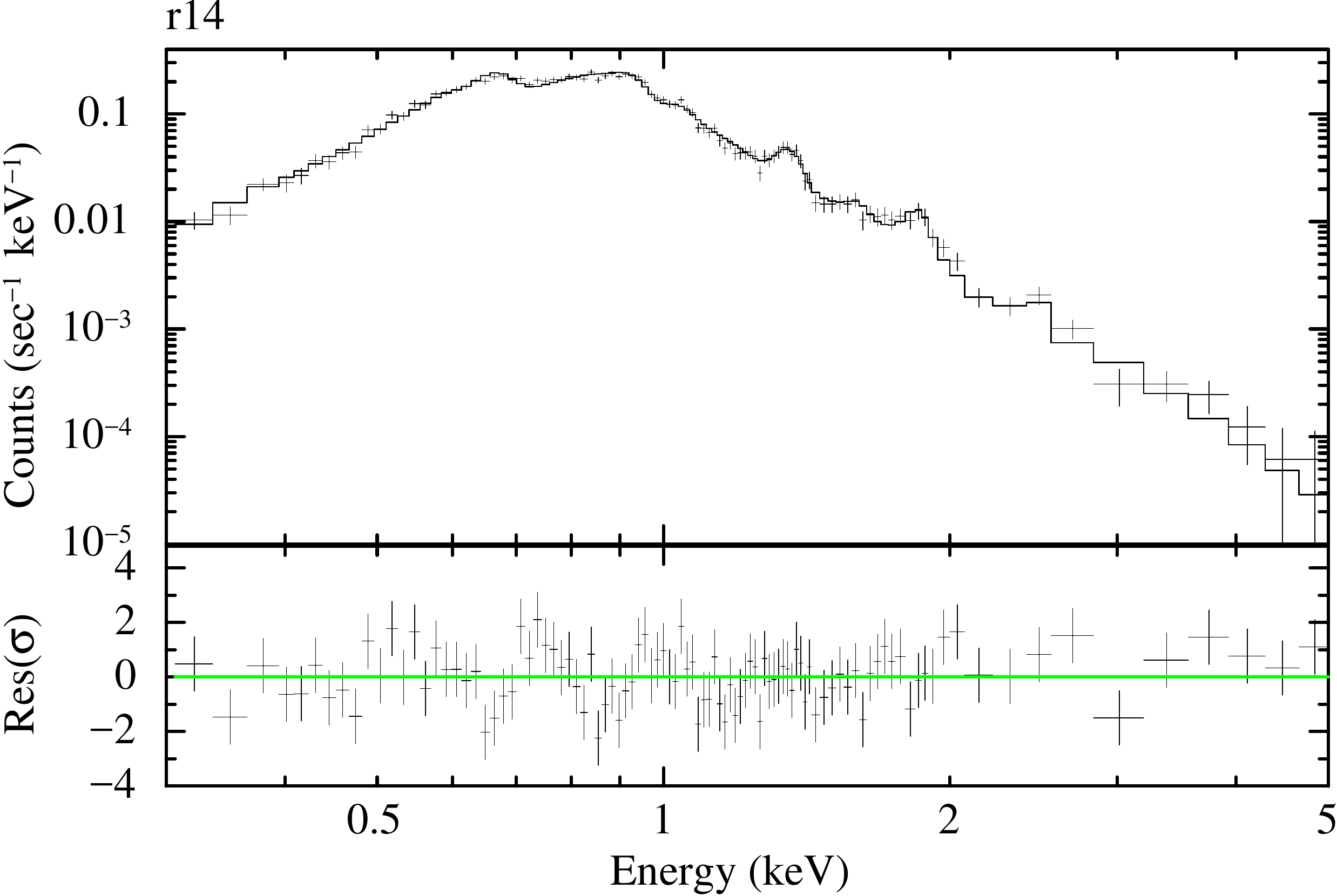}
\includegraphics[width=0.3\linewidth, angle=270]{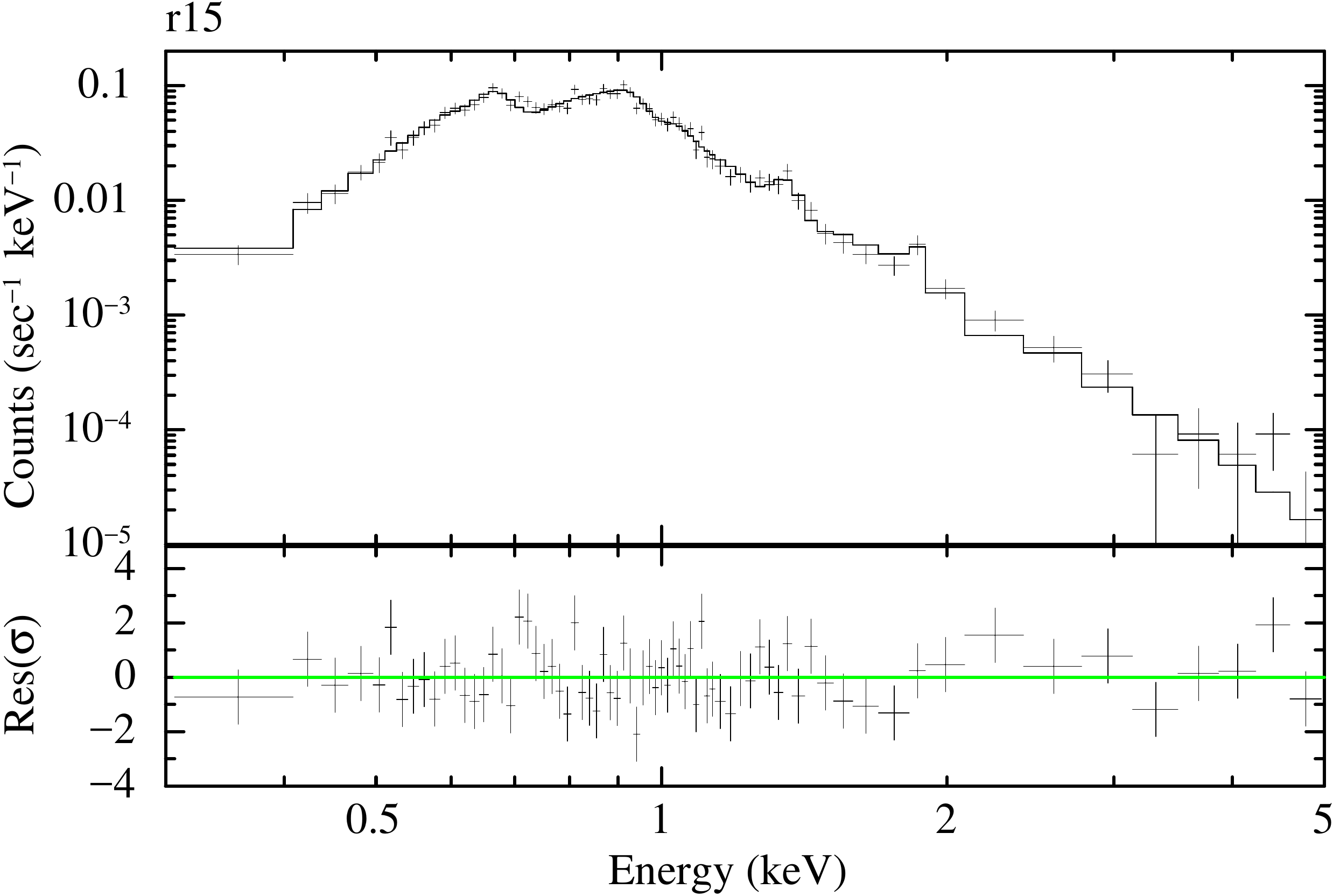}
\caption{Fits for regions $\mathrm{r14}$ and $\mathrm{r15}$ with the single \texttt{vpshock}.}
\label{fig:rims1415}
\end{figure}

\begin{figure}
\includegraphics[width=0.3\linewidth, angle=270]{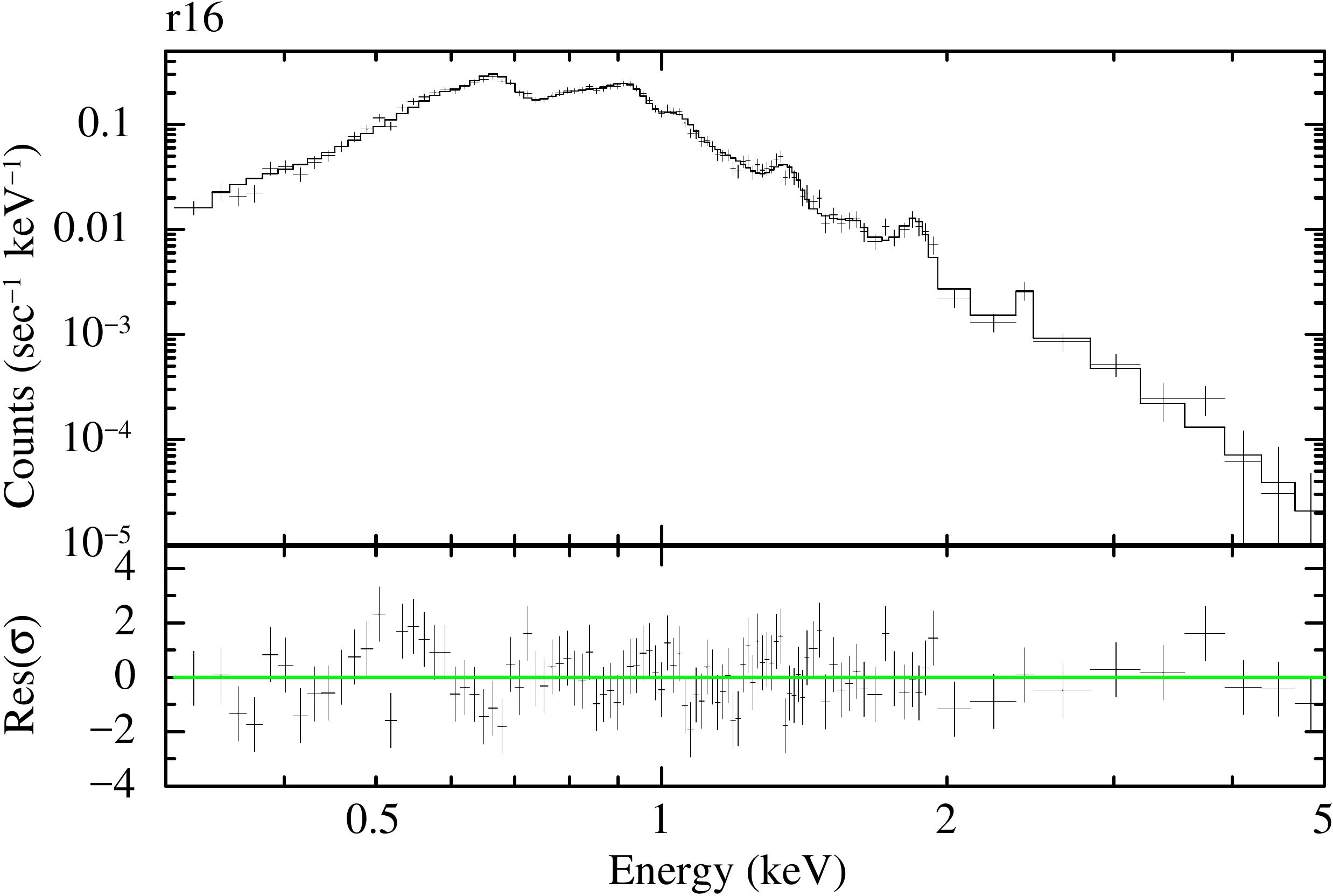}
\includegraphics[width=0.3\linewidth, angle=270]{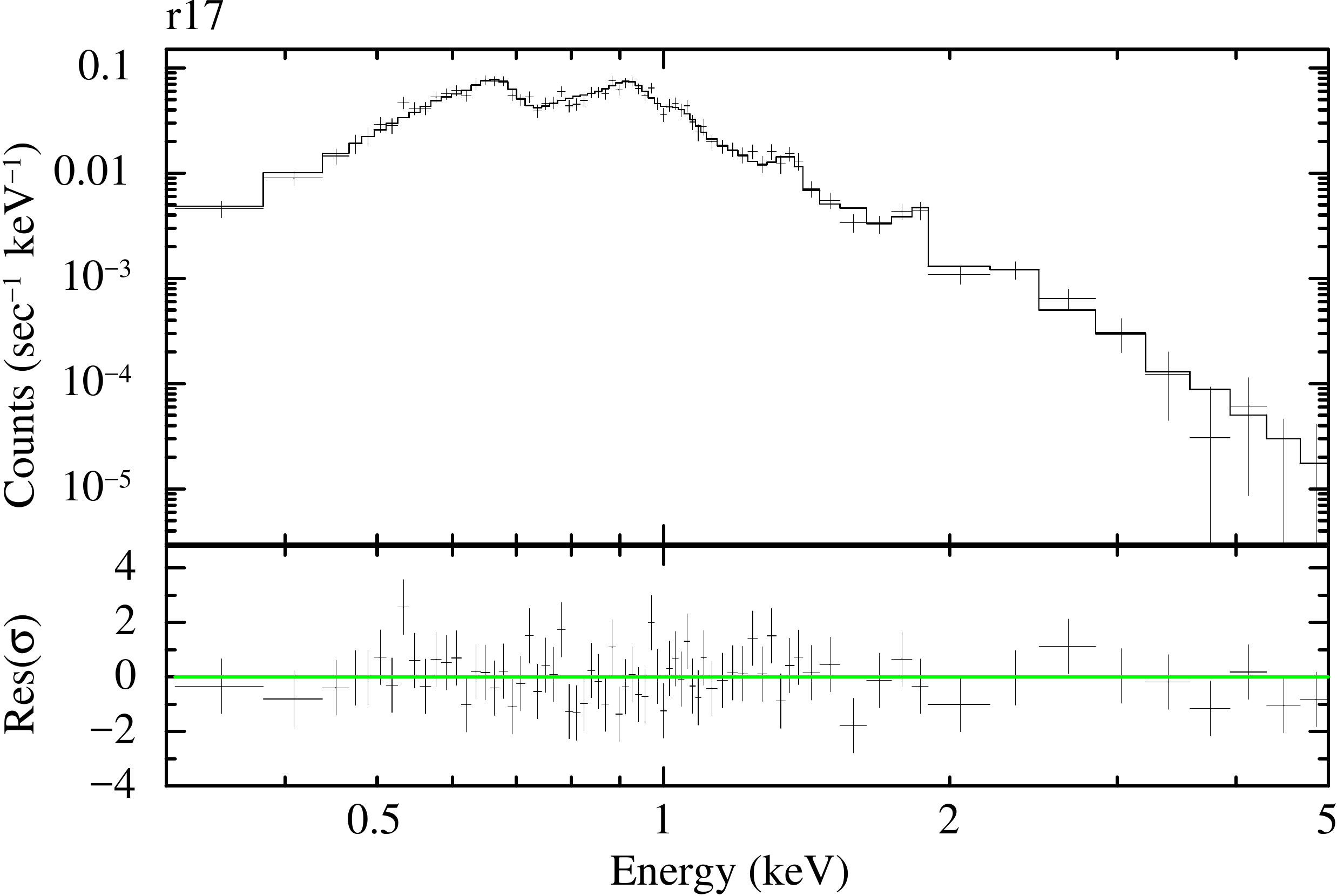}
\caption{Fits for regions $\mathrm{r16}$ and $\mathrm{r17}$ with the single \texttt{vpshock}.}
\label{fig:rims1617}
\end{figure}

\begin{figure}
\includegraphics[width=0.3\linewidth, angle=270]{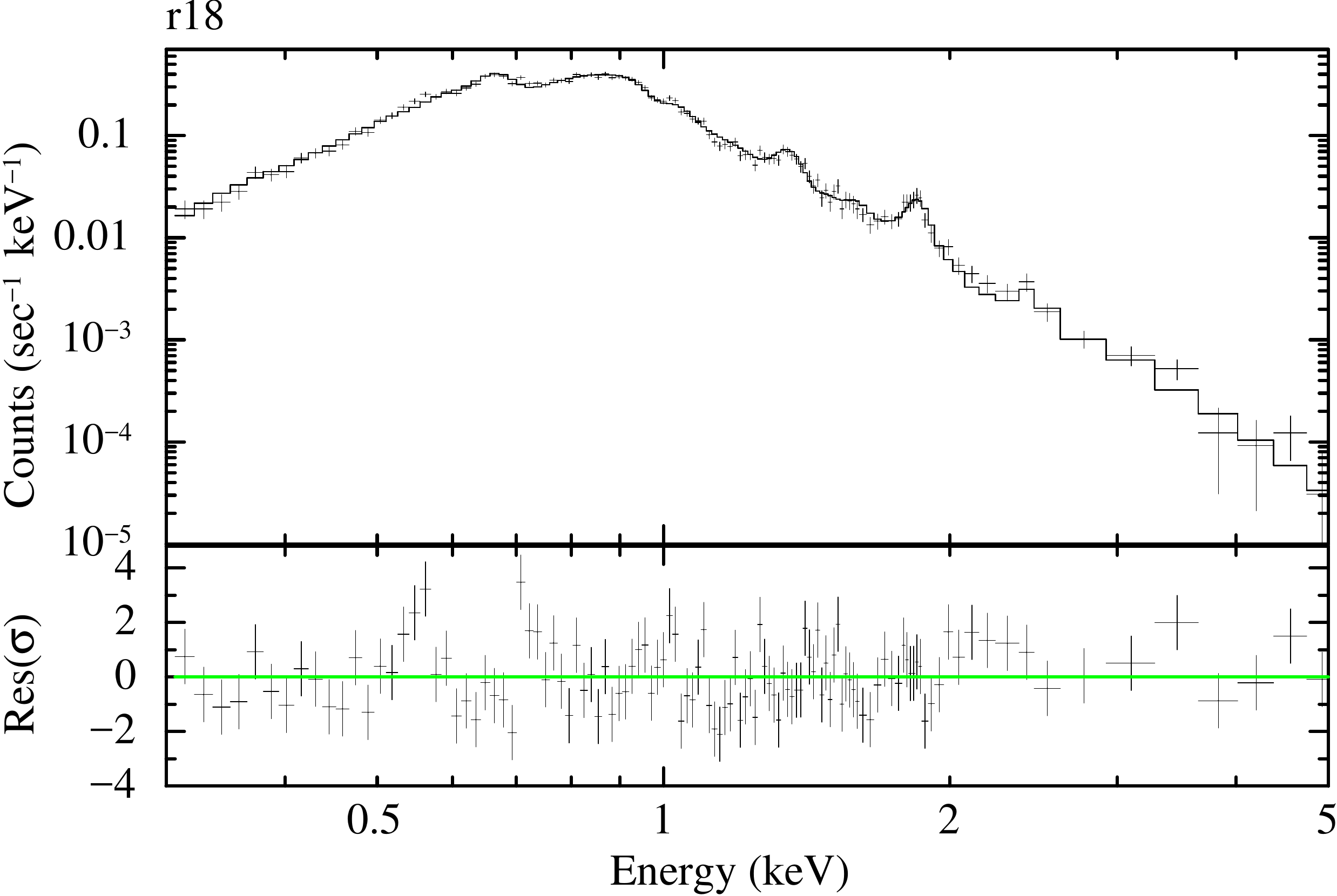}
\includegraphics[width=0.3\linewidth, angle=270]{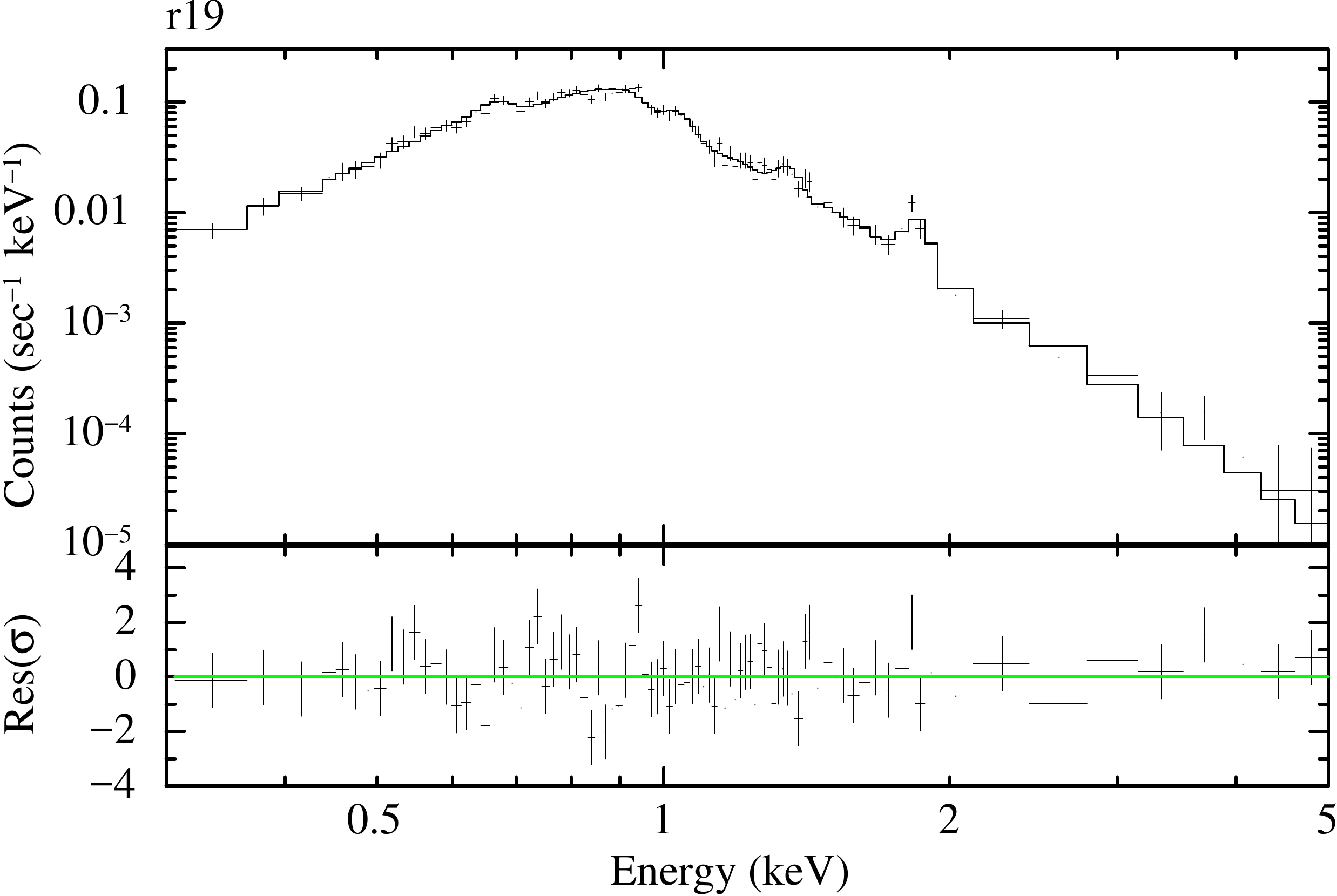}
\caption{Fits for regions $\mathrm{r18}$ and $\mathrm{r19}$ with the single \texttt{vpshock}.}
\label{fig:rims1819}
\end{figure}

\begin{figure}
\includegraphics[width=0.3\linewidth, angle=270]{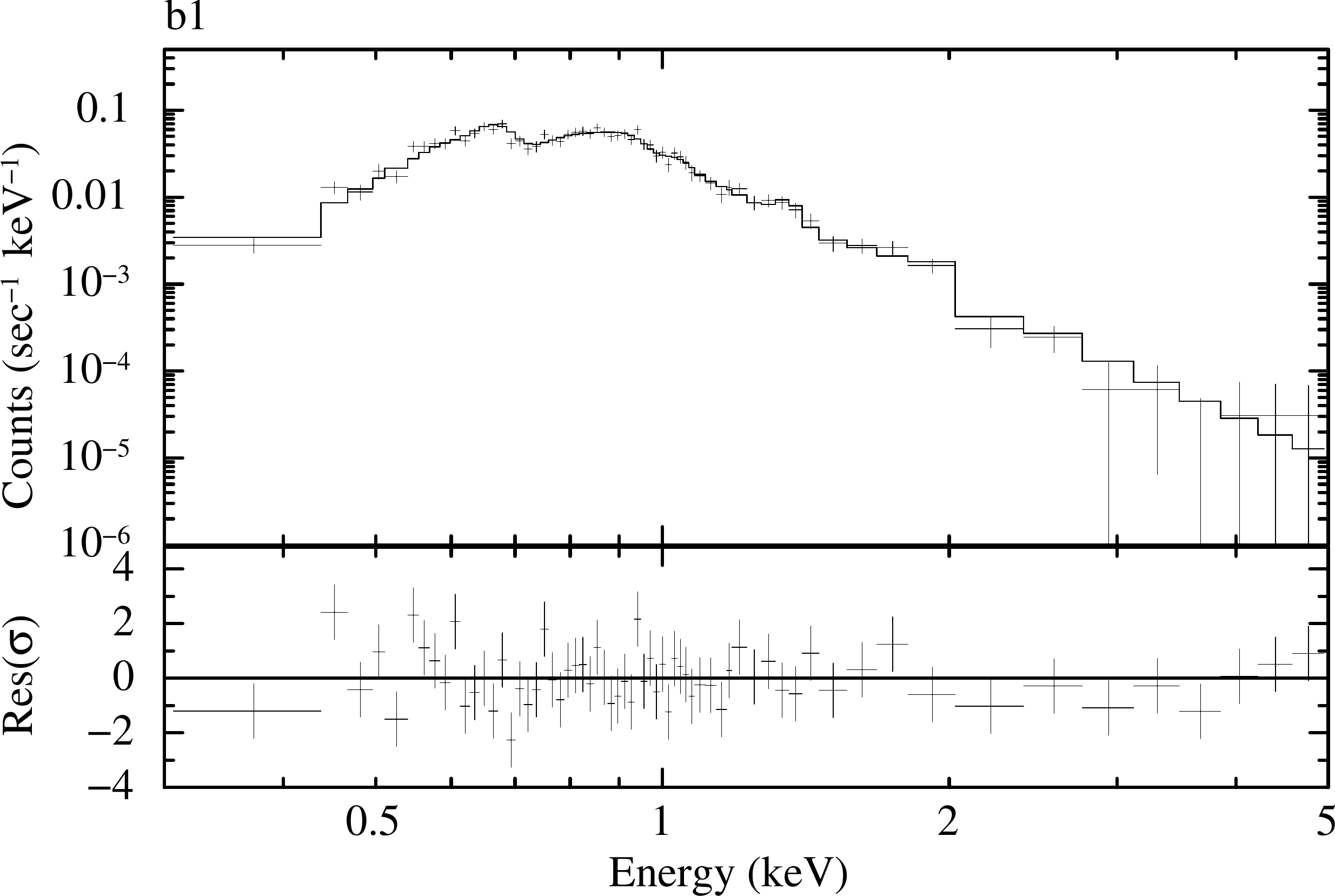}
\includegraphics[width=0.3\linewidth, angle=270]{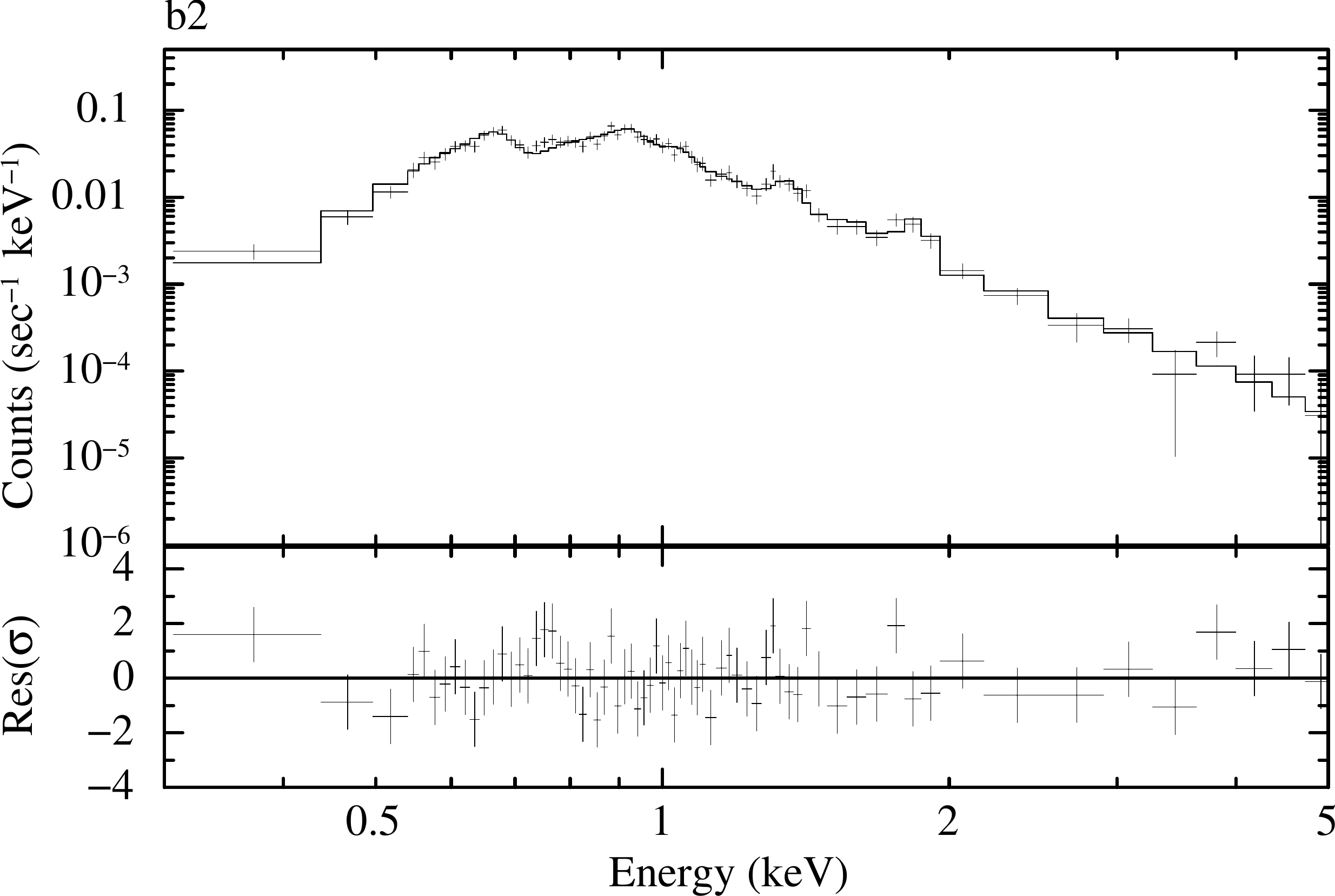}
\caption{Fits for blobs $\mathrm{b1}$ and $\mathrm{b2}$ protruding ahead of the western rim with the single \texttt{vpshock} model.}
\label{fig:blobs12}
\end{figure}

\begin{figure*}
{\includegraphics[angle=270, width=0.45\linewidth]{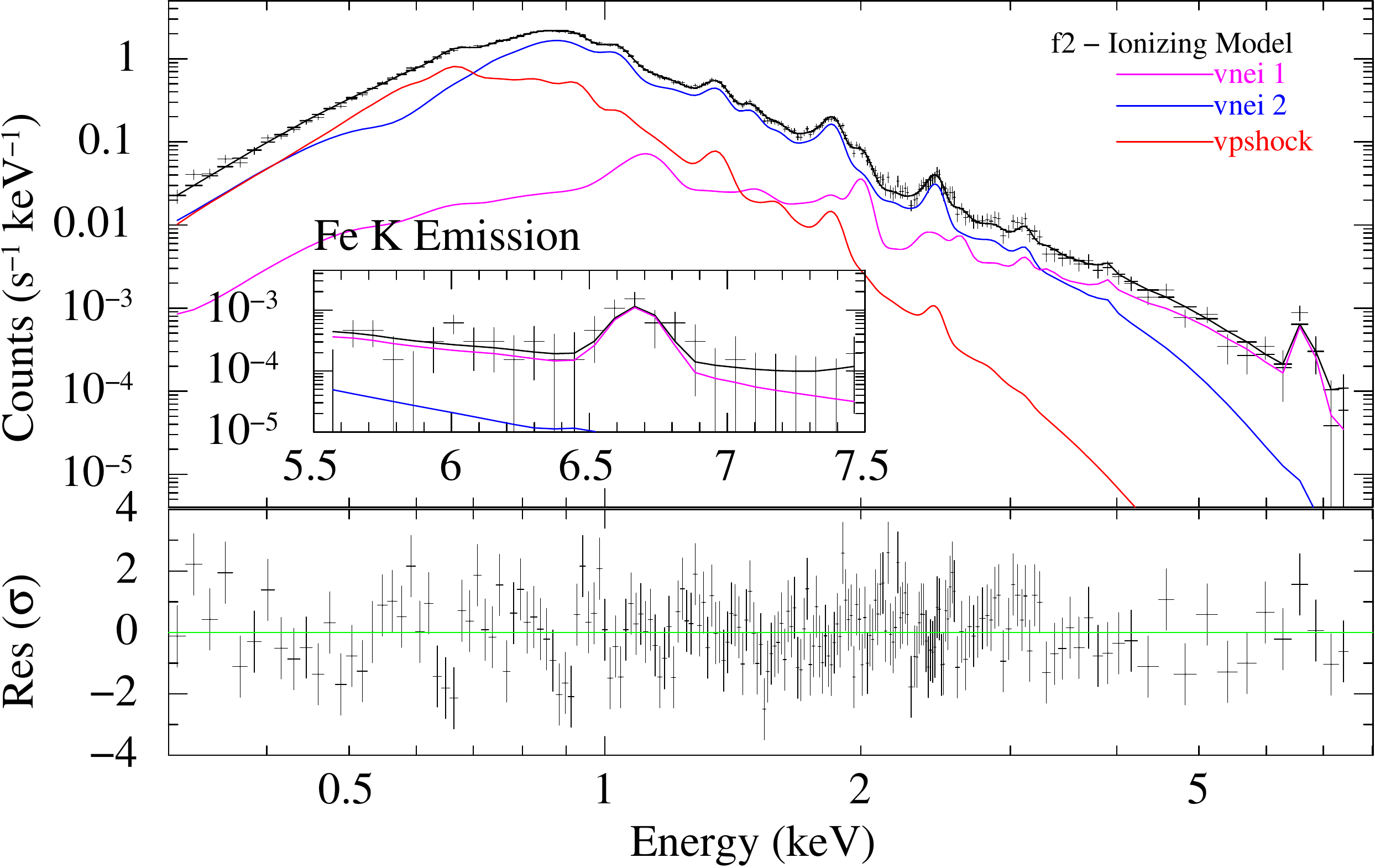}}
{\includegraphics[angle=270, width=0.45\linewidth]{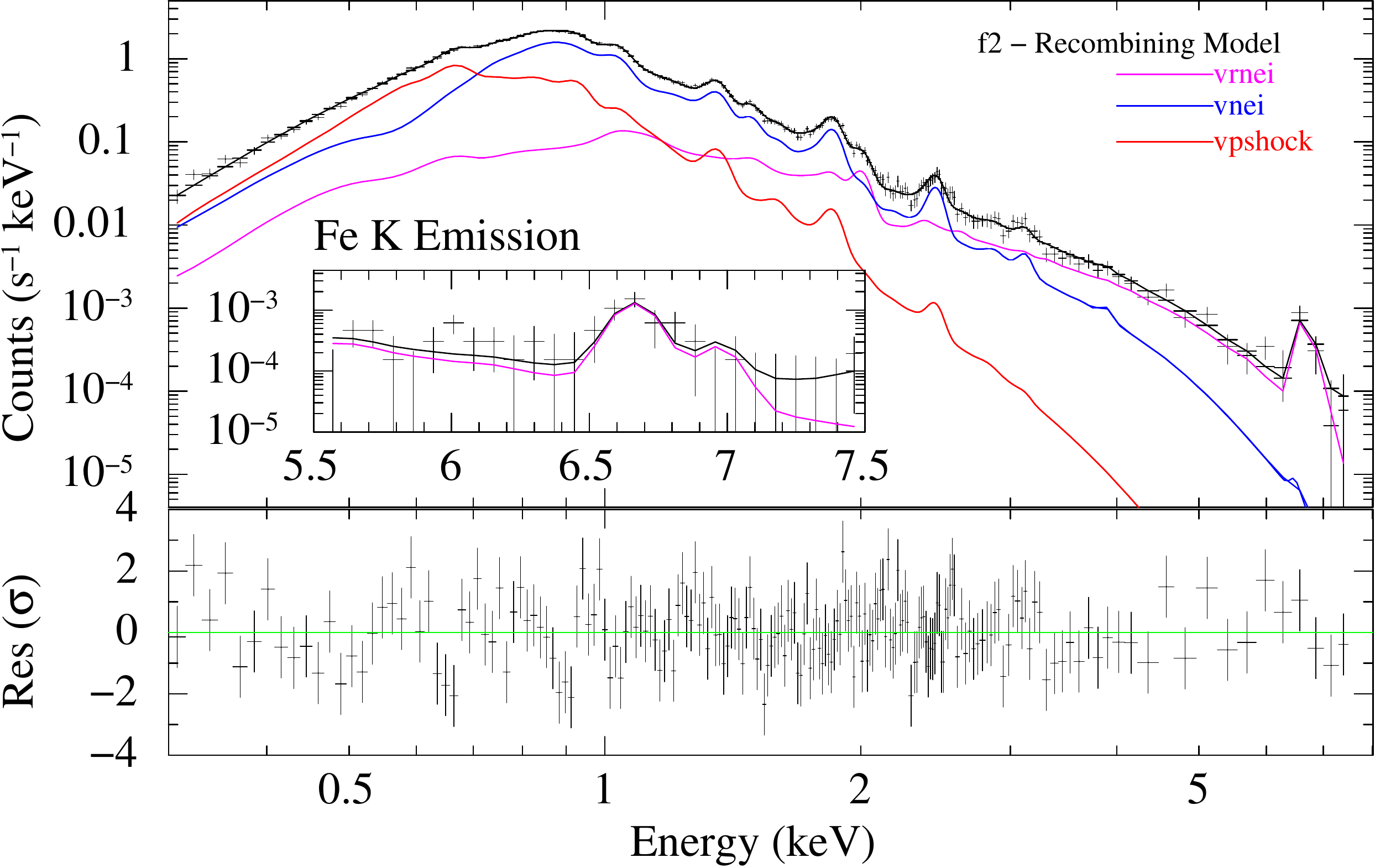}}
\caption{Same as Figure \ref{fig:fek_f1}, but for the region $\mathrm{f2}$.}
\label{fig:fek_f2}
\end{figure*}

\begin{figure*}
{\includegraphics[angle=270, width=0.45\linewidth]{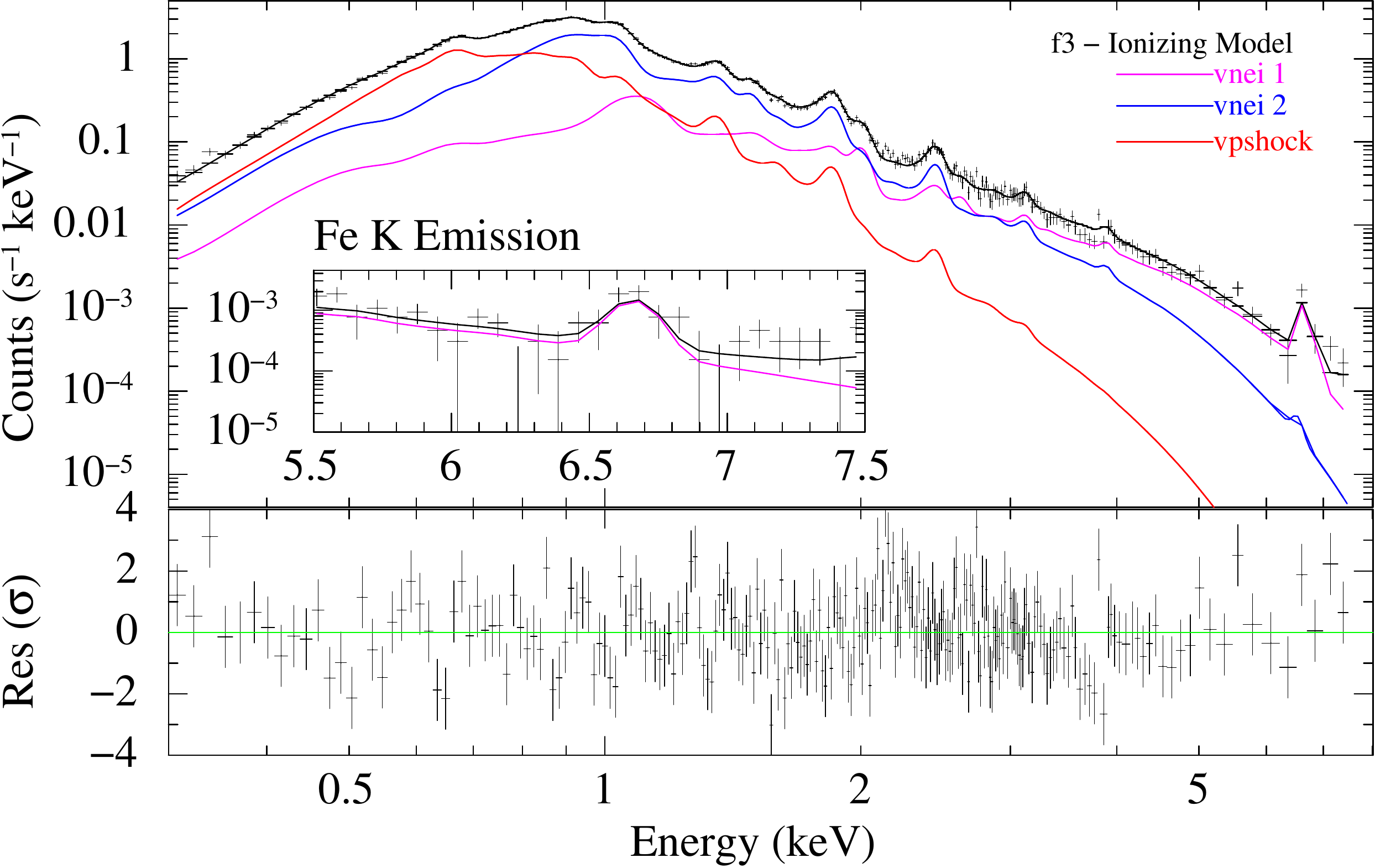}}
{\includegraphics[angle=270, width=0.45\linewidth]{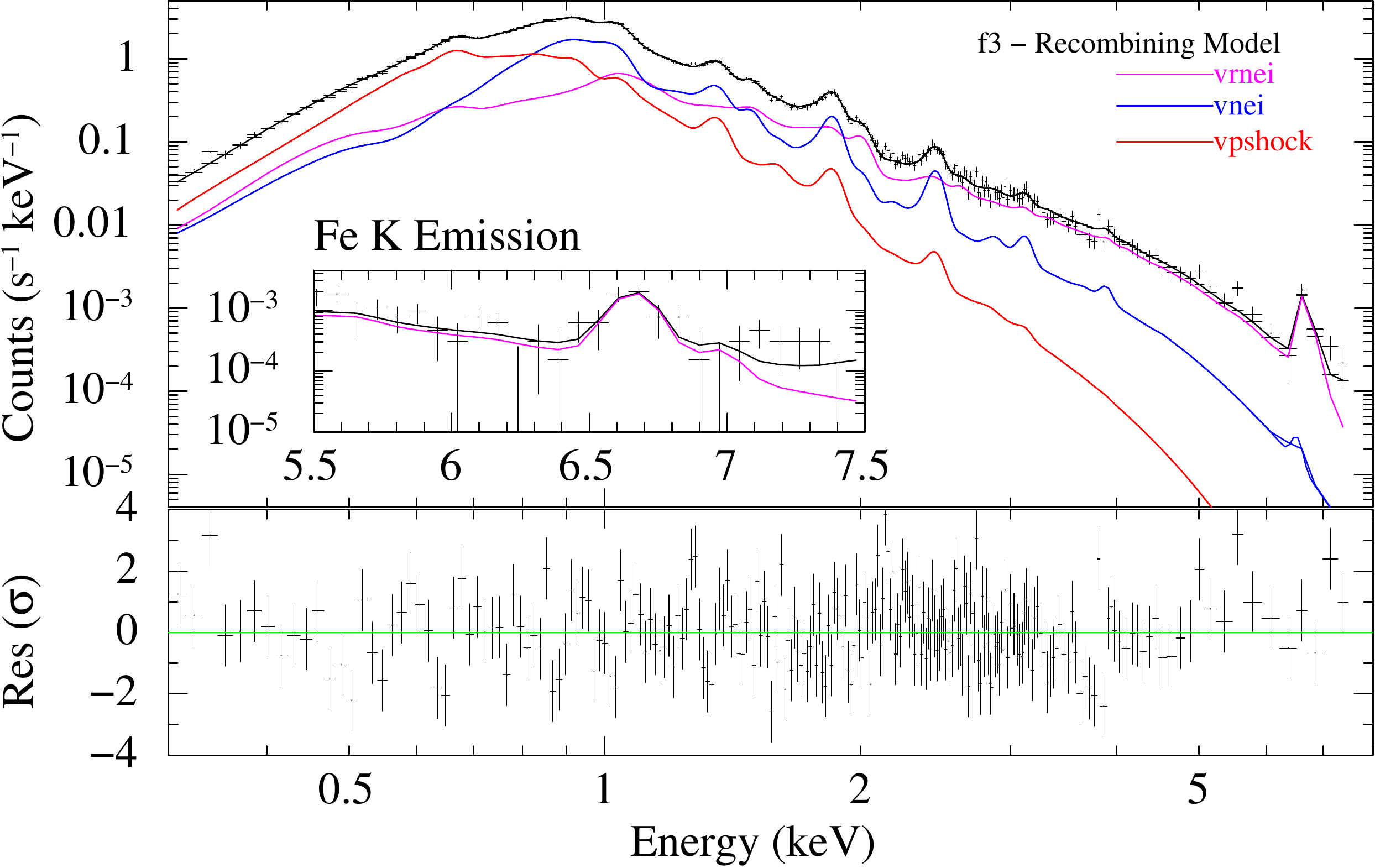}}
\caption{Same as Figure \ref{fig:fek_f1}, but for the region $\mathrm{f3}$.}
\label{fig:fek_f3}
\end{figure*}

\begin{figure*}
{\includegraphics[angle=270, width=0.45\linewidth]{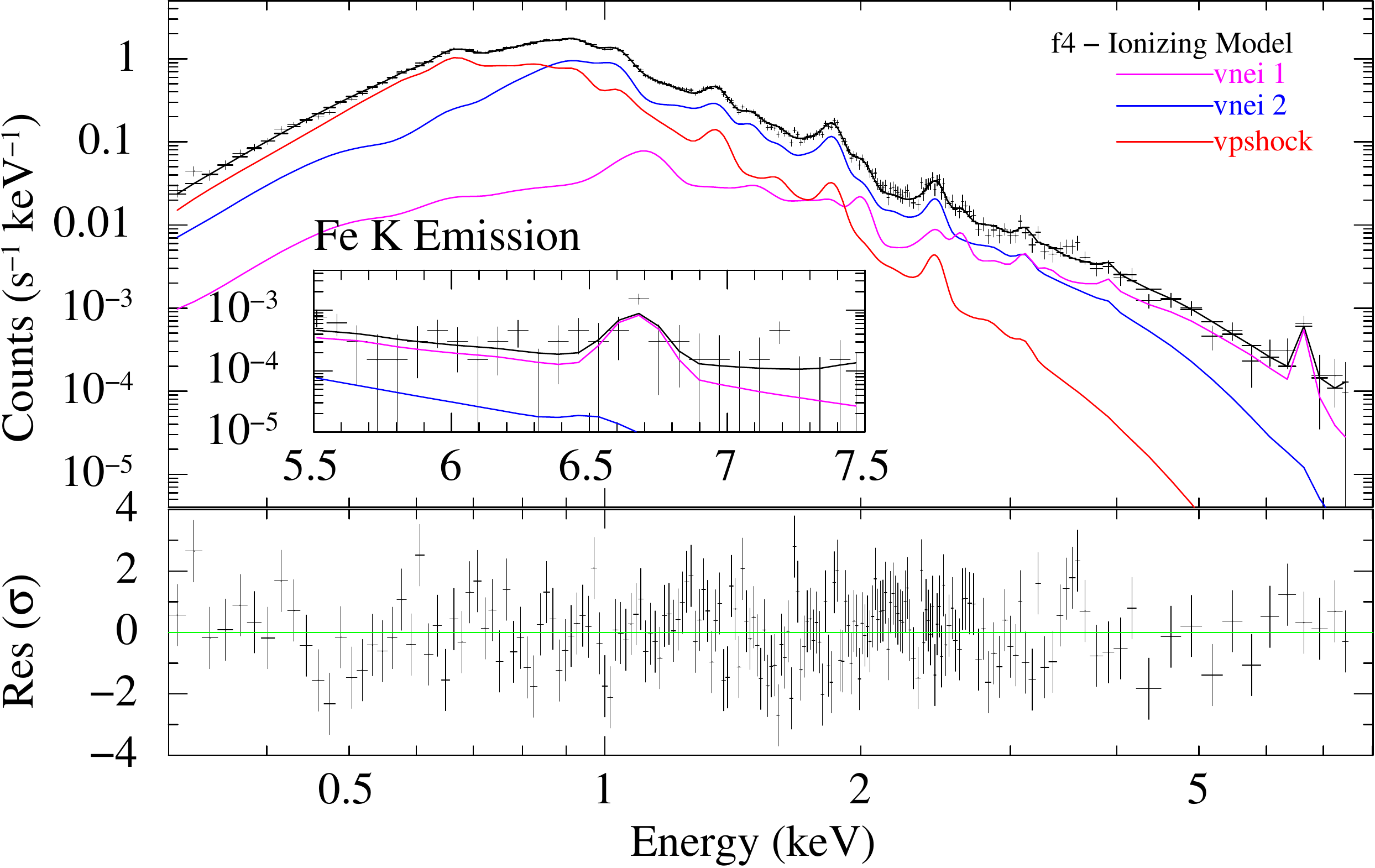}}
{\includegraphics[angle=270, width=0.45\linewidth]{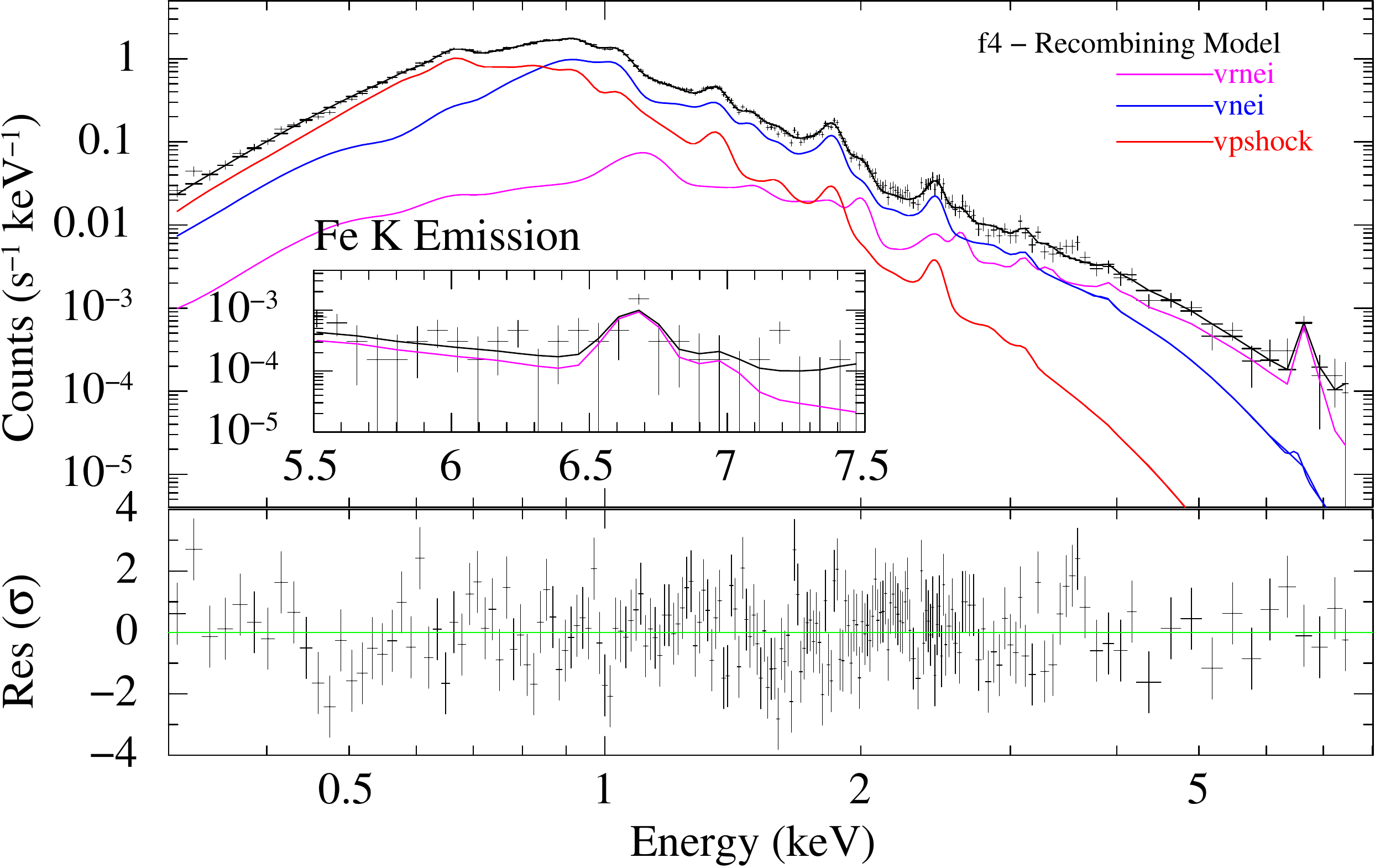}}
\caption{Same as Figure \ref{fig:fek_f1}, but for the region $\mathrm{f4}$.}
\label{fig:fek_f4}
\end{figure*}

\begin{figure*}
{\includegraphics[angle=270, width=0.45\linewidth]{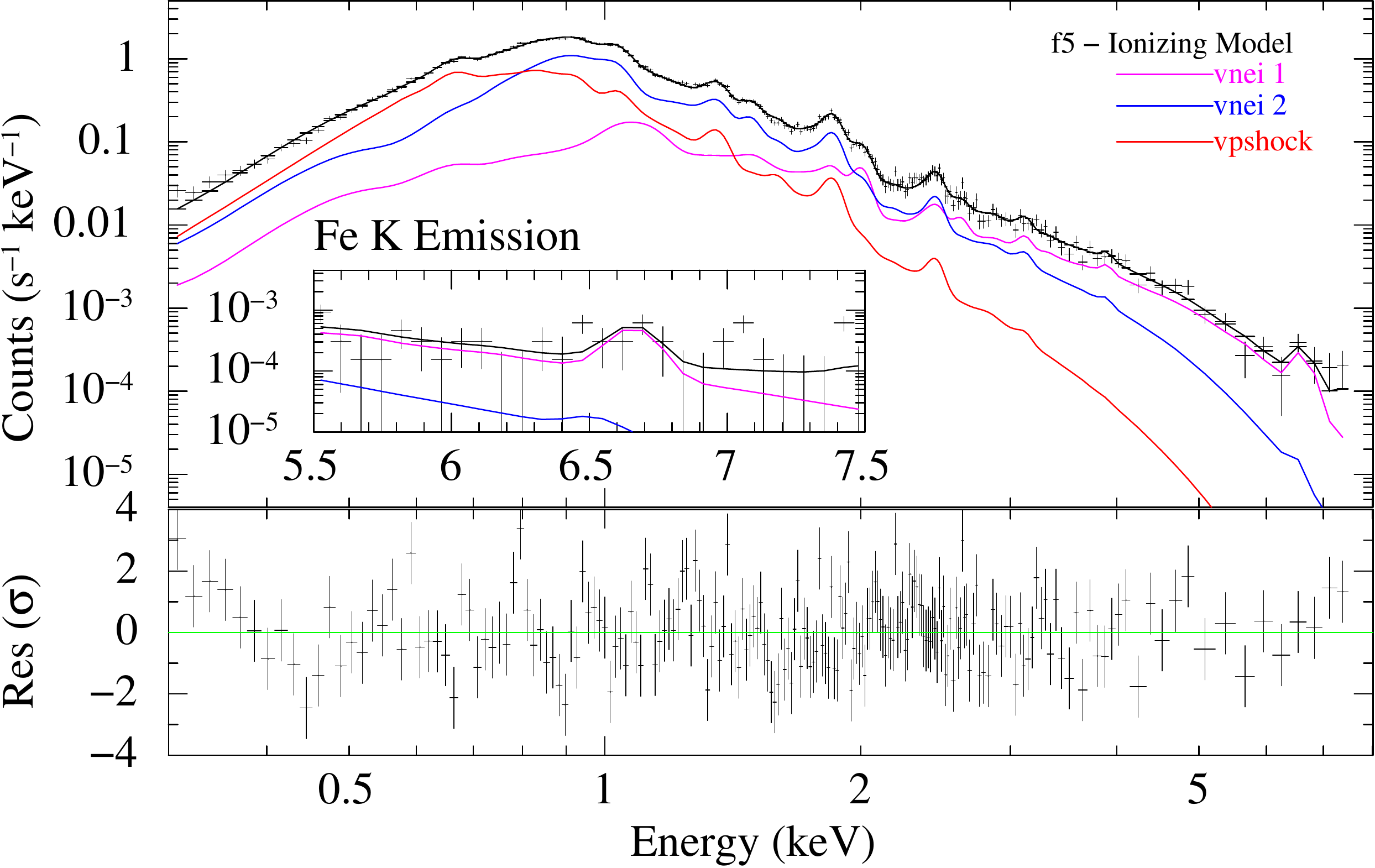}}
{\includegraphics[angle=270, width=0.45\linewidth]{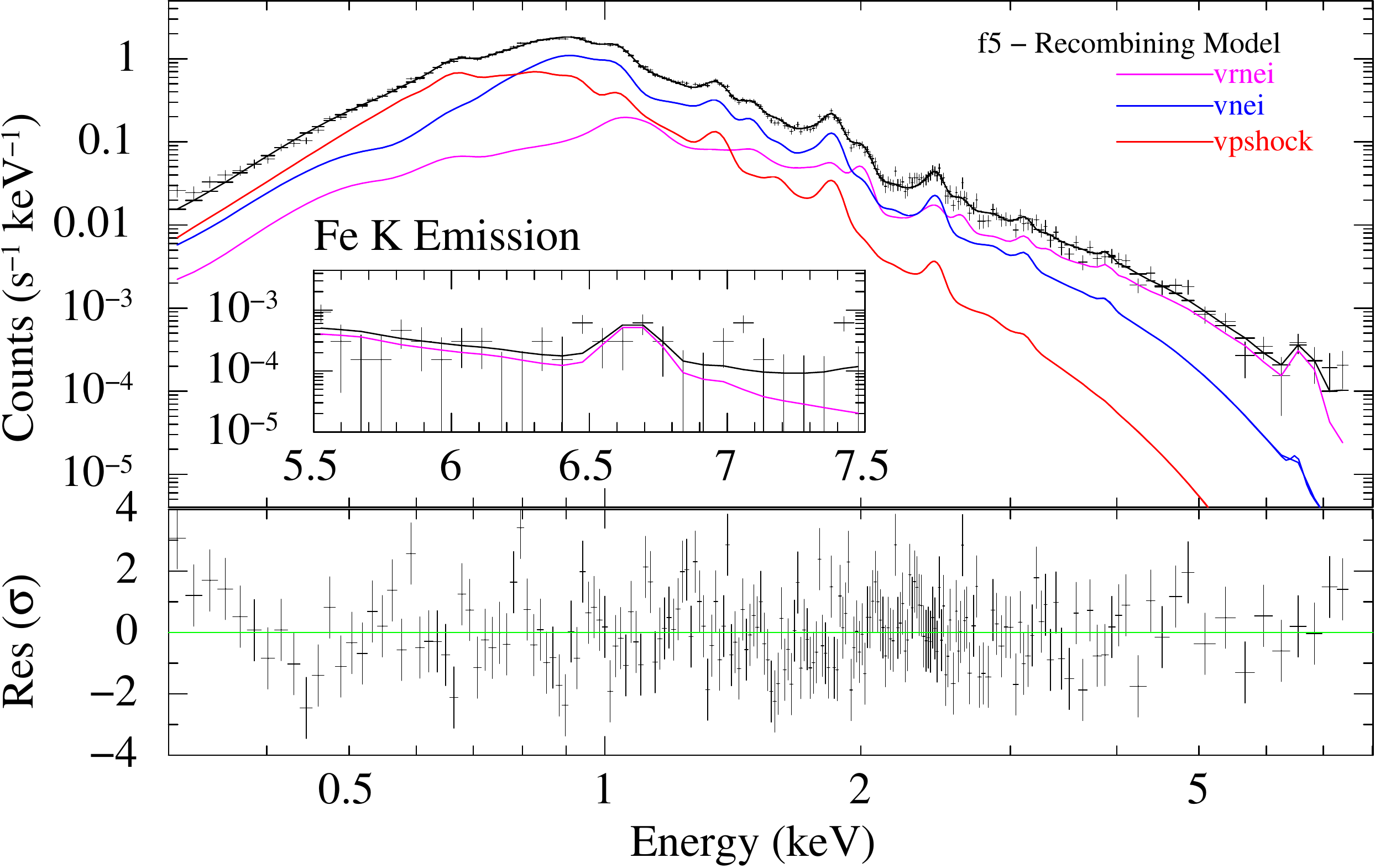}}
\caption{Same as Figure \ref{fig:fek_f1}, but for the region $\mathrm{f5}$.}
\label{fig:fek_f5}
\end{figure*}

\begin{figure*}
{\includegraphics[angle=270, width=0.5\linewidth]{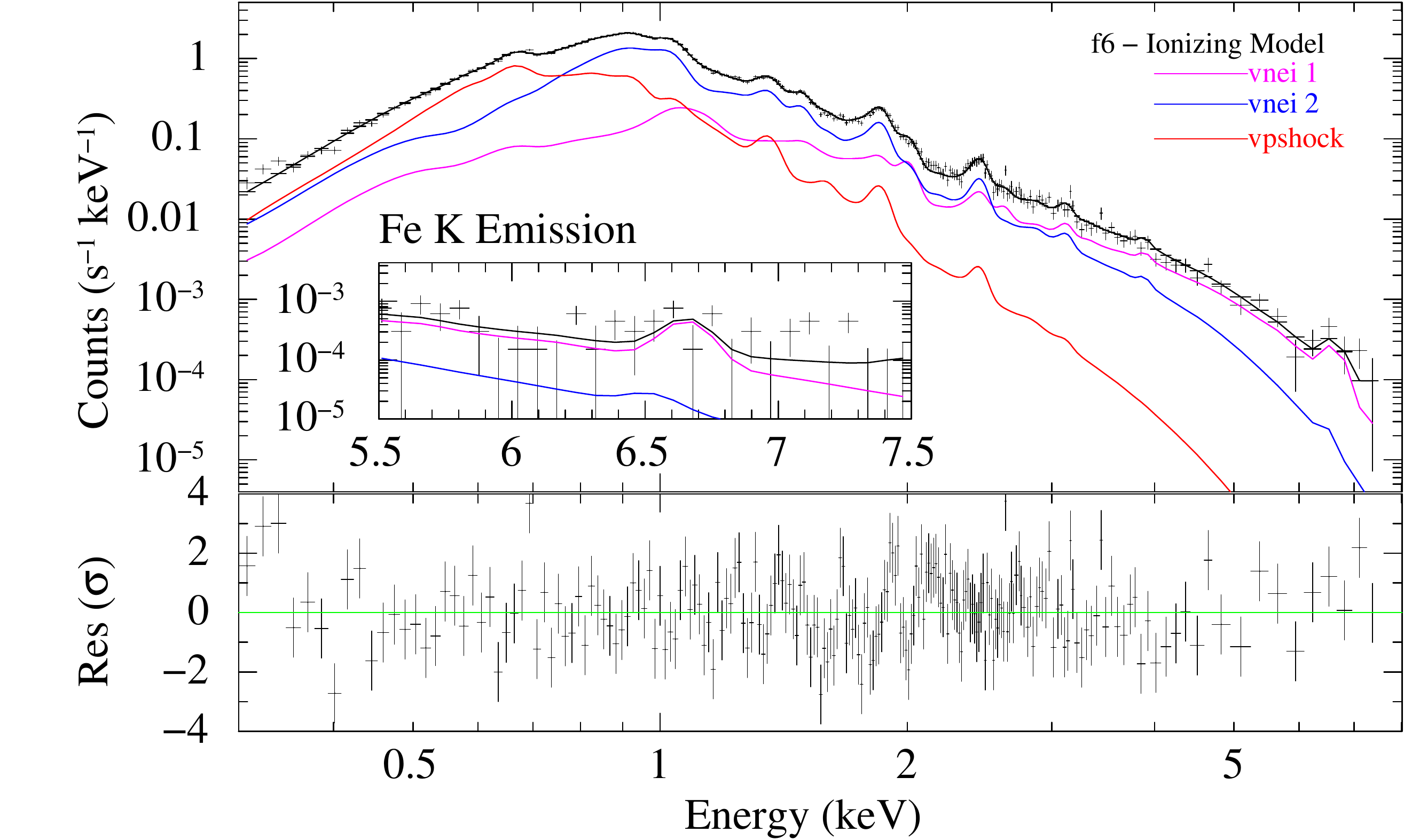}}
{\includegraphics[angle=270, width=0.5\linewidth]{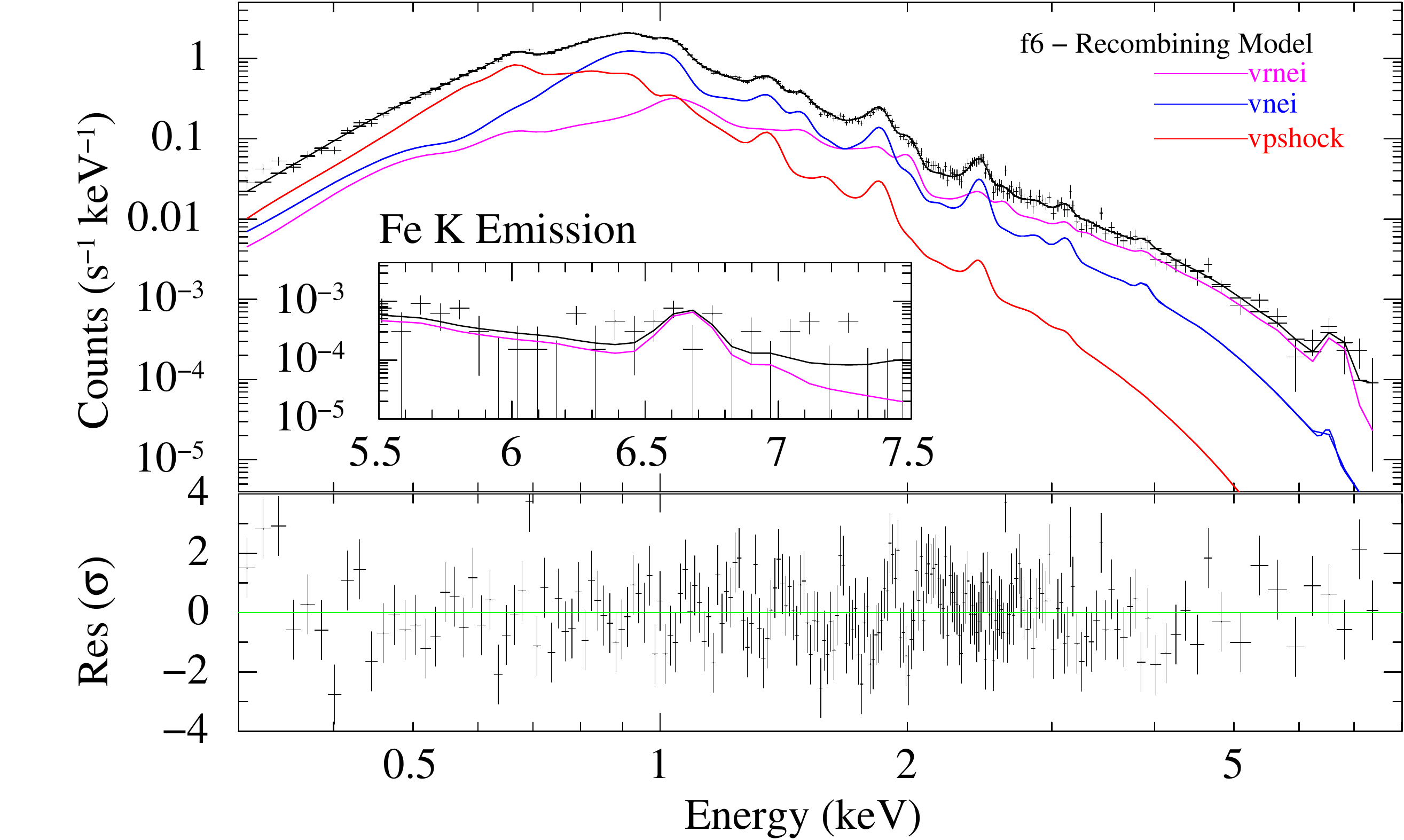}}
\caption{Same as Figure \ref{fig:fek_f1}, but for the region $\mathrm{f6}$.}
\label{fig:fek_f6}
\end{figure*}

Figures \ref{fig:rims23} to \ref{fig:blobs12} display the spectral fits and residuals for rim regions with the \texttt{vpshock} model (regions $\mathrm{r2}$, $\mathrm{r4-r9}$, $\mathrm{r11-r19}$, $\mathrm{b1}$ and $\mathrm{b2}$) and the \texttt{vnei+vpshock} model (regions $\mathrm{r3}$ and $\mathrm{r10}$). Figures \ref{fig:fek_f2} to \ref{fig:fek_f6} present the spectral fits for the Fe K regions. Note that the spectral counts in all the plots have been rebinned for display purposes.

\section{Electron density calculation}
\label{s:append_geometry}
\begin{figure}[bht!]
\centering
\includegraphics[width=0.5\linewidth]{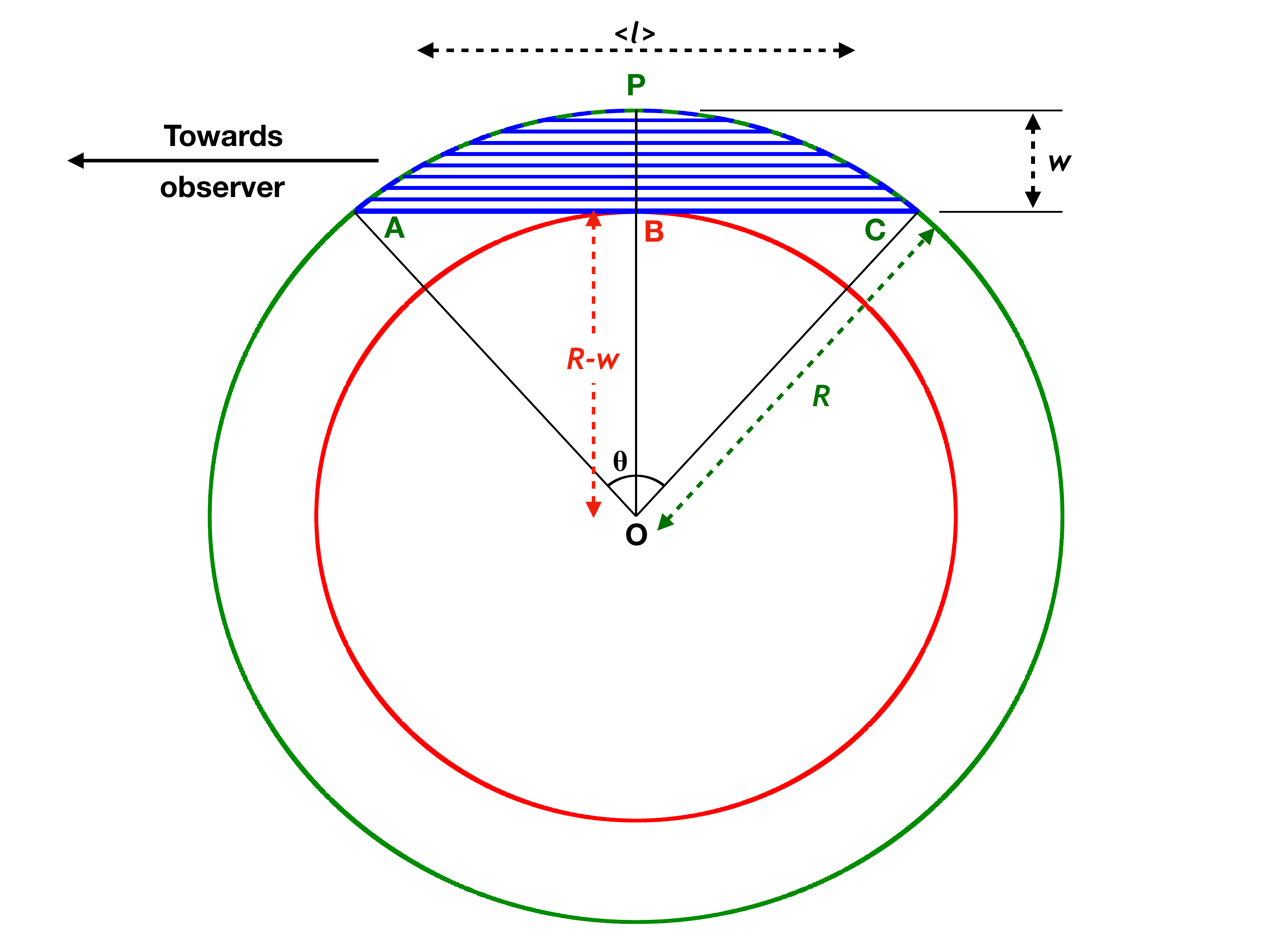}
\caption{Sketch of the geometry assumed for the calculation of the depth $\langle l \rangle$ of the regions used for spectral fitting along the rim. The blue shaded area, enclosed between the arc \texttt{APC} and the horizontal line \texttt{AC}, depicts the region along the line of sight, pictured in cross-section as part of a circular cylindrical shell defined by the green outer circle and the red inner circle. The opening angle \texttt{AOC}$\equiv\theta$ is determined by the width $w$ and the distance of the region from the center of the remnant $R$. The shaded area is the difference between the pie-shaped region \texttt{AOCPA} and the triangle \texttt{AOCBA}.}
\label{fig:sketchslicevolume}
\end{figure}

For thermal plasma models, the \xspec normalization, $\mathtt{norm}$, is proportional to emission measure as $\mathtt{norm} \propto \int dV n_\mathit{e} n_\mathrm{H} \sim \langle n_\mathit{e}^2 \rangle f \Delta V$. Here, we assume $n_\mathit{e} \sim 1.2\,n_\mathrm{H}$ (see, for example, \citealt{2014ApJ...791...50S}; however, this will be a lower limit if considerable quantity of metals is present), $\langle n_e^2 \rangle$ is an average $n_e^2$, $f$ is the volume filling factor for the emitting region, and $\Delta V$ is the volume corresponding to the extraction region, \textit{i.e.}, the projected area, $A = w h$, where $w$ and $h$ are the width and height of the extraction region, times an average line-of-sight depth, $\langle l \rangle$. To estimate the volume, some assumption is needed about the local three-dimensional structure. If the extraction region is assumed to be locally a projection through a figure of revolution with axis in the plane of the sky (see Figure \ref{fig:sketchslicevolume}), the cross-sectional area can be expressed as
\begin{equation}
A_{\mathrm{c}} =   R^2 \mathrm{cos}^{-1}\left({\frac{R-w}{w}}\right) - (R-w)~\sqrt{2Rw-w^2} \,,
\label{eqn:areadepth}
\end{equation}
where $R$ is the radius of the circular segment of width $w$. The average line-of-sight depth $\langle l \rangle$ can then be estimated by dividing this area by the width $w$ of the extraction region, and the volume is then estimated by $\langle l \rangle$ times the area of the extraction region. Thus, $n_\mathit{e} \sim \langle 1.2\,n_\mathit{e}^2 \rangle^{1/2} \sim 1106 \left(\mathtt{norm}/f\Delta V\right)^{1/2} \mathrm{cm}^{-3}$, where $\texttt{norm}$ is in units of $\mathrm{cm}^{-5}$, and $\Delta V$ is in units of $\mathrm{pc}^3$.

\end{document}